\documentclass[fleqn,usenatbib]{mnras}

\usepackage[utf8]{inputenc}
\usepackage{graphicx}
\usepackage{txfonts}
\usepackage[para,online,flushleft]{threeparttable}
\usepackage{xcolor}
\usepackage{xfrac}
\usepackage{subcaption}
\definecolor{redak}{rgb}{0.9,0.15,0.05}

\usepackage{svg}
\newcommand{\orcid}[1]{\href{https://orcid.org/#1}{\includesvg[width=10pt]{orcid}}}

\defcitealias{deJager1988}{dJ88}
\defcitealias{NL00}{NL00}
\defcitealias{Vink2001}{V01}
\defcitealias{Vink2017}{V17}

\title[Hydrogen in progenitors of Type Ib/IIb SNe]{How much hydrogen is in Type Ib and IIb supernova progenitors?}
\author[Gilkis \& Arcavi]{Avishai~Gilkis$^{\orcid{0000-0001-8949-5131}}$ $^{1}$\thanks{\href{agilkis@tauex.tau.ac.il}{agilkis@tauex.tau.ac.il}} and Iair~Arcavi$^{\orcid{0000-0001-7090-4898}}$ $^{1,2}$\thanks{\href{arcavi@tauex.tau.ac.il}{arcavi@tauex.tau.ac.il}}
\\
$^{1}$ The School of Physics and Astronomy, Tel Aviv University, Tel Aviv 6997801, Israel\\
$^{2}$ CIFAR Azrieli Global Scholars program, CIFAR, Toronto, M5G 1M1, Canada \\
}

\pubyear{2022}

\begin{document}
\label{firstpage}
\pagerange{\pageref{firstpage}--\pageref{lastpage}}

\maketitle

\begin{abstract}
     Core-collapse supernovae showing little or no hydrogen (denoted by Type~IIb and Ib, respectively) are the explosions of massive stars that have lost some or most of their outer envelopes. How they lose their mass is unclear, but it likely involves binary interaction. So far, seven progenitors of such supernovae have been identified in pre-explosion imaging (five for Type~IIb events and two for Type~Ib events). Here, we evolve detailed binary stellar evolution models in order to better understand the nature of these progenitors. We find that the amount of hydrogen left in the envelope at the time of explosion greatly depends on the post-interaction mass-loss rate. The leftover hydrogen, in turn, strongly affects progenitor properties, such as temperature and photospheric radius, in non-trivial ways. Together with extinction and distance uncertainties in progenitor data, it is difficult to deduce an accurate progenitor hydrogen mass from pre-explosion imaging. We quantify this uncertainty and find that available data are consistent with a proposed Type Ib--IIb hydrogen mass threshold of $\approx 0.033\,\mathrm{M}_\odot$, implying that even Type Ib progenitors are not pure helium stars. These results alleviate the proposed tension between the Type Ib classification of SN 2019yvr and its candidate progenitor properties. We also estimate the brightness of a surviving 2019yvr progenitor companion, which might be detected in future observations.
\end{abstract}


\section{Introduction}
\label{sec:intro}

Type~IIb supernovae (SNe) are explosive transients in which broad hydrogen lines are initially detected but then disappear, leading to a Type~Ib SN appearance, for which no hydrogen is detected at all \citep{Filippenko1988,Nomotoetal1993}. The explanation for this observed phenomenon is that a hydrogen envelope of a very low mass is present in Type~IIb progenitors, while for Type~Ib progenitors the envelope contains even less hydrogen, or none at all \citep{Dessartetal2011}.

A likely mechanism for removing the hydrogen envelope, or part of it, is the interaction between a massive star in a binary system and its companion (\citealt*{Podsiadlowskietal1992,Yoonetal2010}; \citealt{Claeysetal2011}; \citealt*{Yoonetal2017}; \citealt{Lohevetal2019}, \citealt*{Sravanetal2019}, \citealt{Naimanetal2020}). The minimal hydrogen mass which would give rise to a Type~IIb appearance (vs. a Type~Ib) is uncertain, with estimates varying between even a hydrogen mass of $0.001\,\mathrm{M}_\odot$ giving rise to a IIb appearance \citep{Dessartetal2011} to a Ib--IIb threshold mass of $M_\mathrm{H,min,IIb}\approx 0.033\,\mathrm{M}_\odot$ \citep{Hachinger2012}.

The nature of the progenitors of Type~Ib and Type~IIb SNe can be constrained by using  pre-explosion photometry, when available. A total of five Type~IIb SN progenitors have been identified, from SN~1993J (\citealt{Podsiadlowskietal1993}; \citealt*{Alderingetal1994}) to the more recent SN~2016gkg \citep{Arcavietal2017,Kilpatricketal2017,Tartagliaetal2017,Berstenetal2018}. The progenitors of Type~IIb SNe are consistent with cool supergiants (CSGs) with low-mass hydrogen envelopes \citep{Yoonetal2017}.

The first Type~Ib SN with an identified progenitor was iPTF13bvn \citep{Caoetal2013}. Although initially thought to be a Wolf-Rayet (WR) star (\citealt{Caoetal2013}; \citealt*{Grohetal2013}), follow-up studies soon favored a star stripped by binary interaction \citep{Fremlingetal2014,Eldridgeetal2015,Folatellietal2016}, with a final mass lower than typical WR stars \citep{Berstenetal2014}. According to binary evolution models, the progenitor of iPTF13bvn is consistent with a helium giant \citep{Berstenetal2014,Eldridgeetal2015,EldridgeMaund2016}, which is hotter and contains less hydrogen than the progenitors of Type~IIb SNe (or no hydrogen at all).

A recently identified progenitor candidate for the Type~Ib SN~2019yvr \citep{Kilpatricketal2021} somewhat complicates matters (though post-supernova photometry is still required to confirm the progenitor identification and its properties). While no hydrogen features are observed in the SN itself, the progenitor candidate is cool and large, similar to progenitors of Type~IIb SNe. According to \cite{Kilpatricketal2021}, no hydrogen-free progenitor models can account for the pre-explosion observed photometry, and all models which do fit the pre-SN observations have more hydrogen than would be enough to present type~II SN features according to the simulations of both \cite{Dessartetal2011} and \cite{Hachinger2012}.

Here we reevaluate the analysis of the progenitor of SN~2019yvr and all Type~Ib and Type~IIb SN progenitors in a unified framework. We use detailed binary stellar-evolution simulations combined with synthetic photometry to find the best-fitting models for all progenitors in a uniform way, and we compare their properties.

In Section \ref{sec:method} we describe the main aspects of the stellar evolution simulations and the generation of synthetic photometry from the evolutionary endpoints. In Section \ref{sec:obsdat} we describe the observational data that we fit our computed models to. In Section \ref{sec:results} we present our main findings, and discuss them in comparison to earlier works in Section \ref{sec:discussion}. We summarize in Section \ref{sec:summary}.

\section{Numerical method}
\label{sec:method}

\subsection{Stellar evolution}
\label{subsec:StellarEvolution}

We use the Modules for Experiments in Stellar Astrophysics code (\textsc{mesa}, version 10398, \citealt{Paxton2011,Paxton2013,Paxton2015,Paxton2018}) to evolve stellar models. The methodology is the same as that of \cite{GilkisVinkEldridgeTout2019}, where more details can be found. In this work we expand the parameter space of the initial conditions, and describe the main aspects of the evolution.

\begin{table}
\centering
\caption{Initial masses for stellar evolution calculations.}
\begin{threeparttable}
\begin{tabular}{c|cccc}
\hline
$M_1/ \mathrm{M}_\odot$ & $M_2/ \mathrm{M}_\odot$ &  $M_2/ \mathrm{M}_\odot$ & $M_2/ \mathrm{M}_\odot$ & $M_2/ \mathrm{M}_\odot$ \\
\hline
$11$ & $10$ & $9$  & $7$  & $4$ \\
$12$ & $11$ & $10$  & $8$  & $5$ \\
$13$ & $12$ & $11$  & $8$  & $5$ \\
$14$ & $13$ & $12$  & $9$  & $5$ \\
$16$ & $15$ & $14$  & $10$ & $6$ \\
$19$ & $18$ & $16$  & $12$ & $7$ \\
$22$ & $21$ & $19$  & $14$ & $8$ \\
$25$ & $24$ & $22$  & $16$ & $9$ \\
\hline
\hline
\end{tabular}
\footnotesize
\begin{tablenotes}
Note. The second column lists the companion masses for simulations with a mass ratio within $0.9 < Q < 1$, the third column lists the companion masses for simulations with $0.8 < Q < 0.9$, the fourth column lists the companion masses for $0.6 < Q < 0.7$ and the fifth column lists the companion masses for simulations with a mass ratio within $0.35 < Q < 0.45$.
\end{tablenotes}
\end{threeparttable}
\label{tab:masses}
\end{table}
We evolved stellar models with initial primary masses of $M_1 / \mathrm{M}_\odot \in \{11,12,13,14,16,19,22,25\}$, initial orbital periods of $P_\mathrm{i} / \mathrm{d} \in \{5,10,18,33,60,110,201,367,669,1219,2223\}$, and four mass ratio ranges resulting in companion masses as listed in Table \ref{tab:masses}. The metallicity in our models is $Z=0.019$, which is the Solar value according to \cite{AndersGrevesse1989}. While later studies revised the Solar metallicity downward \citep{Asplundetal2009}, the differences are small and would not impact our results and conclusions.

Both stars are evolved until the primary reaches the end of core carbon burning. We do not follow the evolution of the companion afterwards. We only consider SN progenitors which started as the more massive star in the binary system. When considering the post-SN properties of the remaining companion we assume that it is unchanged, and that it can be described by its properties at the time of the SN.

\subsubsection{Microphysics}
\label{subsubsec:mesa}

The equation of state (EOS) employed by \textsc{mesa} is a blend of the following equations of state: OPAL \citep{Rogers2002}, SCVH \citep*{Saumon1995}, HELM \citep{Timmes2000}, and PC \citep{Potekhin2010}. Radiative opacities are taken primarily from OPAL \citep{Iglesias1993, Iglesias1996}, with low-temperature data taken from \citet{Ferguson2005} and the high-temperature, Compton-scattering dominated regime, calculated according to \citet{Buchler1976}. Electron conduction opacities follow \citet{Cassisi2007}.

We  use  the  built-in \textsc{mesa} nuclear reaction network \texttt{approx21}. Nuclear reaction rates are a combination of the Nuclear Astrophysics Compilation of Reaction rates \citep[NACRE, ][]{Angulo1999} and the Joint Institute for Nuclear Astrophysics (JINA) REACLIB reaction rates \citep{Cyburt2010}, with additional tabulated weak reaction rates (\citealt*{Fuller1985}; \citealt{Oda1994, Langanke2000}) and screening via the prescriptions of \citet{Salpeter1954}, \citet*{Dewitt1973}, \citet{Alastuey1978} and \citet{Itoh1979}. The formulae of \citet{Itoh1996} are used for thermal neutrino loss rates.

\subsubsection{Wind mass loss}
\label{subsubsec:winds}

For hot (effective surface temperatures of $T_\mathrm{eff} \ge 11000\,\mathrm{K}$) phases of the evolution, wind mass loss follows the theoretical prescription of \citet*[][hereafter V01]{Vink2001} if the surface hydrogen mass fraction $X_\mathrm{s}$ is high, $X_\mathrm{s} \ge 0.4$. For hydrogen-deficient envelopes with $X_\mathrm{s} < 0.4$ we use either the empirical mass-loss rate relation of \citet[][hereafter NL00]{NL00} or the theoretical recipe provided by \citet[][hereafter V17]{Vink2017}, so that each evolution track which reaches $X_\mathrm{s} < 0.4$ is simulated twice, once with each of the prescriptions.

For cool ($T_\mathrm{eff} \le 10000\,\mathrm{K}$) phases of the evolution, the empirical relation given by \citet*[][hereafter dJ88]{deJager1988} is employed. For $10000\,\mathrm{K} < T_\mathrm{eff} < 11000\,\mathrm{K}$ the wind mass-loss rate is interpolated between the hot and cool prescriptions.

\subsubsection{Mass transfer efficiency}
\label{subsubsec:masstransfer}

Rather than assume an arbitrary constant mass transfer efficiency, we employ a physically motivated prescription which continuously updates the mass transfer efficiency during the stellar evolution computation, according to the ability of the companion star to accrete mass. More details are given by \cite{GilkisVinkEldridgeTout2019}.

\subsubsection{Mixing}
\label{subsubsec:mixing}

The Ledoux stability criterion is used to define convective regions, where mixing is treated according to mixing-length theory \citep*[MLT;][]{MLT1958,MLT} with a mixing-length parameter of $\alpha_\mathrm{MLT}=1.5$. Overshooting above convective regions follows the exponentially decaying prescription of \cite{Herwig2000}, with a decay scale of $f_\mathrm{ov} H_P$, where  $f_\mathrm{ov}=0.016$ and $H_P$ is the pressure scale height. We employ the MLT++ treatment of \textsc{mesa} for superadiabatic convection \citep{Paxton2013}.

\subsection{Synthetic photometry generation}
\label{subsec:SyntheticPhotometry}

In order to compare the endpoints of our stars to pre-SN observations, we generate synthetic photometry for the combined flux contribution of both components at each endpoint. We do not assume that the star is a blackbody at the end of its life, but instead we associate a spectrum to each stellar endpoint. For endpoints with associated spectra, we determine the synthetic photometry using \textsc{synphot} \citep{synphot2018}. To associate a spectrum to each stellar endpoint, we divide the stellar endpoints into three different regimes - cool stars, hot stars, and Wolf-Rayet (WR) stars. 

For the cool-star regime, defined as $T_\mathrm{eff} < 15000\,\mathrm{K}$, we use the stellar spectral flux library presented by \cite{Pickles1998}. Because the endpoints of our stellar evolution simulations all have high luminosity ($L > 30000\,\mathrm{L}_\odot$; Section \ref{subsec:EvolutionEndPoints}, Figure \ref{fig:HRD1}), we only use the spectra of luminosity class I, with the addition of the M10~III spectrum to cover the lowest temperatures. This selection results in a sequence of spectra which is monotonic in temperature\footnote{\cite{Pickles1998} assigned effective temperatures using a colour-temperature relation.}. For endpoints with temperatures not in the library, we determine the synthetic photometry using a one-dimensional interpolation between the photometry generated from spectra of the nearest available temperatures. 

For the hot-star regime ($15000\,\mathrm{K} \le T_\mathrm{eff} \le 55000\,\mathrm{K}$) we use the synthetic spectra computed with the \textsc{tlusty} code \citep{TlustyO2003,TlustyB2007}. This regime uses two \textsc{tlusty} grids. For $15000\,\mathrm{K} \le T_\mathrm{eff} \le 30000\,\mathrm{K}$ the BSTAR2006 Galactic metallicity grid with a microturbulent velocity of $V_\mathrm{t} = 2\,\mathrm{km}\,\mathrm{s}^{-1}$ is used \citep{TlustyB2007}. For $30000\,\mathrm{K} < T_\mathrm{eff} \le 55000\,\mathrm{K}$ the OSTAR2002 Galactic metallicity grid with a microturbulent velocity of $V_\mathrm{t} = 10\,\mathrm{km}\,\mathrm{s}^{-1}$ is used \citep{TlustyO2003}. Each grid contains computed spectra for several different effective temperatures and surface gravity values, $g$. Synthetic photometry for stellar evolution endpoints with no corresponding spectrum in the grid are calculated using a two-dimensional interpolation between the nearest four spectra with pairs of $T_\mathrm{eff}$ and $g$ values in the grid. Some of the endpoints in our simulations have a surface gravity slightly lower than the lowest value for which a \textsc{tlusty} synthetic spectrum is available. For these cases we perform a one-dimensional interpolation between nearby available points in the grid (see Appendix \ref{sec:appendixa}).

For the WR regime ($55000\,\mathrm{K} < T_\mathrm{eff}$) we use synthetic spectra computed with the \textsc{powr} code (\citealt*{PoWR2002}; \citealt{PoWR2003,PoWR2015}). We use the MW~WNE and MW~WNL-H20 grids presented by \cite{Todt2015}. The former are used for models with $X_\mathrm{s} < 0.05$, and the latter for $X_\mathrm{s} \ge 0.05$. Each grid contains synthetic spectra for several effective temperatures and transformed radii \citep*{Schmutz1989} defined as
\begin{equation}
R_\mathrm{t}=R_* \left(\frac{v_\infty / 2500\,\mathrm{km}\,\mathrm{s}^{-1}}{\sqrt{D} \dot{M} / 10^{-4}\,\mathrm{M}_\odot\,\mathrm{yr}^{-1}}\right)^{2/3},
    \label{eq:Rt}
\end{equation}
where $R_*$ is the stellar radius, $v_\infty$ the terminal wind velocity, $\dot{M}$ the wind mass-loss rate and $D$ the so-called clumping factor which allows for an inhomogeneous wind density. Models with $X_\mathrm{s} < 0.05$ (MW~WNE grid) have $v_\infty = 1600\,\mathrm{km}\,\mathrm{s}^{-1}$, while models with $X_\mathrm{s} \ge 0.05$ (MW~WNL-H20 grid) have $v_\infty = 1000\,\mathrm{km}\,\mathrm{s}^{-1}$. All models use $D=4$. The stellar evolution endpoints are assigned magnitudes by a two-dimensional interpolation between the nearest four pairs of $T_\mathrm{eff}$ and $R_\mathrm{t}$.

Each spectrum is subjected to extinction according to the reddening model given by \cite*{CCM1989}. We vary the extinction parameter $A_V$ from $0.035$ to $3.5$ in steps of $0.035$, and the reddening-law parameter $R_V$ from $2$ to $6$ in steps of $0.1$. In total, each stellar evolution endpoint is assigned synthetic magnitudes for $4100$ combinations of $A_V$ and $R_V$ in every relevant filter (later, when comparing to observations, we choose only the subset of extinction values consistent with the $E\left(B-V\right)$ ranges presented in the literature for each progenitor, see below). We apply this extinction to account for the combined contributions of the Milky Way and the supernova host.

\section{Observational data}
\label{sec:obsdat}

\begin{table*}
\centering
\caption{Pre-explosion magnitudes, dust extinction and host galaxy distance.}
\begin{threeparttable}
\begin{tabular}{c|ccccccc}
\hline
SN & \textit{U} & \textit{B} & \textit{V} & \textit{R} & \textit{I} & $E\left(B-V\right)/ \mathrm{mag}$ & $d / \mathrm{Mpc}$\\
\hline
\vspace{0.1cm}
1993J & $21.45 \pm 0.2$ & $21.73 \pm 0.07$ & $20.6 \pm 0.16$ & $19.87 \pm 0.11$ & $19.43 \pm 0.17$ & $0.1935^{+0.1291}_{-0.129}$ & $3.63\pm 0.14$\\
\vspace{0.1cm}
2008ax & $> 22.9$ & $24.14 \pm 0.22$ & & $23.85 \pm 0.42$ & $23.61 \pm 0.22$ & $0.3^{+0.1}_{-0.1}$ & $7.77\pm 1.54$\\
\vspace{0.1cm}
2011dh & $23.39\pm 0.25$ & $22.36\pm 0.02$ & $21.83\pm 0.04$ & $21.28\pm 0.04$ & $21.2 \pm 0.03$ & $0.07^{+0.07}_{-0.04}$ & $7.9\pm 1.0$\\
2013df & & $> 25.65$ & $24.535\pm 0.071$ & & $23.144 \pm 0.055$ & $0.0968^{+0.0161}_{-0.0161}$ & $16.6\pm 0.4$\\
\vspace{0.1cm}
\vspace{0.1cm}
$\mathrm{iPTF13bvn}_\mathrm{E}$ & & $25.8\pm 0.12$ & $25.8\pm 0.11$ & & $25.88 \pm 0.24$ & $0.1237^{+0.07}_{-0.04}$ & $26.7\pm 2.5$\\
$\mathrm{iPTF13bvn}_\mathrm{F}$ & & $25.99\pm 0.14$ & $26.06\pm 0.13$ & & $25.82 \pm 0.12$ & $0.1237^{+0.07}_{-0.04}$ & $26.7\pm 2.5$\\
\vspace{0.1cm}
2016gkg & & $24.46 \pm 0.22$ & & $24.31 \pm 0.18$ & $24.02 \pm 0.2$ & $0.1071^{+0.06}_{-0.03}$ & $26.4\pm 5.3$\\
\vspace{0.1cm}
2019yvr & & $26.2882\pm 0.1622$ & $25.3812\pm 0.0319$ & $24.7471\pm 0.0221$ & $23.8333 \pm 0.0319$ & $0.53^{+0.27}_{-0.16}$ & $14.4\pm 1.3$\\
\hline
\hline
\end{tabular}
\footnotesize
\begin{tablenotes}
\textit{Notes.} \textit{HST} \textit{U} filters: F300W (2008ax), F336W (2011dh); \textit{HST} \textit{B} filters: F435W (2011dh, iPTF13bvn), F438W (2019yvr), F439W (2013df), F450W (2008ax, 2016gkg); \textit{HST} \textit{V} filter: F555W (2011dh, 2013df, iPTF13bvn, 2019yvr); \textit{HST} \textit{R} filters: F606W (2008ax, 2016gkg), F625W (2019yvr), F658N (2011dh); \textit{HST} \textit{I} filter: F814W (2008ax, 2011dh, 2013df, iPTF13bvn, 2016gkg, 2019yvr). The pre-explosion photometry of SN~1993J is taken from ground-based observations, predating the launch of the \textit{HST}. The values in all filters for all SN progenitors are given in Vega magnitudes.
\end{tablenotes}
\end{threeparttable}
\label{tab:magnitudes}
\end{table*}
\begin{table*}
\centering
\caption{Upper limits we consider from post-explosion data.}
\begin{threeparttable}
\begin{tabular}{c|ccccccccccc}
\hline
SN & \textit{F218W} & \textit{F275W} & \textit{F336W} & \textit{F438W} & \textit{F555W} & \textit{F625W} & \textit{F814W} & \textit{F850LP} & \textit{F105W} & \textit{F125W} & \textit{F160W} \\
 & \textit{F225W} & & & \textit{F435W} & & \textit{F606W} & & & & \\
\hline
\vspace{0.1cm}
1993J & $ > 21.62$ & $ > 21.599$ & $ > 22.234$ & $ > 22.563$ & $ > 21.493$ & $ > 21.366$ & $ > 20.876$ & $ > 21.371$ & $ > 18.686$ & $ > 19.519$ & $ > 19.374$ \\
2008ax & & $> 25.6$ & $> 25.7$ & $> 26.6$ & $> 26.9$ & & $> 25.6$ & & & & \\
\vspace{0.1cm}
2011dh & $ > 24.61$ & & $ > 25$ & $> 26.3$ & & & & & & & \\
\vspace{0.1cm}
$\mathrm{iPTF13bvn}_\mathrm{E}$ & & & & $> 26.4$ & $> 26.28$ & & & & & & \\
$\mathrm{iPTF13bvn}_\mathrm{F}$ & $> 26.4$ & & & $> 26.48$ & $> 26.64$ & & $> 25.88$ & & & & \\
2016gkg & & $> 24.49$ & & $> 26.49$ & & $> 24.95$ & & & & & \\
\hline
\hline
\end{tabular}
\footnotesize
\begin{tablenotes}
\textit{Note.} All values are given in Vega magnitudes.
\end{tablenotes}
\end{threeparttable}
\label{tab:postmagnitudes}
\end{table*}

Here we list the sources of observational data for pre- and-post explosion photometry (when available) which we fit our models to. We summarise the observed pre-explosion magnitudes, the dust reddening parameter $E\left(B-V\right)$ and host galaxy distance of the SN progenitors in Table \ref{tab:magnitudes}. We do not attempt to fit post-explosion photometry to a surviving companion model because of the various possible contributions to the post-explosion flux, such as from the SN itself, its remnant, or a light echo (see, for example \citealt{Foxetal2014} regarding SN~1993J). We do require that the companion star in our models not violate any upper limits derived from post-explosion observations. In Table \ref{tab:postmagnitudes} we summarise the post-explosion upper limits which we adopt for five SNe.

\subsection*{SN 1993J (IIb)}
\label{subsec:SN1993J}

We use the pre-explosion photometry of SN~1993J from \cite{Alderingetal1994}. We follow \cite{Maundetal2004} and take the distance to the host galaxy of SN~1993J, M81, to be $d=3.63\pm 0.14\,\mathrm{Mpc}$ from the Cepheid distance modulus \citep{Ferrareseetal2000}. We set the range of the dust reddening parameter $E\left(B-V\right)$ according to $0.2\,\mathrm{mag} \le A_V \le 1\,\mathrm{mag}$ \citep{Mathesonetal2000} and $R_V=3.1$.

Post-explosion observations of SN~1993J \citep{Maundetal2004,Foxetal2014} can supply additional information on the surviving companion. \cite{Foxetal2014} discuss the various contributions to the flux at the SN site $20$ years after the explosion, and suggest that the companion might be observed in the far UV, while the flux in longer wavelengths results from the fading SN. As mentioned above, we do not try to fit the post-explosion UV data, but we rather take the brighter bounds from the magnitudes reported by \cite{Foxetal2014} as upper limits on the flux contribution of the surviving companion (Table \ref{tab:postmagnitudes}).

\subsection*{SN 2008ax (IIb)}
\label{subsec:2008ax}

The progenitor of SN~2008ax has been studied by \cite{Crockettetal2008} and \cite{Folatellietal2015}. We take the pre-explosion magnitudes from \cite{Folatellietal2015}, who revised the analysis of \cite{Crockettetal2008} by using high-resolution post-explosion images to subtract nearby contaminating stellar sources. We adopt the distance of $d=7.77\pm 1.54\,\mathrm{Mpc}$ to the host galaxy of SN~2008ax, NGC~4490, following \cite{Folatellietal2015}. We take a host galaxy reddening of $E\left(B-V\right)_\mathrm{host} = 0.3\pm 0.1\,\mathrm{mag}$ \citep{Crockettetal2008,Folatellietal2015} and neglect the small Milky Way contribution of $E\left(B-V\right)_\mathrm{MW} = 0.019\,\mathrm{mag}$ \citep{SchlaflyFinkbeiner2011}. \cite{Folatellietal2015} provide post-explosion upper limits on the remaining companion (listed in Table \ref{tab:postmagnitudes}), which we include in our analysis.

\subsection*{SN 2011dh (IIb)}
\label{subsec:2011dh}

The pre-explosion photometry for SN~2011dh is taken from \cite{Maundetal2011}. We follow \cite{Ergonetal2014} in adopting a distance of $d=7.8^{+1.1}_{-0.9}\,\mathrm{Mpc}$ to the host galaxy of SN~2011dh, the Whirlpool Galaxy, and in taking a total dust reddening of $E\left(B-V\right) = 0.07^{+0.07}_{-0.04}\,\mathrm{mag}$. \cite{Maund2019} used a light echo to isolate the flux contribution of a surviving companion. We take the upper limit in the \textit{F435W} filter, and treat the brighter bounds reported by \cite{Maund2019} in the \textit{F225W} and \textit{F336W} filters as upper limits (Table \ref{tab:postmagnitudes}).

\subsection*{SN 2013df (IIb)}
\label{subsec:2013df}

The pre-explosion photometry for SN~2013df is taken from \cite{VanDyketal2014}. The Cepheid-based distance, $d=16.6\pm 0.4\,\mathrm{Mpc}$, to the host galaxy of SN~2013df, NGC~4414, is taken from \cite{Freedmanetal2001}. We follow \cite{VanDyketal2014} and take a total dust extinction of $A_V=0.3\pm 0.05\,\mathrm{mag}$, which translates to $E\left(B-V\right)=0.0968\pm 0.0161\,\mathrm{mag}$ for $R_V=3.1$.

\subsection*{iPTF13bvn (Ib)}
\label{subsec:iPTF13bvn}

We consider two sets of photometry estimates for iPTF13bvn, one from \cite{Eldridgeetal2015} and one from \cite{Folatellietal2016}. \cite{EldridgeMaund2016} report post-explosion observations of iPTF13bvn, from which we take the brighter bounds as upper limits on the flux contribution of the companion star that we include in our analysis (Table \ref{tab:postmagnitudes}). \cite{Folatellietal2016} also report post-SN observations of iPTF13bvn, from which we take the upper limit in the \textit{F225W} filter and treat the brighter bounds in the \textit{F438W}, \textit{F555W} and \textit{F814W} filters as upper limits (Table \ref{tab:postmagnitudes}). Our analysis of iPTF13bvn is performed once with the combination of the pre-SN photometry form \cite{Eldridgeetal2015} and the post-SN photometry of \cite{EldridgeMaund2016} (denoted $\mathrm{iPTF13bvn}_\mathrm{E}$) and a second time with the pre-SN and post-SN photometry reported by \cite{Folatellietal2016} (denoted $\mathrm{iPTF13bvn}_\mathrm{F}$). 

We follow \cite{Fremlingetal2016} in taking a host galaxy reddening of $E\left(B-V\right)_\mathrm{host} = 0.08^{+0.07}_{-0.04}\,\mathrm{mag}$ and a Milky Way contribution of $E\left(B-V\right)_\mathrm{MW} = 0.0437\,\mathrm{mag}$ \citep{SchlaflyFinkbeiner2011} for a total of $E\left(B-V\right) = 0.1237^{+0.07}_{-0.04}\,\mathrm{mag}$. We also follow \cite{Fremlingetal2016} in adopting the distance of $d=26.8^{+2.6}_{-2.4}\,\mathrm{Mpc}$ to the host galaxy of iPTF13bvn, NGC~5806, from \cite{Tullyetal2013}. We note that this distance is larger than that considered by most of the early studies of iPTF13bvn \citep{Caoetal2013,Berstenetal2014,Fremlingetal2014,Eldridgeetal2015}.

\subsection*{SN 2016gkg (IIb)}
\label{subsec:2016gkg}

We take the pre-explosion photometry for SN~2016gkg from \cite{Kilpatricketal2022}, transformed from their AB magnitudes to Vega magnitudes. Similarly to \cite{Kilpatricketal2017}, we take from \cite*{Nasonovaetal2011} the distance $d=26.4\pm 5.3\,\mathrm{Mpc}$ to the host galaxy of SN~2016gkg, NGC~613. We follow \cite{Arcavietal2017} and adopt a host galaxy reddening of $E\left(B-V\right)_\mathrm{host}=0.09^{+0.06}_{-0.03}\,\mathrm{mag}$, which together with a Milky Way extinction of $A_{V\mathrm{,MW}}=0.053\,\mathrm{mag}$ \citep{SchlaflyFinkbeiner2011} and $R_V=3.1$ gives $E\left(B-V\right)=0.1071^{+0.06}_{-0.03}\,\mathrm{mag}$. \cite{Kilpatricketal2022} report late-time observations of SN~2016gkg from which we take the upper limit in the \textit{F275W} filter, and treat the brighter bounds in the \textit{F438W} and \textit{F606W} filters as upper limits (Table \ref{tab:postmagnitudes}), transformed from their AB magnitudes to Vega magnitudes.

\subsection*{SN 2019yvr (Ib)}
\label{subsec:2019yvr}

We take the pre-explosion photometry from \cite{Kilpatricketal2021}\footnote{While writing this paper, new photometry for SN~2019yvr was published by \cite{Sunetal2021}. This photometry is very similar to that of \cite{Kilpatricketal2021} in the three shorter-wavelength filters, and is slightly fainter in \textit{F814W}. This difference will have a negligible effect on our results and conclusions and therefore we stay with the \cite{Kilpatricketal2021} data.}, transformed from their AB magnitudes to Vega magnitudes. We follow \cite{Kilpatricketal2021} and adopt a distance of $d=14.4\pm 1.3\,\mathrm{Mpc}$ to the host galaxy of SN~2019yvr, NGC~4666, derived from the light curve of the Type~Ia SN ASASSN-14lp which occurred in the same galaxy \citep{Shappeeetal2016}. \cite{Kilpatricketal2021} find from the colour curves of SN~2019yvr a host reddening of $E\left(B-V\right)_\mathrm{host}=0.51^{+0.27}_{-0.16}\,\mathrm{mag}$. Together with the Milky Way contribution of $E\left(B-V\right)_\mathrm{MW} = 0.02\,\mathrm{mag}$ \citep{SchlaflyFinkbeiner2011} we have $E\left(B-V\right) = 0.53^{+0.27}_{-0.16}\,\mathrm{mag}$.

\section{Results}
\label{sec:results}

\subsection{Endpoints of stellar evolution simulations}
\label{subsec:EvolutionEndPoints}

\begin{figure*}
\centering
\includegraphics[width=1.0\textwidth]{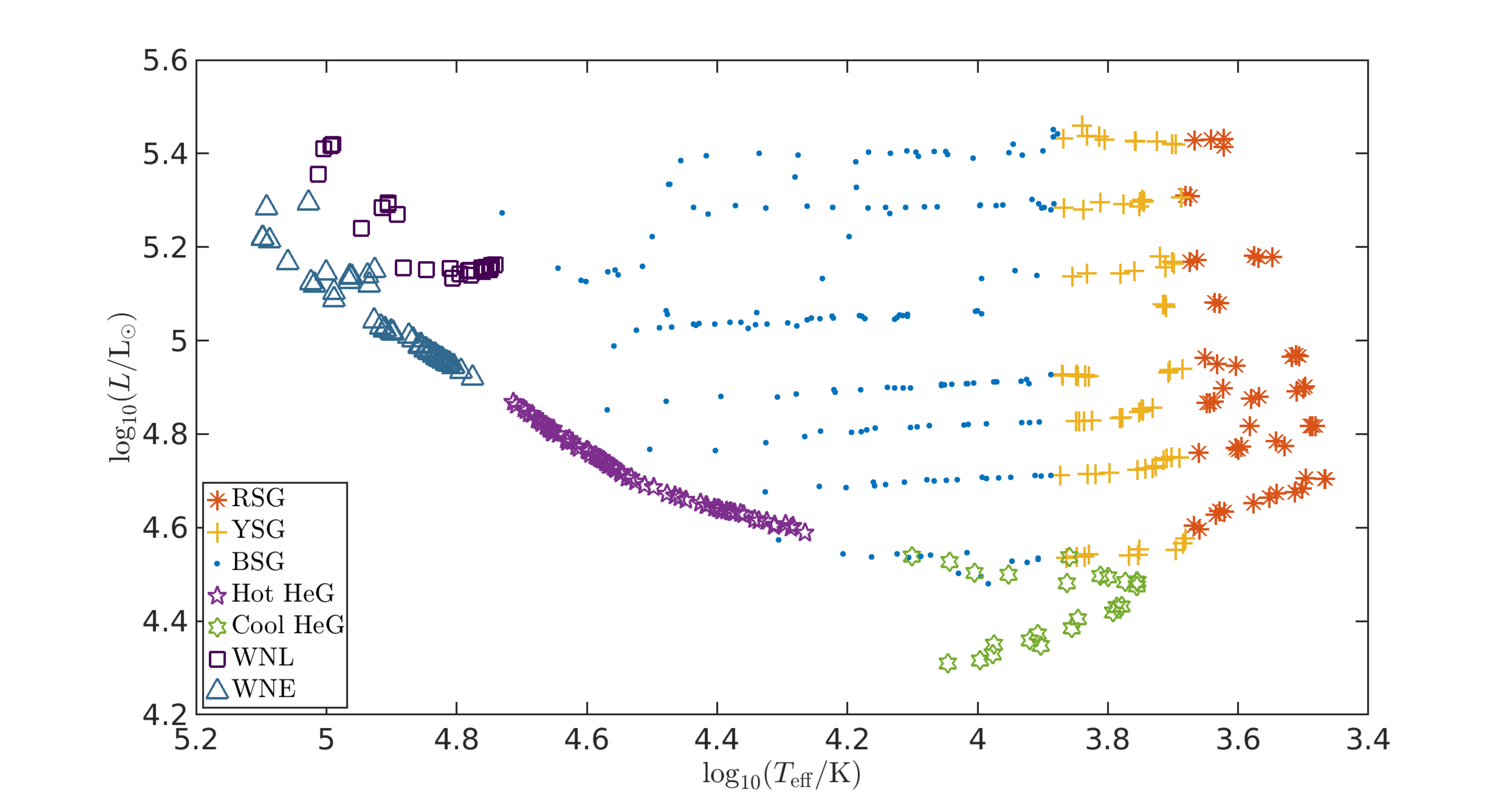} \\
\caption{Hertzsprung--Russell diagram for the stellar evolution endpoints of the primary star in all computed models.}
\label{fig:HRD1}
\end{figure*}
\begin{figure}
\centering
\includegraphics[width=0.48\textwidth]{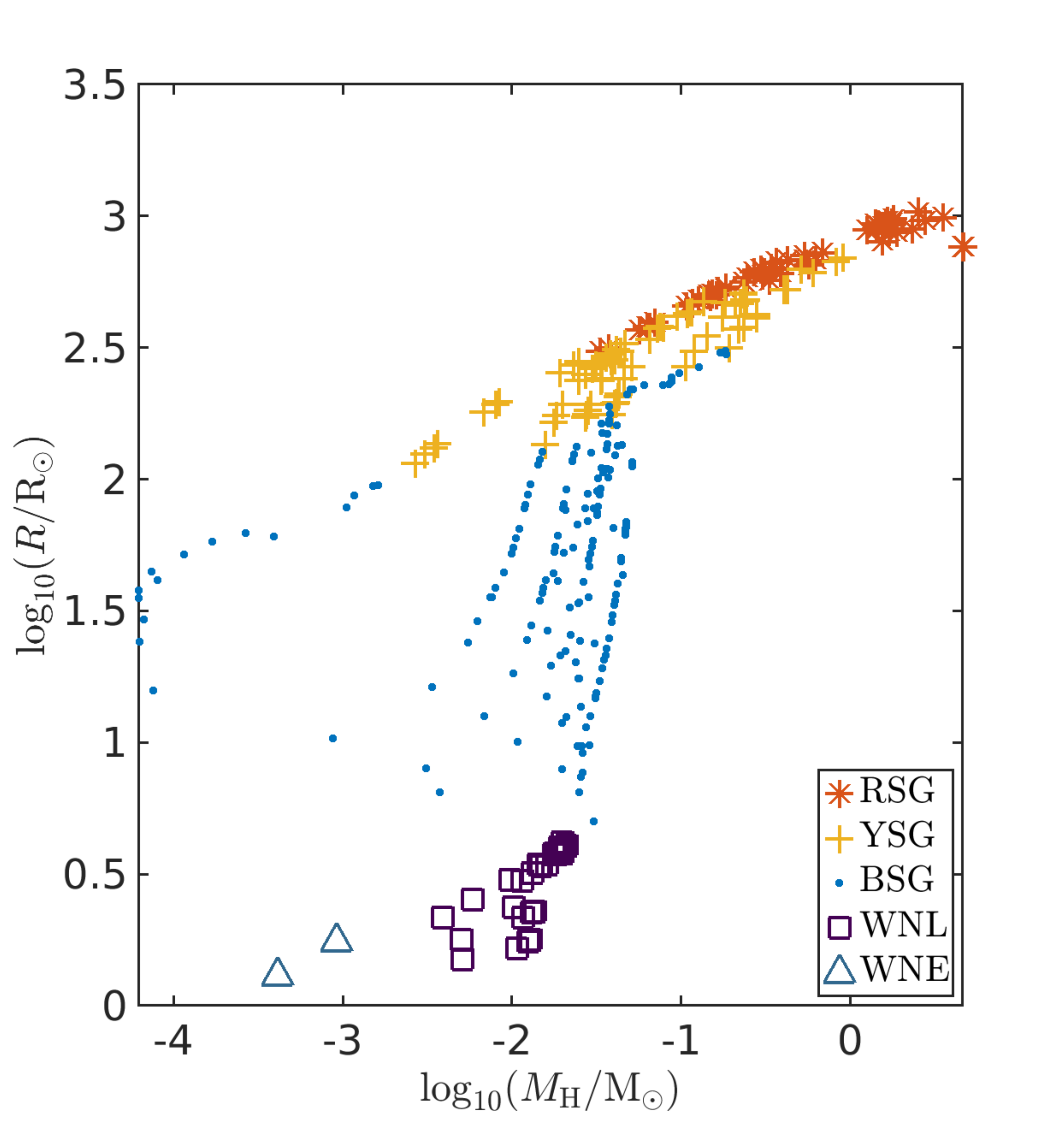} \\
\caption{Stellar radius as a function of total hydrogen mass for the stellar evolution endpoints of the primary star in all computed models.}
\label{fig:MHR1}
\end{figure}
\begin{figure}
\centering
\includegraphics[width=0.48\textwidth]{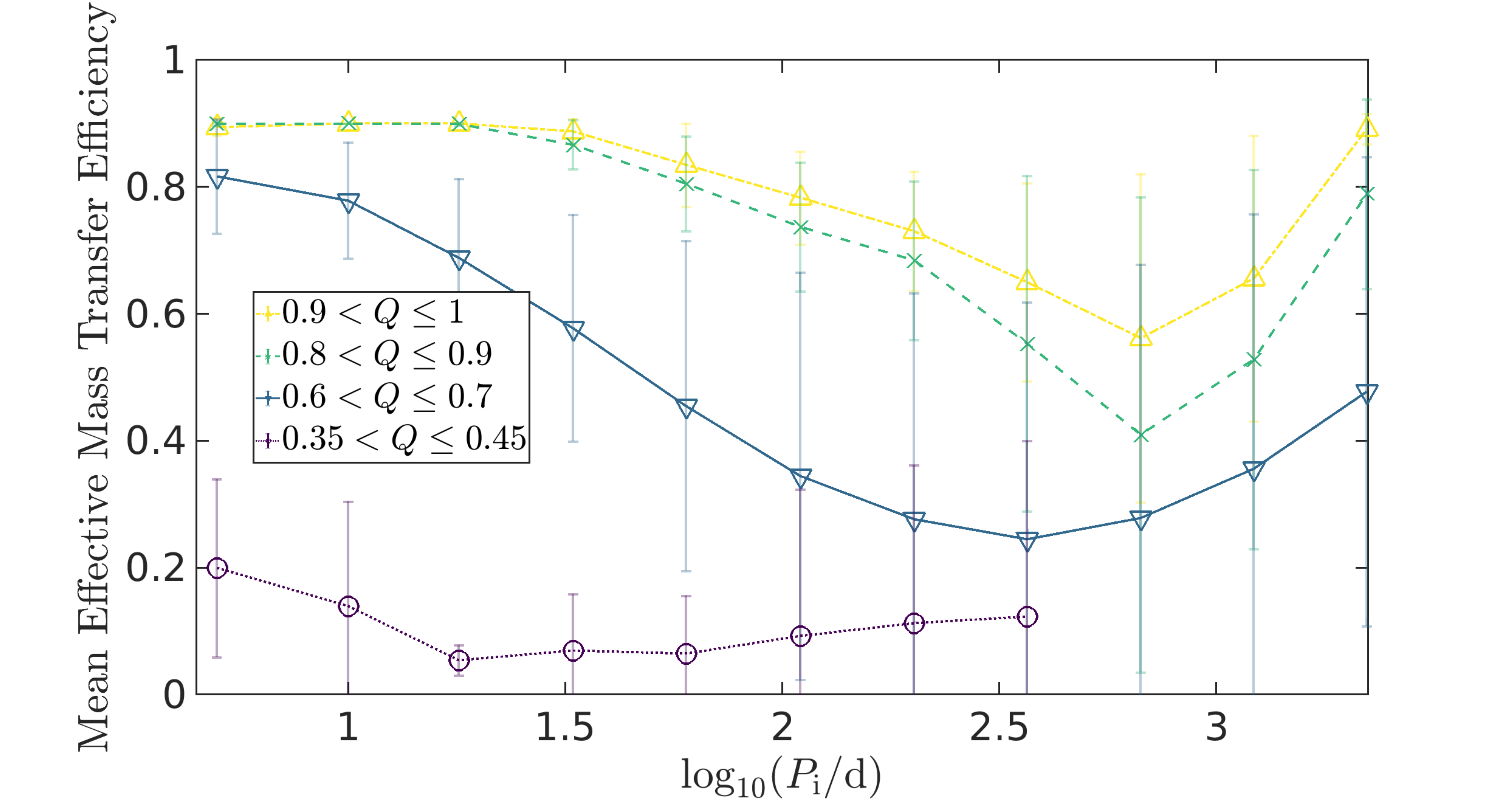} \\
\caption{Mean effective mass transfer efficiency as a function of initial orbital period for the different mass ratio regimes, where $Q\equiv M_2 / M_1$ is the mass ratio. Each point is an average over all initial primary masses and both wind schemes and the error bars denote the standard deviation. The points for the longest initial orbital periods and lowest companion masses are not shown because of a very low success rate of simulations for those initial conditions, with most of the evolutionary tracks entering CEE phases.}
\label{fig:PMT1}
\end{figure}
Of the $352$ combinations of initial conditions (detailed in Section \ref{sec:method}), $191$ reached a point where $T_\mathrm{eff} > 10000\,\mathrm{K}$ and $X_\mathrm{s} < 0.4$ along their evolution, and were therefore simulated twice, for the two hot hydrogen-deficient wind schemes\footnote{While both wind schemes cannot be correct simultaneously, we pool all tracks together so that our analysis will cover as much of the progenitor property parameter space as possible, and also to allow us to directly compare the suitableness of the two schemes.}. Of the $543$ simulations that we ran, $49$ encountered numerical problems or entered common envelope evolution (CEE; where the orbital separation becomes smaller than the sum of the two stellar radii) before the end of the simulation and were discounted. In total, we have $494$ useful binary evolution tracks, which reached the end of core carbon burning. The endpoints of the stellar evolution are shown in Figure \ref{fig:HRD1}. Models are classified according to their effective surface temperature and surface hydrogen mass fraction, as follows:
\begin{itemize}
    \item Red supergiant (RSG): $T_\mathrm{eff} \le 4.8\,\mathrm{kK}$, $0.01 \le X_\mathrm{s}$;
    \item Yellow supergiant (YSG): $4.8\,\mathrm{kK} < T_\mathrm{eff} < 7.5\,\mathrm{kK}$, $0.01 \le X_\mathrm{s}$;
    \item Blue supergiant (BSG): $7.5\,\mathrm{kK} \le T_\mathrm{eff} \le 55\,\mathrm{kK}$, $0.01 \le X_\mathrm{s}$;
    \item Hot helium giant (HeG): $15\,\mathrm{kK} \le T_\mathrm{eff} \le 55\,\mathrm{kK}$, $X_\mathrm{s} < 0.01$;
    \item Cool helium giant: $T_\mathrm{eff} < 15\,\mathrm{kK}$, $X_\mathrm{s} < 0.01$;
    \item Early nitrogen-sequence WR (WNE): $55\,\mathrm{kK} < T_\mathrm{eff}$, $X_\mathrm{s} < 0.05$;
    \item Late nitrogen-sequence WR (WNL): $55\,\mathrm{kK} < T_\mathrm{eff}$, $0.05 \le X_\mathrm{s}$.
\end{itemize}
Wolf-Rayet (WR) stars can appear also in carbon- or oxygen-sequences if their surface nitrogen mass fractions are low enough, though this does not occur in our models. Although the WR phenomenon is not defined by temperature, but rather by the wind mass loss and corresponding transformed radius (Eq.~\ref{eq:Rt}), for the evolution endpoints our definition by temperature suffices. We consider models with a metallicity close to (or slightly above) Galactic. Galactic WR stars have a minimum luminosity of $L^\mathrm{WR}_\mathrm{min} \approx 10^{4.9}\,\mathrm{L}_\odot$ \citep{Shenar2020}. We do not have models with $L<L^\mathrm{WR}_\mathrm{min}$ and $T_\mathrm{eff} > 55\,\mathrm{kK}$. The reason that our simple temperature threshold for WR stars works is that we are looking at the final evolutionary stage, after the end of core carbon burning, and a significant expansion and cooling of the outer layers. During earlier phases, such as core helium burning, the models ultimately classified as helium giants were more compact and hotter ($T_\mathrm{eff} > 55\,\mathrm{kK}$) but should probably not have been given a WR classification.

The models classified as WNE stars or as helium giants all result from evolutionary tracks which employed the \citetalias{NL00} wind scheme. The vast majority of models classified as WNL stars result from evolutionary tracks which employed the \citetalias{Vink2017} wind scheme. The minimal leftover hydrogen mass among the \citetalias{Vink2017} models is $M_\mathrm{H} \approx 10^{-4}\,\mathrm{M}_\odot$ (with a corresponding surface hydrogen mass fraction of $X_\mathrm{s}\approx 0.01$), while most of the \citetalias{NL00} models have $M_\mathrm{H} \ll 10^{-4}\,\mathrm{M}_\odot$.

The leftover hydrogen mass in the envelope strongly affects the stellar radius, as we show in Figure \ref{fig:MHR1}. For $M_\mathrm{H}\ga 0.1\,\mathrm{M}_\odot$ there is a tight relation between $M_\mathrm{H}$ and $R$, while for lower $M_\mathrm{H}$ the general trend is similar though there is a large spread corresponding to differences in luminosity, with higher luminosity corresponding to smaller radii at a given $M_\mathrm{H}$. The high sensitivity of the stellar radius to the leftover hydrogen mass shown in Figure \ref{fig:MHR1} indicates the importance of covering a large number of models in the relevant parameter space of initial conditions.

As explained by \cite{GilkisVinkEldridgeTout2019}, the mass transfer efficiency during Roche-lobe overflow (RLOF) is computed continuously during the evolution according to the thermal timescale of the accreting star and its size relative to its own Roche lobe. While the efficiency of mass transfer does not impact much the mass lost from the donor star and its subsequent evolution and final characteristics, the companion star is greatly affected. This is of interest if the companion star contributes a non-negligible fraction of the flux in pre-explosion images, or if we have post-explosion photometry (Appendix \ref{sec:appendixb}). The mean effective mass transfer efficiency that results from the custom mass transfer efficiency prescription of \cite{GilkisVinkEldridgeTout2019} is presented in Figure \ref{fig:PMT1}. The main parameter which lowers the mass transfer efficiency is the mass of the companion star, with lower companion masses resulting in low efficiencies because of the limited ability of the relatively low-mass stars to accrete material at the rate it is lost from the primary star during RLOF. Overall, the resulting effective mass transfer efficiency strongly depends on the mass ratio, with the efficiency increasing as the initial mass of the companion approaches that of the primary.

\subsection{Best-fitting progenitor models}
\label{subsec:chi2best}

\begin{figure*}
  \centering
    \begin{subfigure}{0.49\textwidth}
    \centering
    \includegraphics[width=\textwidth]{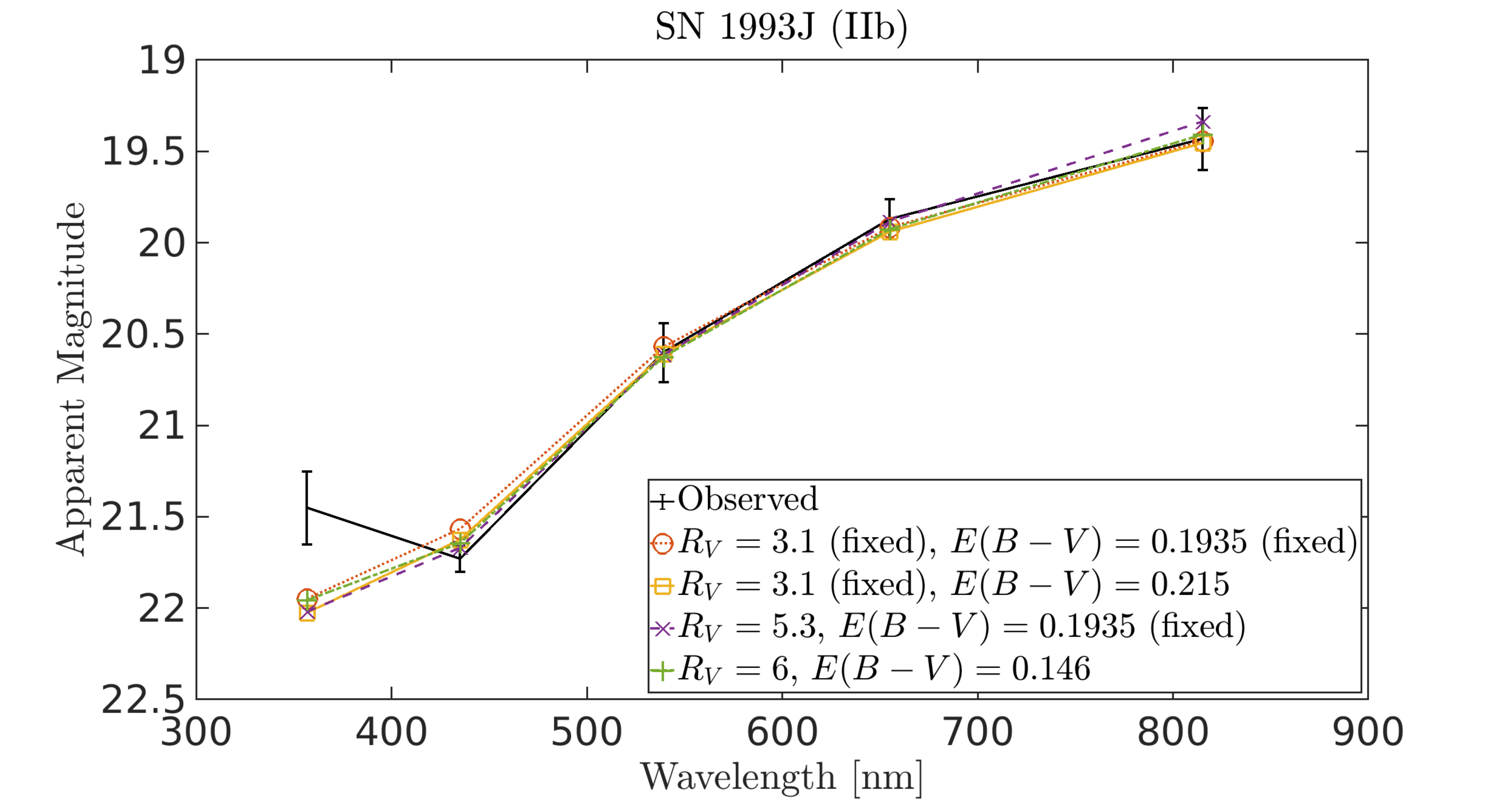}
    \label{fig:chi2SN1993J}
  \end{subfigure}
  \begin{subfigure}{0.49\textwidth}
    \centering
    \includegraphics[width=\textwidth]{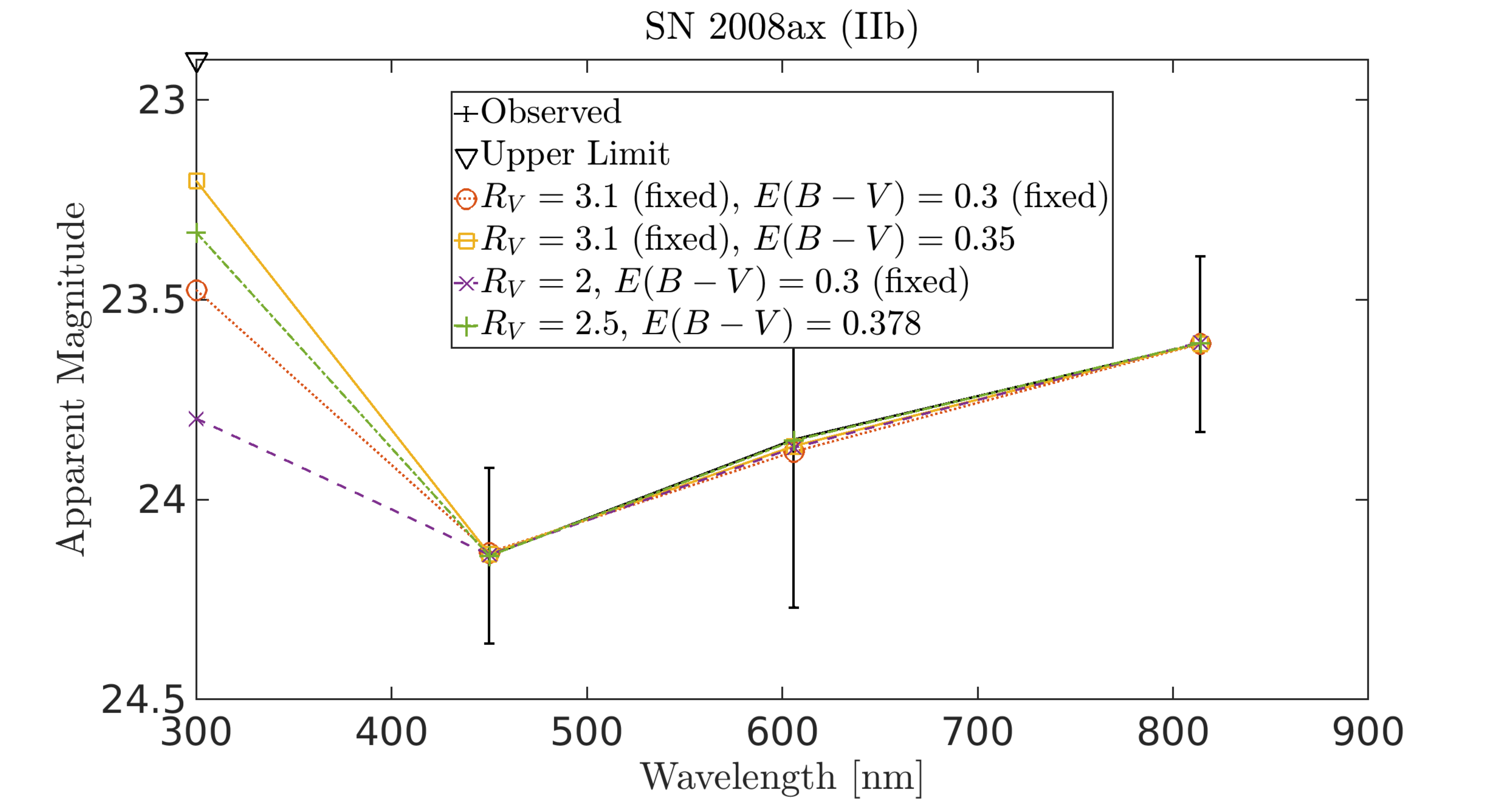}
    \label{fig:chi2SN2008ax}
  \end{subfigure}\\
    \begin{subfigure}{0.49\textwidth}
    \centering
    \includegraphics[width=\textwidth]{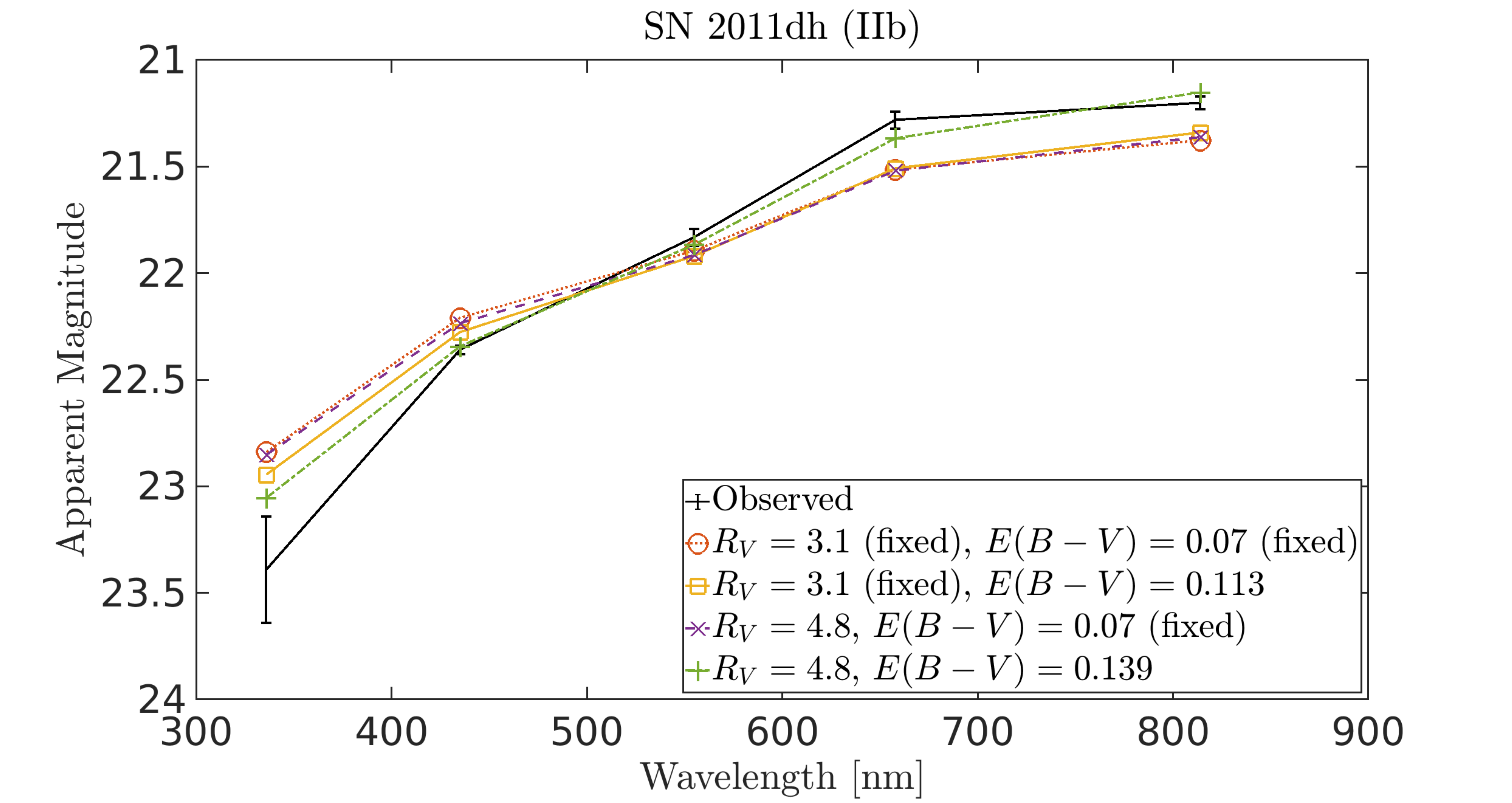}
    \label{fig:chi2SN2011dh}
  \end{subfigure}
  \centering
    \begin{subfigure}{0.49\textwidth}
    \centering
    \includegraphics[width=\textwidth]{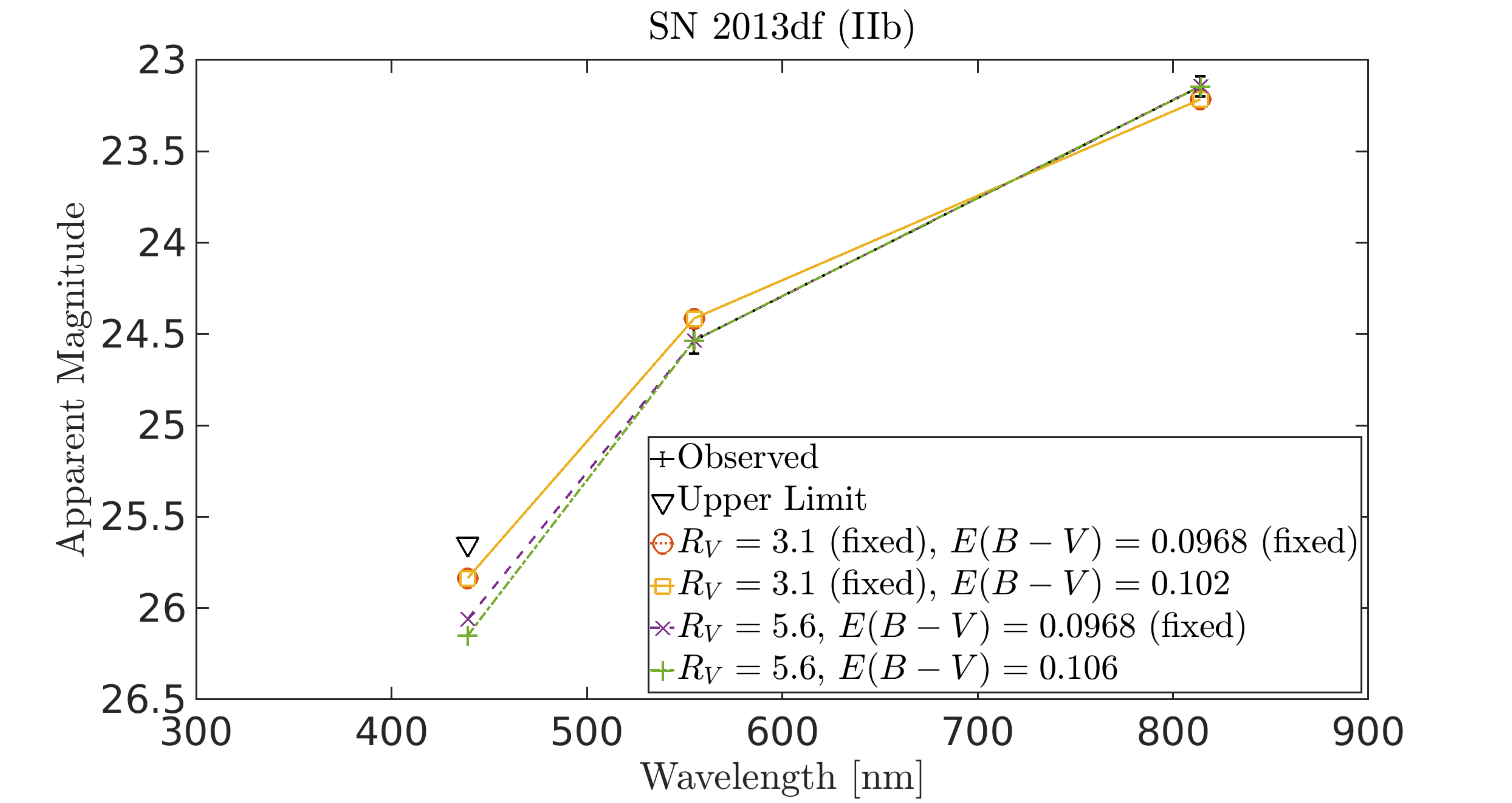}
    \label{fig:chi2SN2013df}
  \end{subfigure}\\
  \begin{subfigure}{0.49\textwidth}
    \centering
    \includegraphics[width=\textwidth]{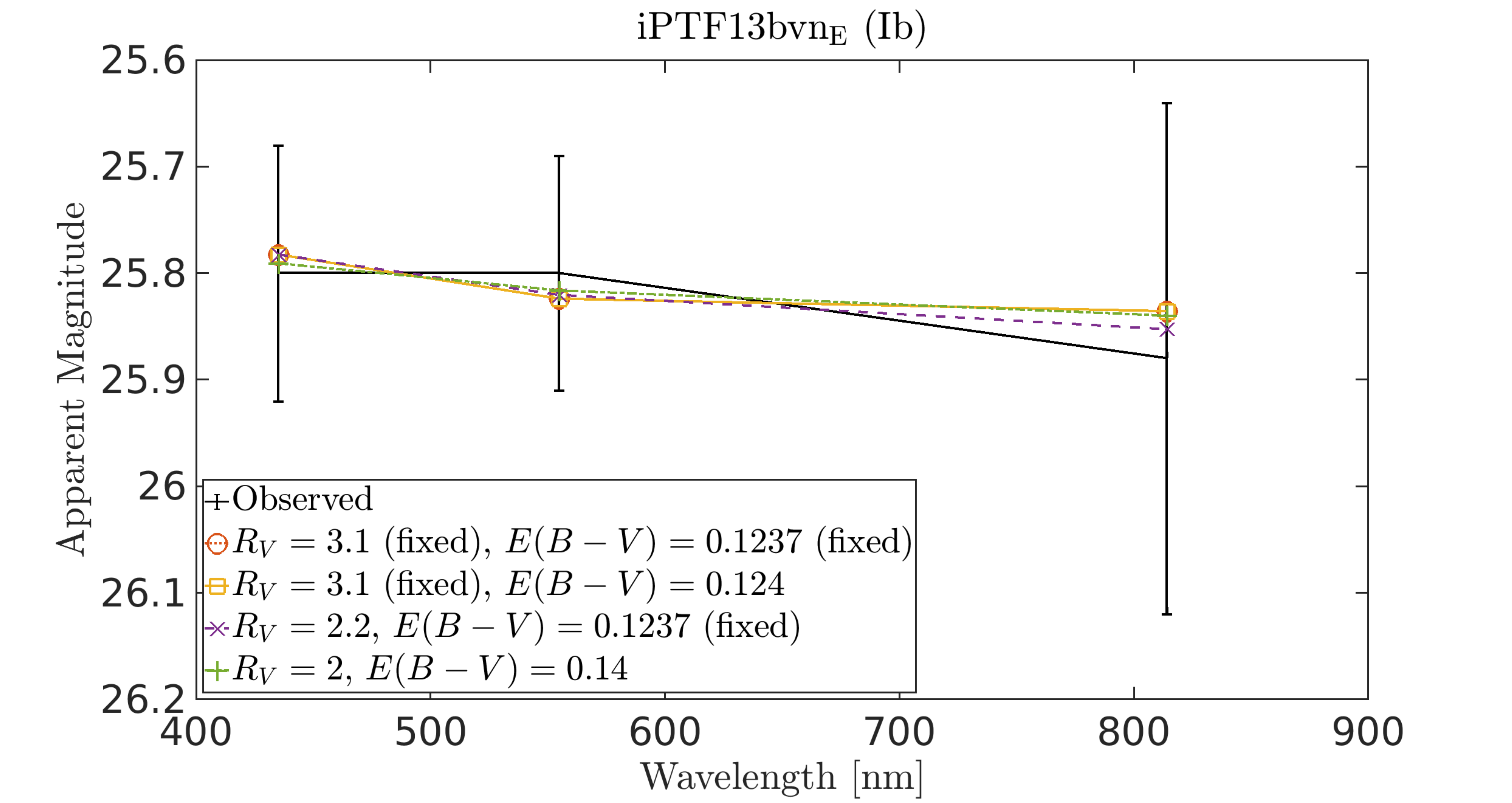}
    \label{fig:chi2iPTF13bvnE}
  \end{subfigure}
    \begin{subfigure}{0.49\textwidth}
    \centering
    \includegraphics[width=\textwidth]{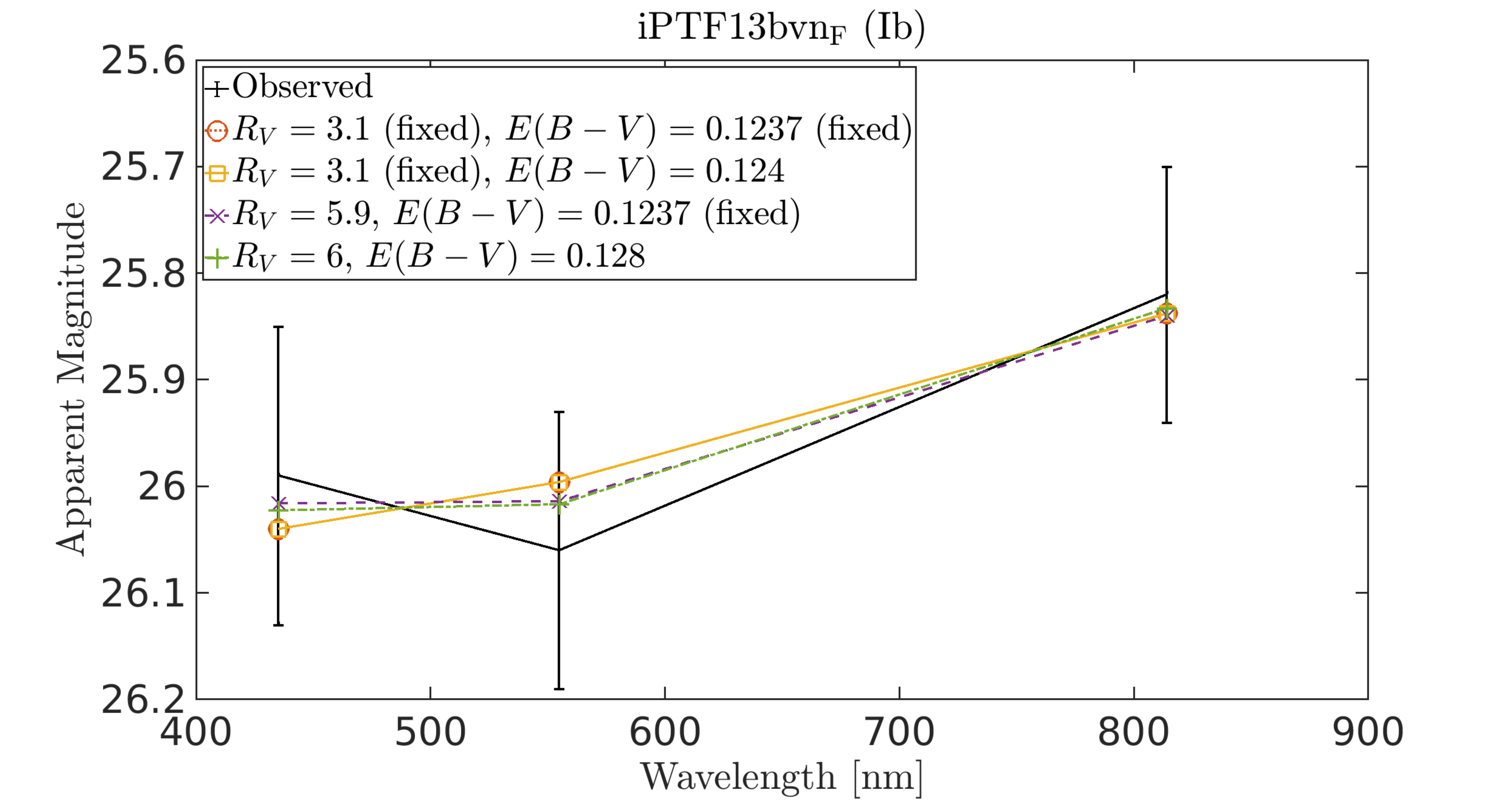}
    \label{fig:chi2iPTF13bvnF}
  \end{subfigure}\\
    \begin{subfigure}{0.49\textwidth}
    \centering
    \includegraphics[width=\textwidth]{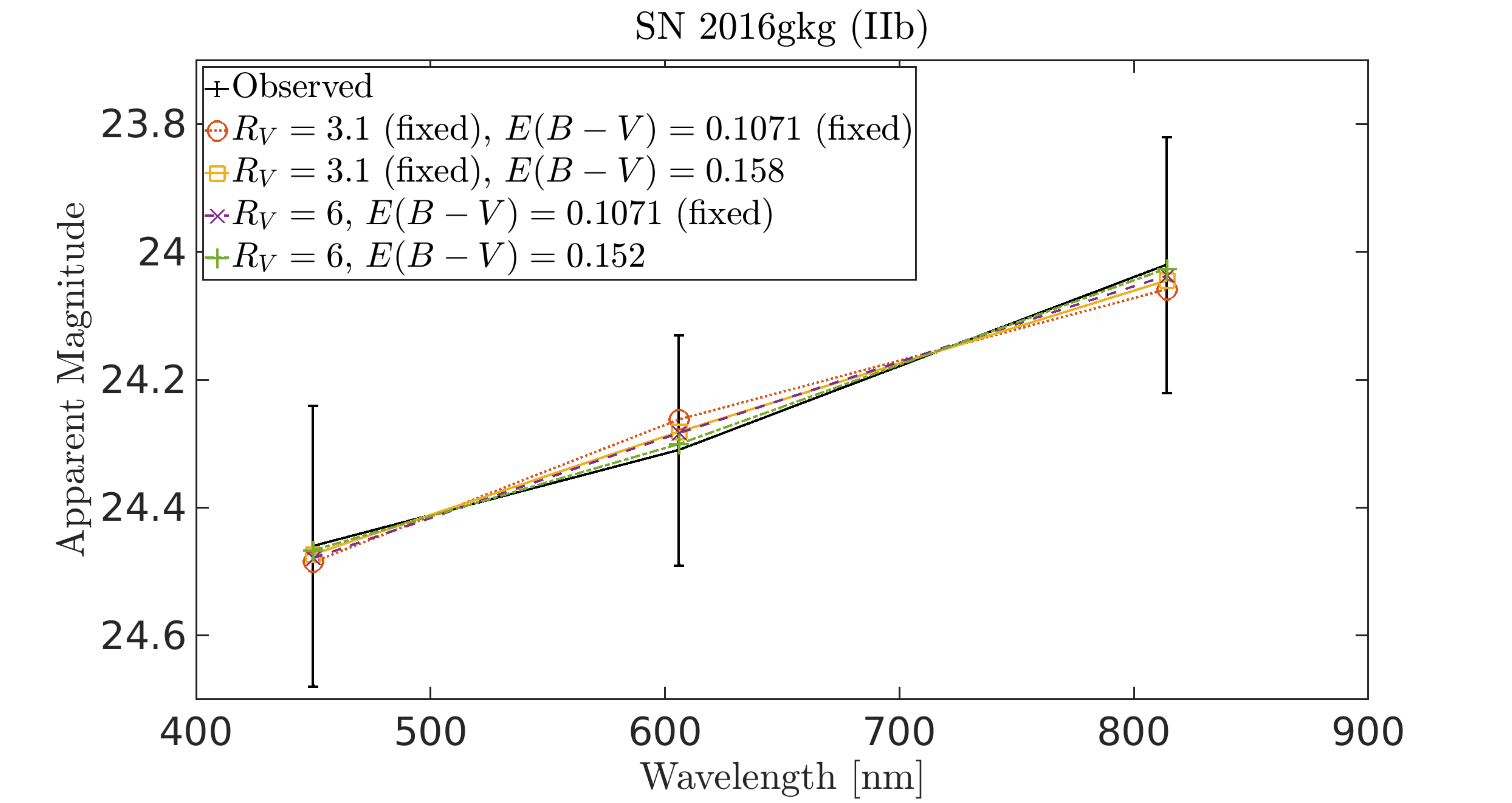}
    \label{fig:chi2SN2016gkg}
  \end{subfigure}
  \begin{subfigure}{0.49\textwidth}
    \centering
    \includegraphics[width=\textwidth]{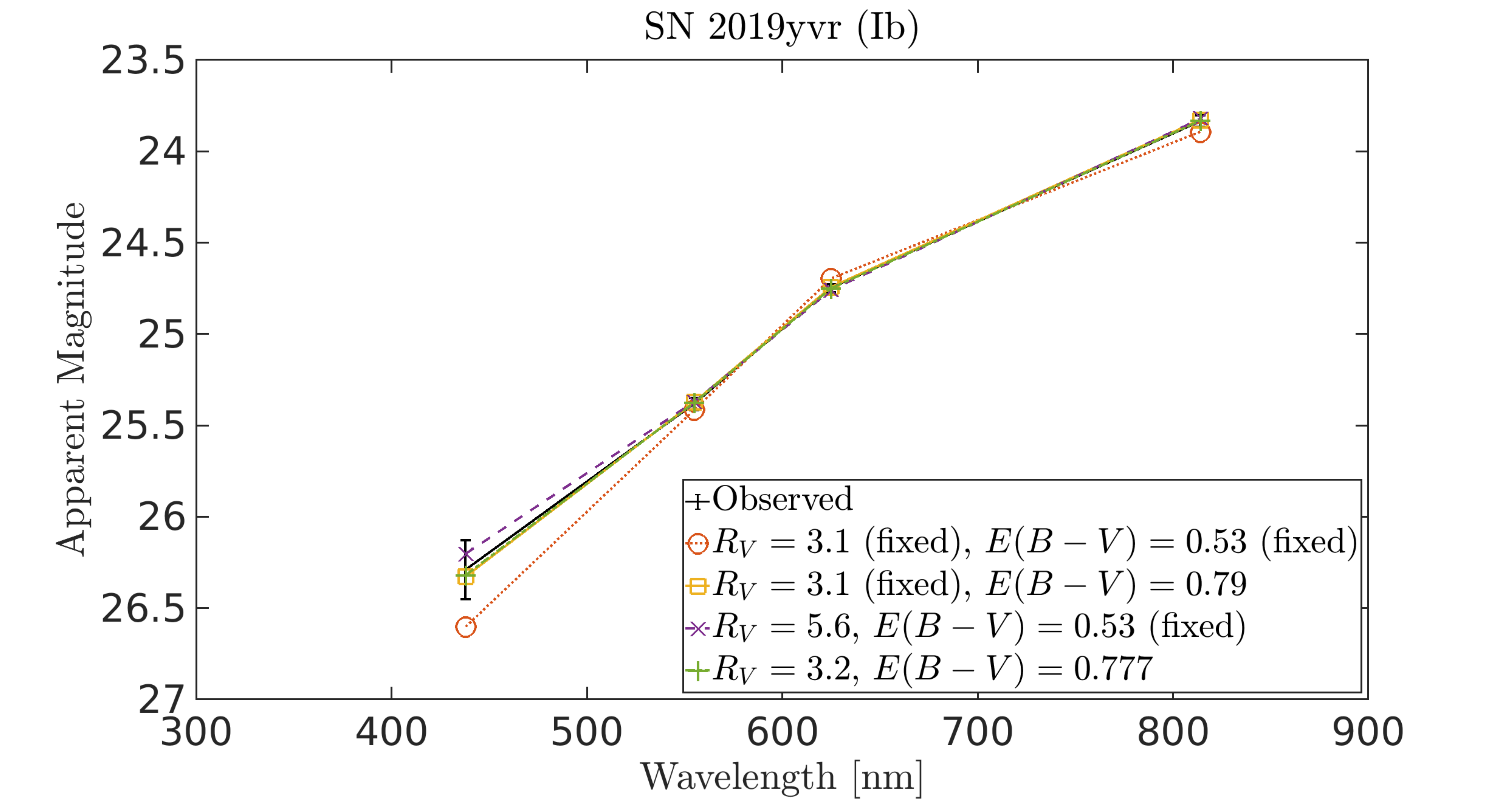}
    \label{fig:chi2SN2019yvr}
  \end{subfigure}\\    
  \caption{Computed magnitudes for best-fitting models compared to the observed magnitudes.} 
  \label{fig:chi2_obs_vs_calc}
\end{figure*}
\begin{figure}
    \centering
    \includegraphics[width=0.48\textwidth]{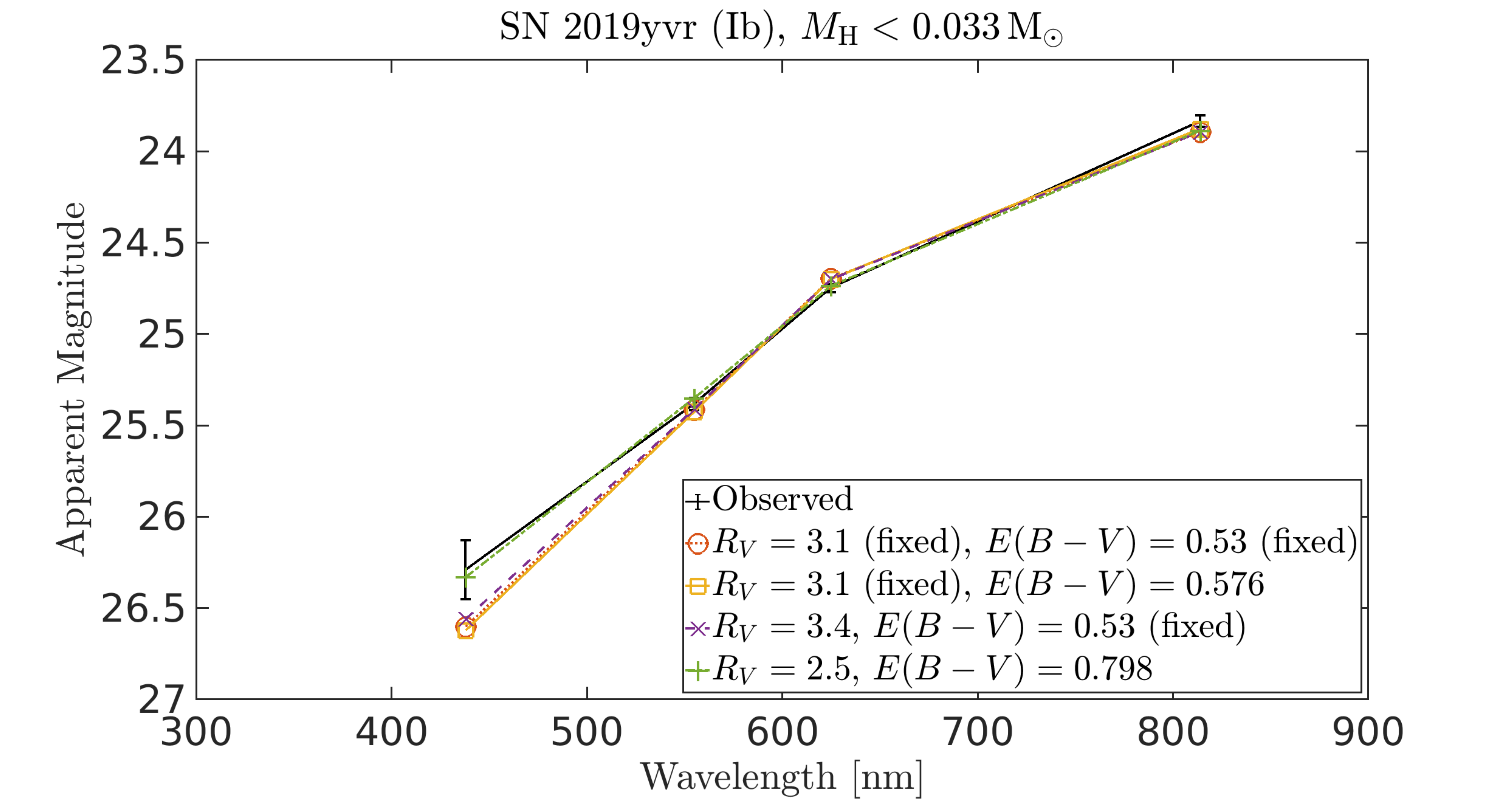}
  \caption{Computed magnitudes for best-fitting models compared to the observed magnitudes, showing only models with $M_\mathrm{H} < 0.033\,\mathrm{M}_\odot$ allowed for SN~2019yvr.} 
    \label{fig:chi2SN2019yvrMH033}
\end{figure}
\begin{figure*}
  \centering
    \begin{subfigure}{0.49\textwidth}
    \centering
    \includegraphics[width=\textwidth]{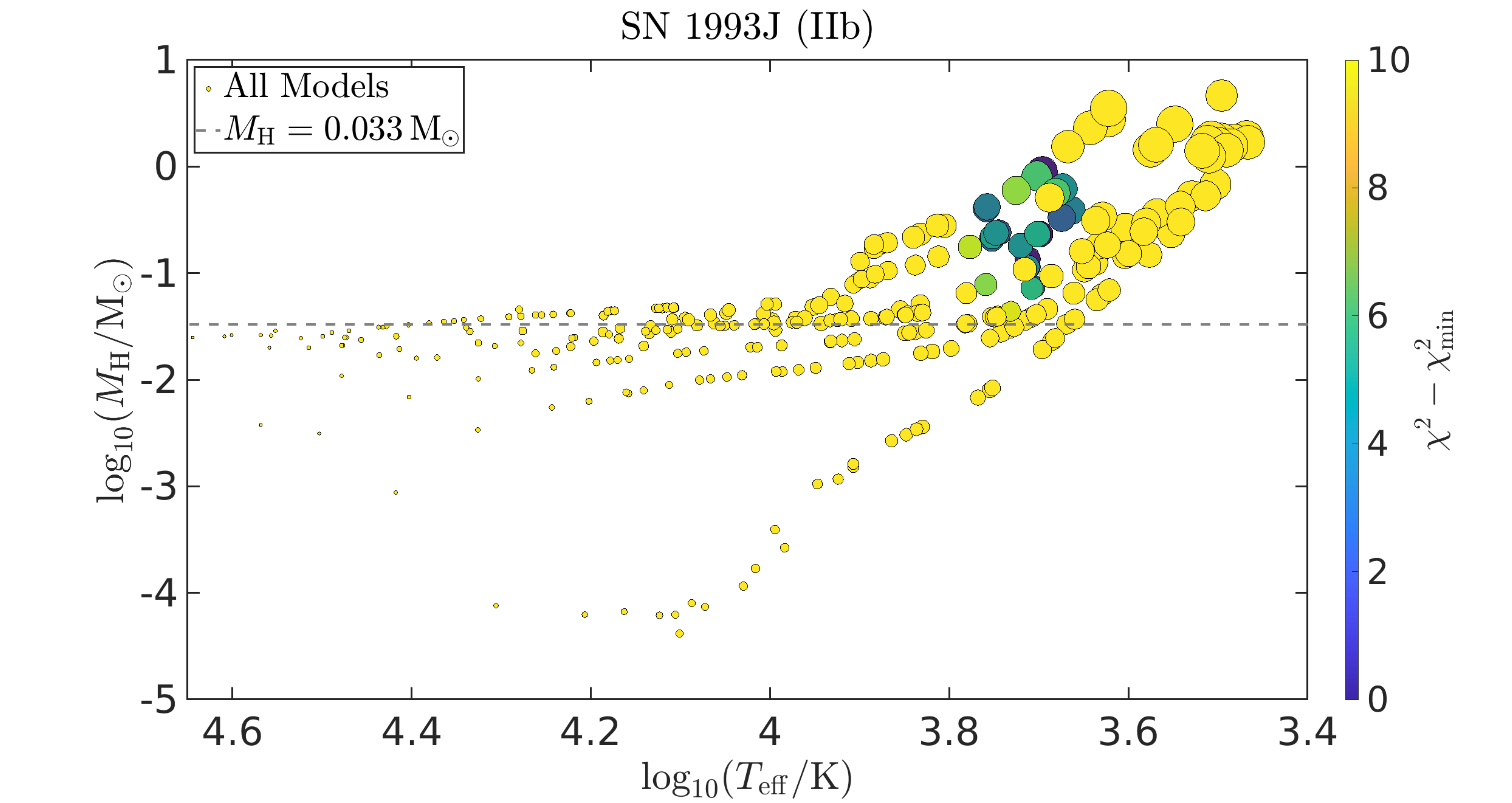}
    \label{fig:MHchi2SN1993J}
  \end{subfigure}
  \begin{subfigure}{0.49\textwidth}
    \centering
    \includegraphics[width=\textwidth]{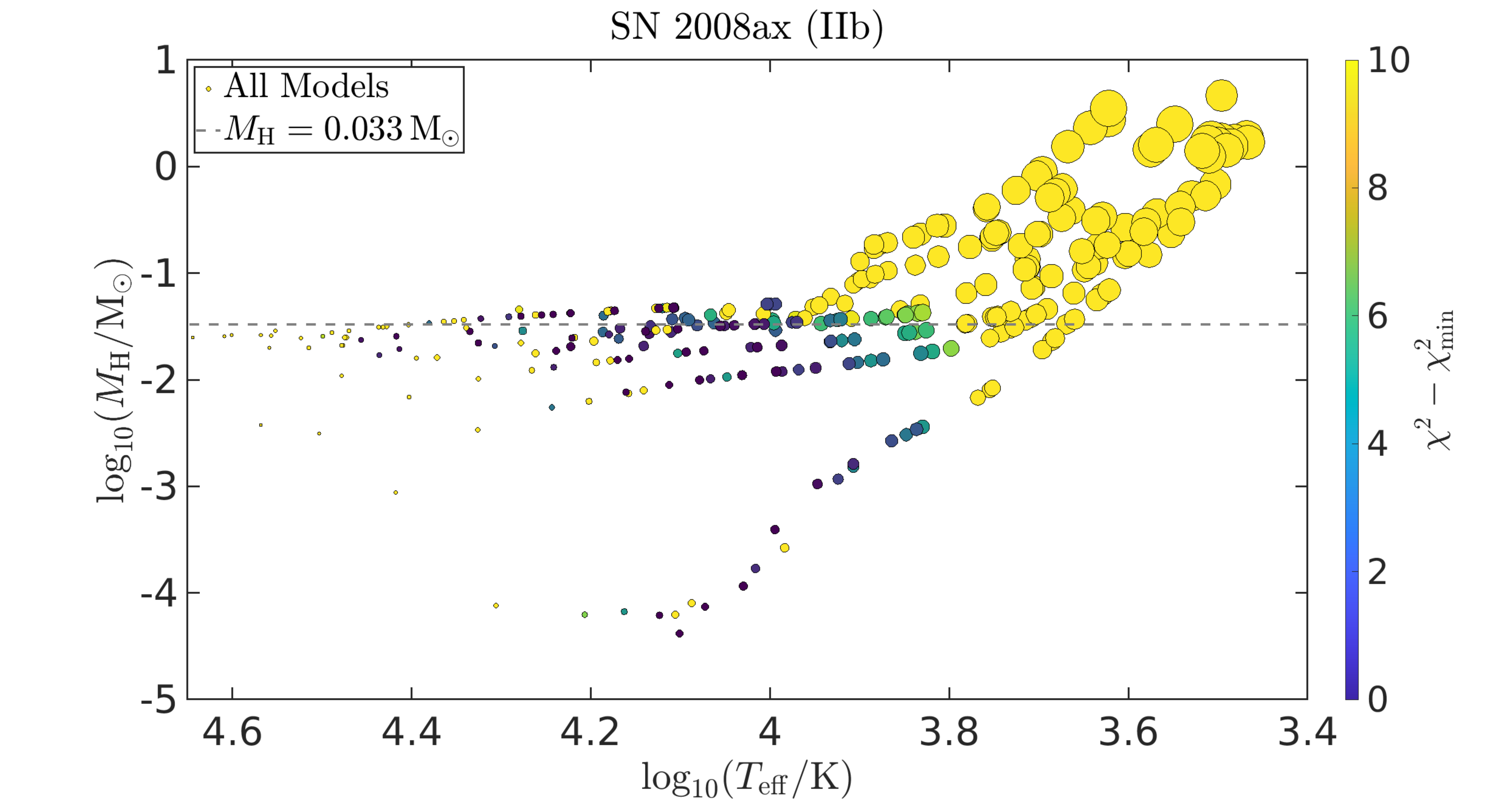}
    \label{fig:MHchi2SN2008ax}
  \end{subfigure}\\
  \centering
    \begin{subfigure}{0.49\textwidth}
    \centering
    \includegraphics[width=\textwidth]{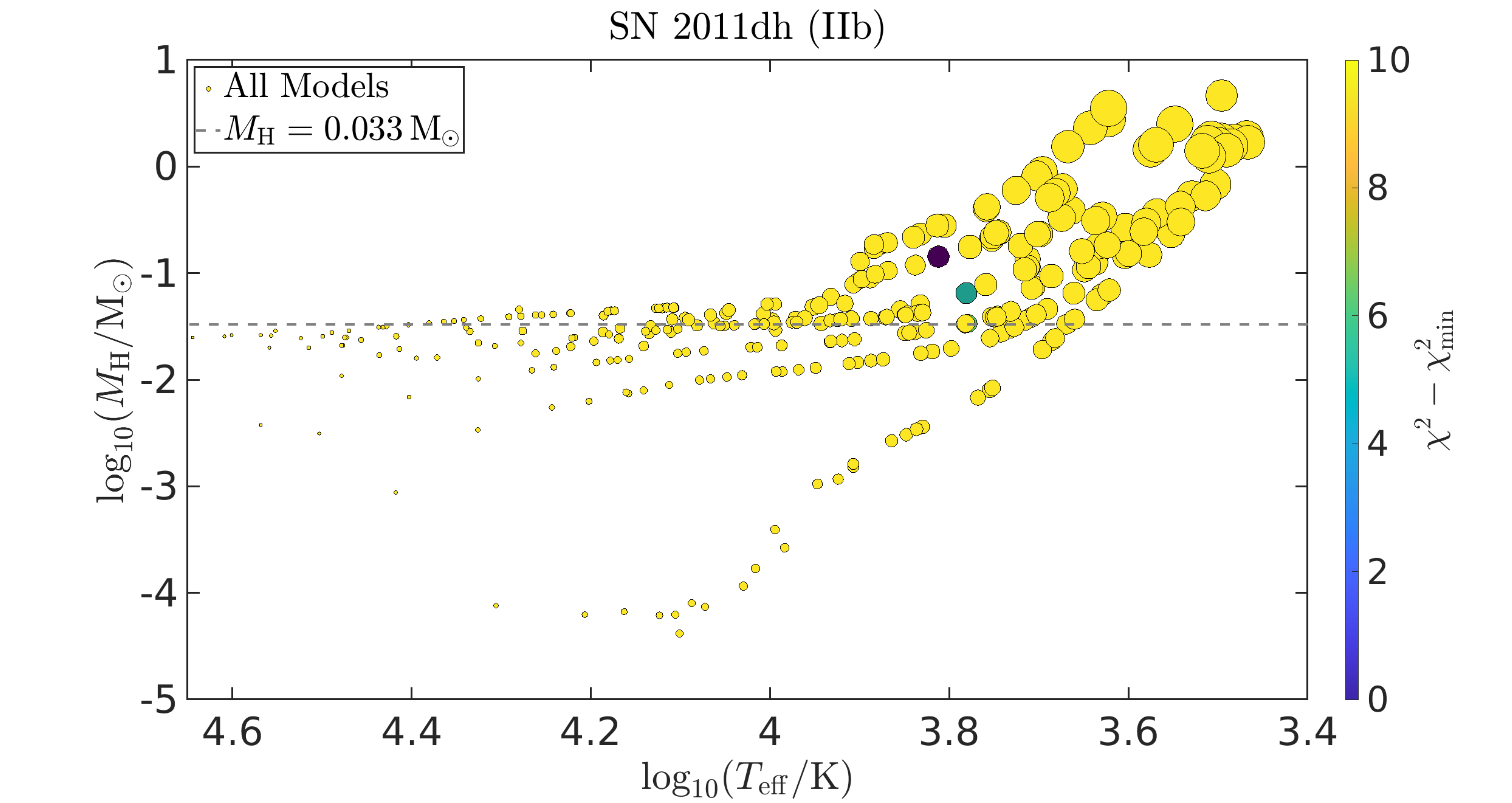}
    \label{fig:MHchi2SN2011dh}
  \end{subfigure}
    \begin{subfigure}{0.49\textwidth}
    \centering
    \includegraphics[width=\textwidth]{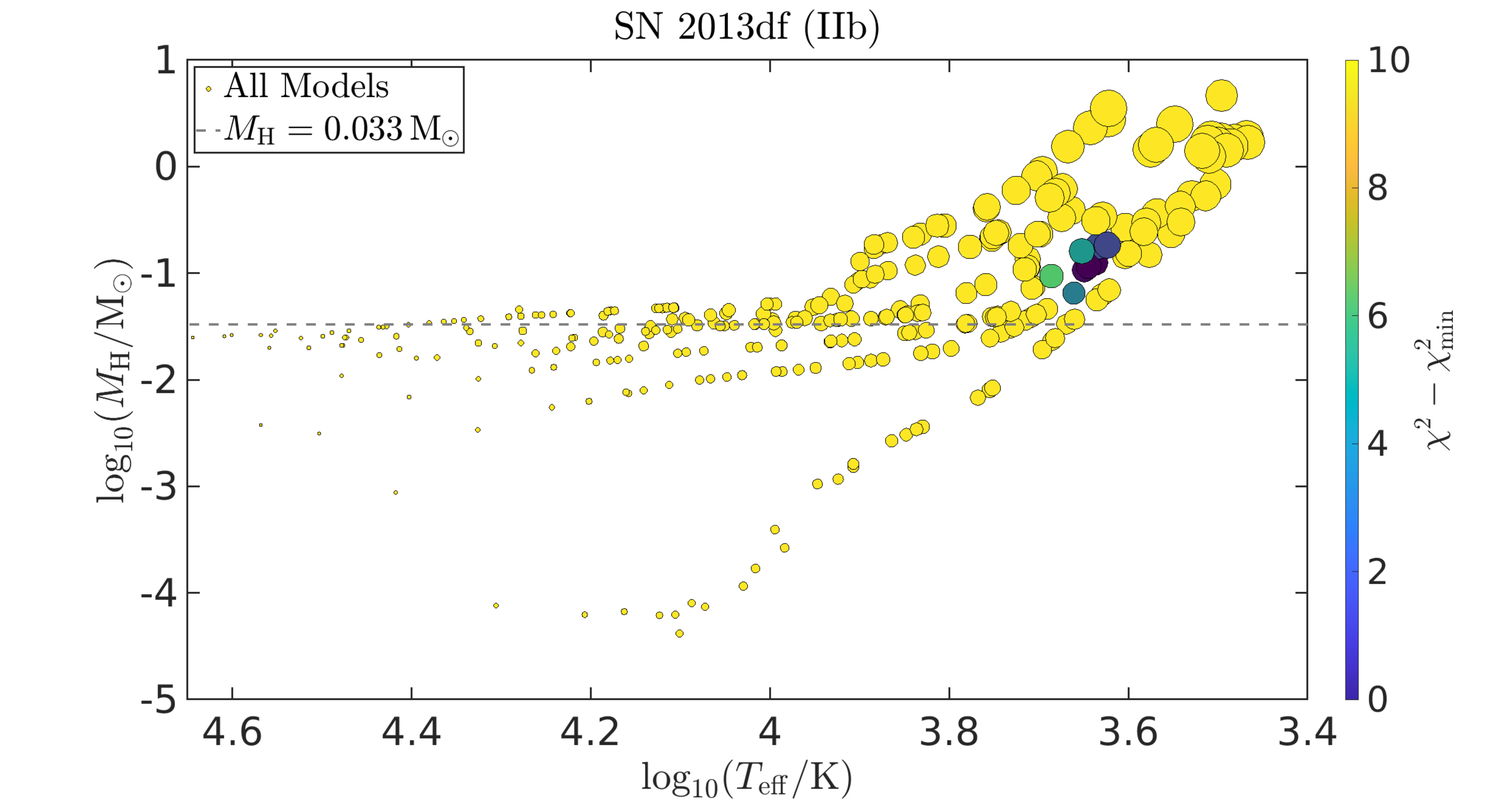}
    \label{fig:MHchi2SN2013df}
  \end{subfigure}\\
  \centering
    \begin{subfigure}{0.49\textwidth}
    \centering
    \includegraphics[width=\textwidth]{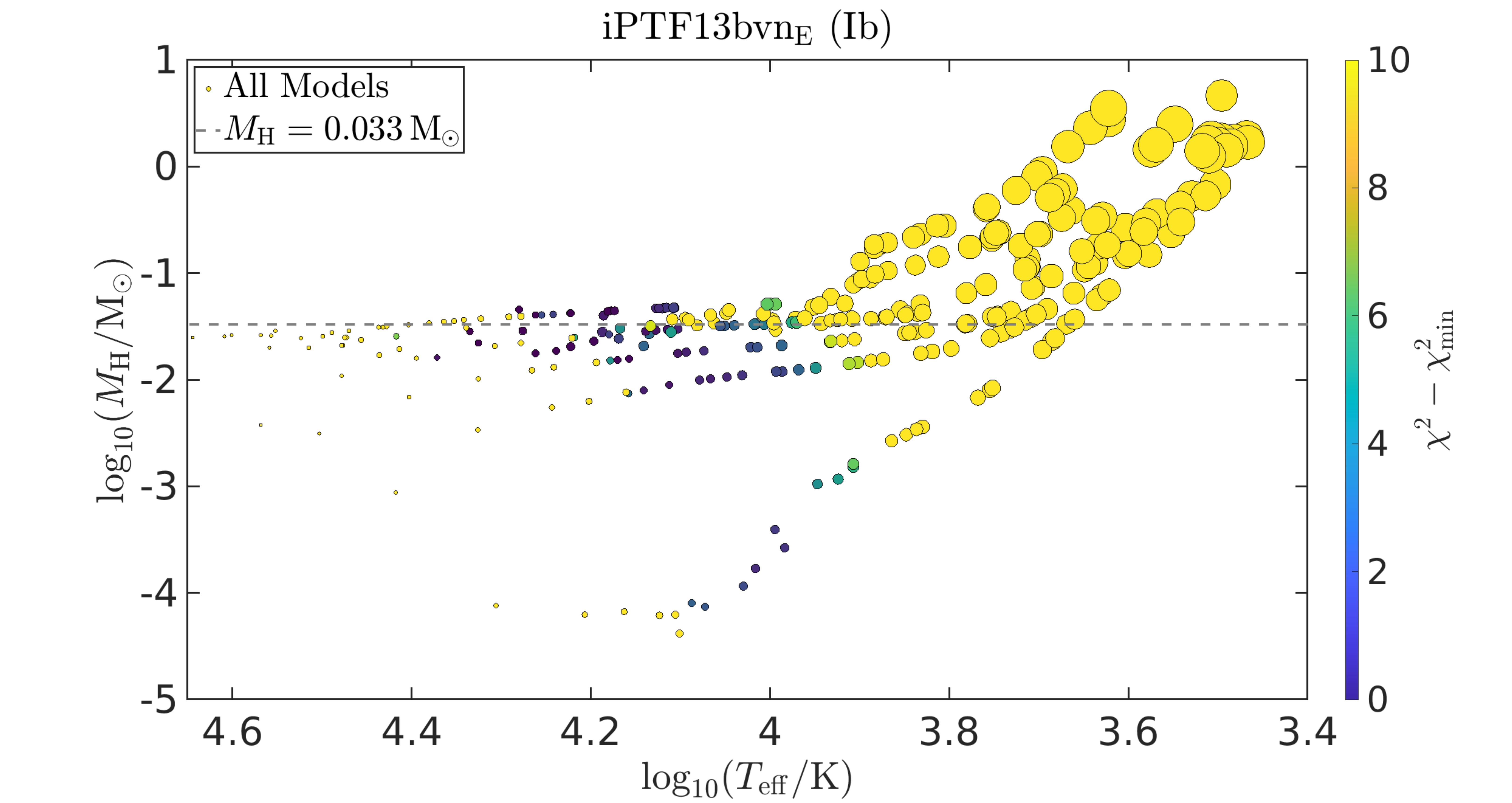}
    \label{fig:MHchi2iPTF13bvnE}
  \end{subfigure}
  \begin{subfigure}{0.49\textwidth}
    \centering
    \includegraphics[width=\textwidth]{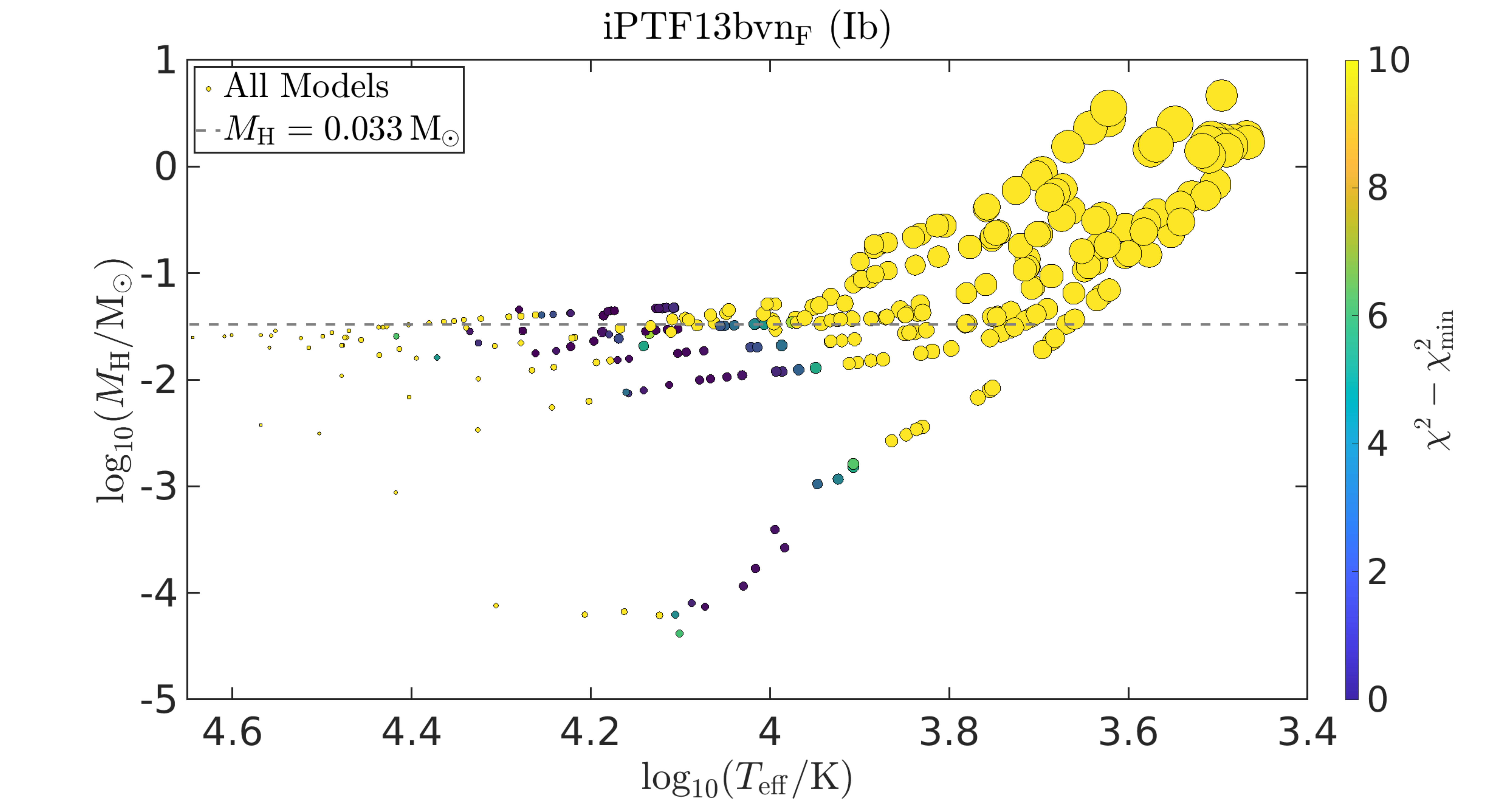}
    \label{fig:MHchi2iPTF13bvnF}
  \end{subfigure}\\
  \begin{subfigure}{0.49\textwidth}
    \centering
    \includegraphics[width=\textwidth]{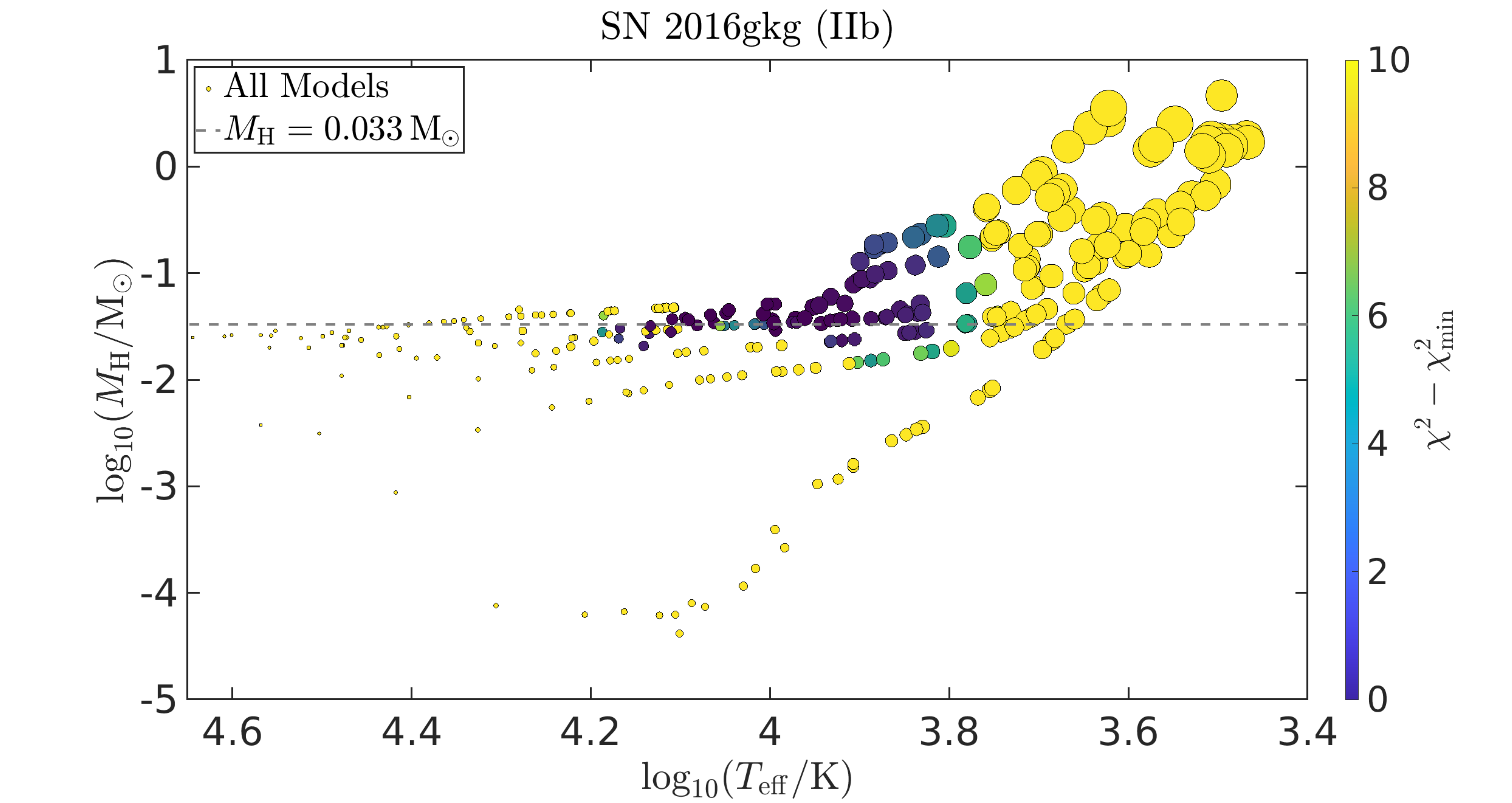}
    \label{fig:MHchi2SN2016gkg}
  \end{subfigure}
  \begin{subfigure}{0.49\textwidth}
    \centering
    \includegraphics[width=\textwidth]{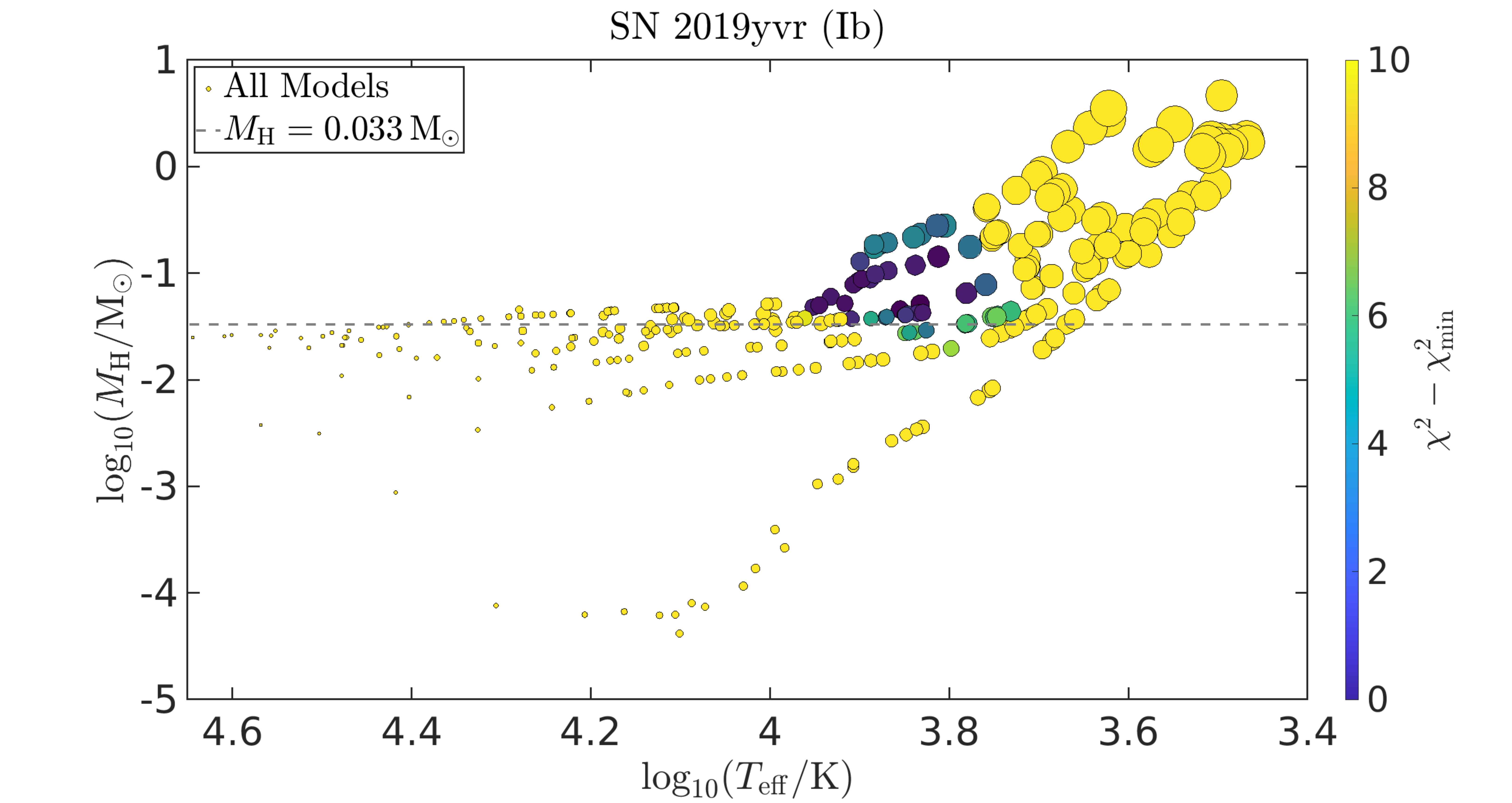}
    \label{fig:MHchi2SN2019yvr}
  \end{subfigure}\\
  \caption{Hydrogen mass as a function of effective surface temperature for the primary star in all models, with colours denoting the value of $\chi^2 - \chi^2_\mathrm{min}$. Each evolutionary endpoint is marked with a circle whose size is proportional to the stellar radius.}
  \label{fig:chi2_MH_vs_Teff}
\end{figure*}
For each SN progenitor, we find the best-fitting progenitor model by finding the minimal $\chi^2$, computed as
\begin{equation}
\chi^2 = \sum\limits_{\lambda} \left(\frac{m^\mathrm{obs}_\lambda - m^\mathrm{calc}_\lambda}{\Delta m^\mathrm{obs}_\lambda}\right)^2,
    \label{eq:chi2}
\end{equation}
where $m^\mathrm{obs}_\lambda$ and $\Delta m^\mathrm{obs}_\lambda$ are the magnitudes and their errors from Table \ref{tab:magnitudes}, $m^\mathrm{calc}_\lambda$ are the computed magnitudes, and $\lambda$ denotes the various filters. We compute $m^\mathrm{calc}_\lambda$ in four different approaches to dust extinction: ($i$) setting the reddening law parameter fixed at $R_V=3.1$ and $E\left(B-V\right)$ fixed at the nominal value from Table \ref{tab:magnitudes} for each SN; ($ii$) setting $R_V=3.1$ and allowing $E\left(B-V\right)$ to vary within the range defined by the errors; ($iii$) allowing the reddening law parameter to vary in $2 \le R_V \le 6$ and keeping $E\left(B-V\right)$ at the nominal value; and ($iv$) allowing both $R_V$ and $E\left(B-V\right)$ to vary within the ranges described above. For each computed evolutionary endpoint and each dust extinction approach we find the distance $d$ (within the distance estimates of each SN) which minimises the $\chi^2$ in Equation~(\ref{eq:chi2}), as long as the upper limits as detailed in Table \ref{tab:postmagnitudes} are not violated.

The comparison between the computed magnitudes for the best-fitting models and the observed magnitudes for all SN progenitors is presented in Figure \ref{fig:chi2_obs_vs_calc}. For the progenitors of SN~2008ax, iPTF13bvn and SN~2016gkg a good fit is easily found for all approaches. The UV excess of the SN~1993J progenitor is not reproduced well in any model. For SN~2011dh, allowing $R_V$ and $E\left(B-V\right)$ to vary significantly improves the best fit, while for SN~2013df just varying $R_V$ helps. For the progenitor of SN~2019yvr, allowing either $R_V$ or $E\left(B-V\right)$, or both, to vary allows a better fit compared to keeping $R_V$ and $E\left(B-V\right)$ fixed.

We repeated the analysis for our two Type Ib's, SN~2019yvr and iPTF13bvn, a couple of times, limiting the model set once to evolutionary endpoints with $M_\mathrm{H} < 0.033\,\mathrm{M}_\odot$, and a second time with $M_\mathrm{H} < 0.001\,\mathrm{M}_\odot$. The result for SN~2019yvr with the $M_\mathrm{H} < 0.033\,\mathrm{M}_\odot$ constraint is shown in Figure \ref{fig:chi2SN2019yvrMH033}, where a good fit is possible if both $R_V$ and $E\left(B-V\right)$ are allowed to vary. When aggravating the constraint to $M_\mathrm{H} < 0.001\,\mathrm{M}_\odot$, no reasonable fits are found for SN~2019yvr. For $\mathrm{iPTF13bvn}_\mathrm{E}$ requiring $M_\mathrm{H} < 0.033\,\mathrm{M}_\odot$ has no effect on the best-fitting models, while taking $M_\mathrm{H} < 0.001\,\mathrm{M}_\odot$ markedly reduces the quality of the best-fitting models, increasing $\chi^2$. For $\mathrm{iPTF13bvn}_\mathrm{F}$ the best-fitting models with fixed $R_V$ have $M_\mathrm{H} < 0.033\,\mathrm{M}_\odot$, while allowing $R_V$ to vary results in better fits for models with $M_\mathrm{H} = 0.044\,\mathrm{M}_\odot$ and higher $R_V$. Requiring $M_\mathrm{H} < 0.001\,\mathrm{M}_\odot$ has a smaller effect on the fit quality for the $\mathrm{iPTF13bvn}_\mathrm{F}$ photometry than the same requirement for $\mathrm{iPTF13bvn}_\mathrm{E}$.

\begin{table*}
\centering
\caption{Details of best-fitting models.}
\begin{threeparttable}
\begin{tabular}{c|ccccccccccc}
\hline
SN & $M_1/ \mathrm{M}_\odot$ & $M_2/ \mathrm{M}_\odot$ & $P_\mathrm{i}/ \mathrm{d}$ & $M_\mathrm{f}/ \mathrm{M}_\odot$ & $M_\mathrm{H}/ \mathrm{M}_\odot$ & $\log_{10}(T_\mathrm{eff}/ \mathrm{K})$ & $\log_{10}(L / \mathrm{L}_\odot)$ & $R_V$ & $E(B-V)/ \mathrm{mag}$ & wind & ST${^a}$ \\ 
\hline
1993J & $16$ & $15$ & $1219$ & $5.43$ & $0.114$ & $3.71$ & $5.07$ & $3.1$ & $0.1935$ & $X_\mathrm{s}\ge 0.4$ & YSG\\ 
1993J & $16$ & $15$ & $1219$ & $5.43$ & $0.114$ & $3.71$ & $5.07$ & $3.1$ & $0.2145$ & $X_\mathrm{s}\ge 0.4$ & YSG\\ 
1993J & $19$ & $18$ & $669$ & $6.47$ & $0.233$ & $3.7$ & $5.17$ & $5.3$ & $0.1935$ & $X_\mathrm{s}\ge 0.4$ & YSG\\ 
1993J & $19$ & $18$ & $669$ & $6.47$ & $0.233$ & $3.7$ & $5.17$ & $6$ & $0.1458$ & $X_\mathrm{s}\ge 0.4$ & YSG\\ 
\hline
2008ax & $13$ & $5$ & $60$ & $3.59$ & $0.016$ & $4.16$ & $4.81$ & $3.1$ & $0.3$ & V17 & BSG\\ 
2008ax & $22$ & $8$ & $110$ & $7.14$ & $0.02$ & $4.22$ & $5.29$ & $3.1$ & $0.35$ & $X_\mathrm{s}\ge 0.4$ & BSG\\ 
2008ax & $11$ & $7$ & $367$ & $2.72$ & $4\times 10^{-5}$ & $4.1$ & $4.54$ & $2$ & $0.3$ & NL00 & HeG\\ 
2008ax & $22$ & $8$ & $110$ & $7.14$ & $0.02$ & $4.22$ & $5.29$ & $2.5$ & $0.378$ & $X_\mathrm{s}\ge 0.4$ & BSG\\ 
\hline
2011dh & $14$ & $9$ & $367$ & $4.3$ & $0.043$ & $3.83$ & $4.93$ & $3.1$ & $0.07$ & V17 & YSG\\ 
2011dh & $14$ & $9$ & $367$ & $4.3$ & $0.043$ & $3.83$ & $4.93$ & $3.1$ & $0.1129$ & V17 & YSG\\ 
2011dh & $14$ & $9$ & $367$ & $4.3$ & $0.043$ & $3.83$ & $4.93$ & $4.8$ & $0.07$ & V17 & YSG\\ 
2011dh & $22$ & $8$ & $2223$ & $7.64$ & $0.142$ & $3.81$ & $5.3$ & $4.8$ & $0.1385$ & $X_\mathrm{s}\ge 0.4$ & YSG\\ 
\hline
2013df & $12$ & $8$ & $669$ & $3.46$ & $0.065$ & $3.66$ & $4.76$ & $3.1$ & $0.0968$ & $X_\mathrm{s}\ge 0.4$ & RSG\\ 
2013df & $12$ & $8$ & $669$ & $3.46$ & $0.065$ & $3.66$ & $4.76$ & $3.1$ & $0.1016$ & $X_\mathrm{s}\ge 0.4$ & RSG\\ 
2013df & $13$ & $12$ & $669$ & $4.04$ & $0.126$ & $3.64$ & $4.87$ & $5.6$ & $0.0968$ & $X_\mathrm{s}\ge 0.4$ & RSG\\ 
2013df & $13$ & $8$ & $1219$ & $4.01$ & $0.116$ & $3.64$ & $4.87$ & $5.6$ & $0.1063$ & $X_\mathrm{s}\ge 0.4$ & RSG\\ 
\hline
$\mathrm{iPTF13bvn}_\mathrm{E}$ & $25$ & $9$ & $60$ & $8.54$ & $0.028$ & $4.34$ & $5.4$ & $3.1$ & $0.1237$ & $X_\mathrm{s}\ge 0.4$ & BSG\\ 
$\mathrm{iPTF13bvn}_\mathrm{E}$ & $25$ & $9$ & $60$ & $8.54$ & $0.028$ & $4.34$ & $5.4$ & $3.1$ & $0.1242$ & $X_\mathrm{s}\ge 0.4$ & BSG\\ 
$\mathrm{iPTF13bvn}_\mathrm{E}$ & $25$ & $9$ & $60$ & $8.54$ & $0.028$ & $4.34$ & $5.4$ & $2.2$ & $0.1237$ & $X_\mathrm{s}\ge 0.4$ & BSG\\ 
$\mathrm{iPTF13bvn}_\mathrm{E}$ & $25$ & $9$ & $60$ & $8.54$ & $0.028$ & $4.34$ & $5.4$ & $2$ & $0.14$ & $X_\mathrm{s}\ge 0.4$ & BSG\\ 
\hline
$\mathrm{iPTF13bvn}_\mathrm{F}$ & $14$ & $12$ & $10$ & $4.1$ & $0.029$ & $4.13$ & $4.9$ & $3.1$ & $0.1237$ & V17 & BSG\\ 
$\mathrm{iPTF13bvn}_\mathrm{F}$ & $14$ & $12$ & $10$ & $4.1$ & $0.029$ & $4.13$ & $4.9$ & $3.1$ & $0.1242$ & V17 & BSG\\ 
$\mathrm{iPTF13bvn}_\mathrm{F}$ & $16$ & $10$ & $201$ & $5.16$ & $0.044$ & $4.17$ & $5.05$ & $5.9$ & $0.1237$ & V17 & BSG\\ 
$\mathrm{iPTF13bvn}_\mathrm{F}$ & $16$ & $10$ & $201$ & $5.16$ & $0.044$ & $4.17$ & $5.05$ & $6$ & $0.1283$ & V17 & BSG\\ \hline
2016gkg & $22$ & $14$ & $110$ & $7.23$ & $0.038$ & $3.96$ & $5.29$ & $3.1$ & $0.1071$ & $X_\mathrm{s}\ge 0.4$ & BSG\\ 
2016gkg & $22$ & $8$ & $367$ & $7.22$ & $0.034$ & $4$ & $5.29$ & $3.1$ & $0.1581$ & $X_\mathrm{s}\ge 0.4$ & BSG\\ 
2016gkg & $22$ & $21$ & $60$ & $7.21$ & $0.037$ & $3.97$ & $5.29$ & $6$ & $0.1071$ & $X_\mathrm{s}\ge 0.4$ & BSG\\ 
2016gkg & $25$ & $22$ & $60$ & $8.55$ & $0.045$ & $4.05$ & $5.4$ & $6$ & $0.1517$ & $X_\mathrm{s}\ge 0.4$ & BSG\\ 
\hline
2019yvr & $12$ & $11$ & $110$ & $3.3$ & $0.03$ & $3.73$ & $4.73$ & $3.1$ & $0.53$ & V17 & YSG\\ 
2019yvr & $19$ & $12$ & $201$ & $5.82$ & $0.051$ & $3.83$ & $5.14$ & $3.1$ & $0.7903$ & $X_\mathrm{s}\ge 0.4$ & YSG\\ 
2019yvr & $22$ & $8$ & $2223$ & $7.64$ & $0.142$ & $3.81$ & $5.3$ & $5.6$ & $0.53$ & $X_\mathrm{s}\ge 0.4$ & YSG\\ 
2019yvr & $19$ & $12$ & $201$ & $5.82$ & $0.051$ & $3.83$ & $5.14$ & $3.2$ & $0.7766$ & $X_\mathrm{s}\ge 0.4$ & YSG\\ 
\hline
\hline
\end{tabular}
\footnotesize
\begin{tablenotes}
\textit{Notes.} For each SN the first line is for fixed $R_V$ and $E\left(B-V\right)$, the second line is for fixed $R_V$ and variable $E\left(B-V\right)$, the third line is for variable $R_V$ and fixed $E\left(B-V\right)$ and the fourth line is for variable $R_V$ and $E\left(B-V\right)$. Models which have $X_\mathrm{s}\ge 0.4$ in the wind column did not have any point in their evolution where $T_\mathrm{eff} > 10000\,\mathrm{K}$ and $X_\mathrm{s} < 0.4$, and therefore neither the \citetalias{NL00} wind scheme nor the \citetalias{Vink2017} was employed. 
${^a}$ Stellar Type (defined in Section \ref{subsec:EvolutionEndPoints}).
\end{tablenotes}
\end{threeparttable}
\label{tab:bestfitdetails}
\end{table*}
To show not only the best-fitting models but all those with relatively low $\chi^2$, we colour-code all models according to their computed $\chi^2$, in Figure \ref{fig:chi2_MH_vs_Teff}, for the seven SNe. For the computation of $\chi^2$ as marked in Figure \ref{fig:chi2_MH_vs_Teff} we chose the fourth approach to dust extinction, as described above, allowing both $R_V$ and $E\left(B-V\right)$ to vary. The details for all best-fitting models are presented in Table \ref{tab:bestfitdetails}.

\subsection{Monte Carlo realisations}
\label{subsec:MCbest}

\begin{figure*}
  \centering
    \begin{subfigure}{0.49\textwidth}
    \centering
    \includegraphics[width=\textwidth]{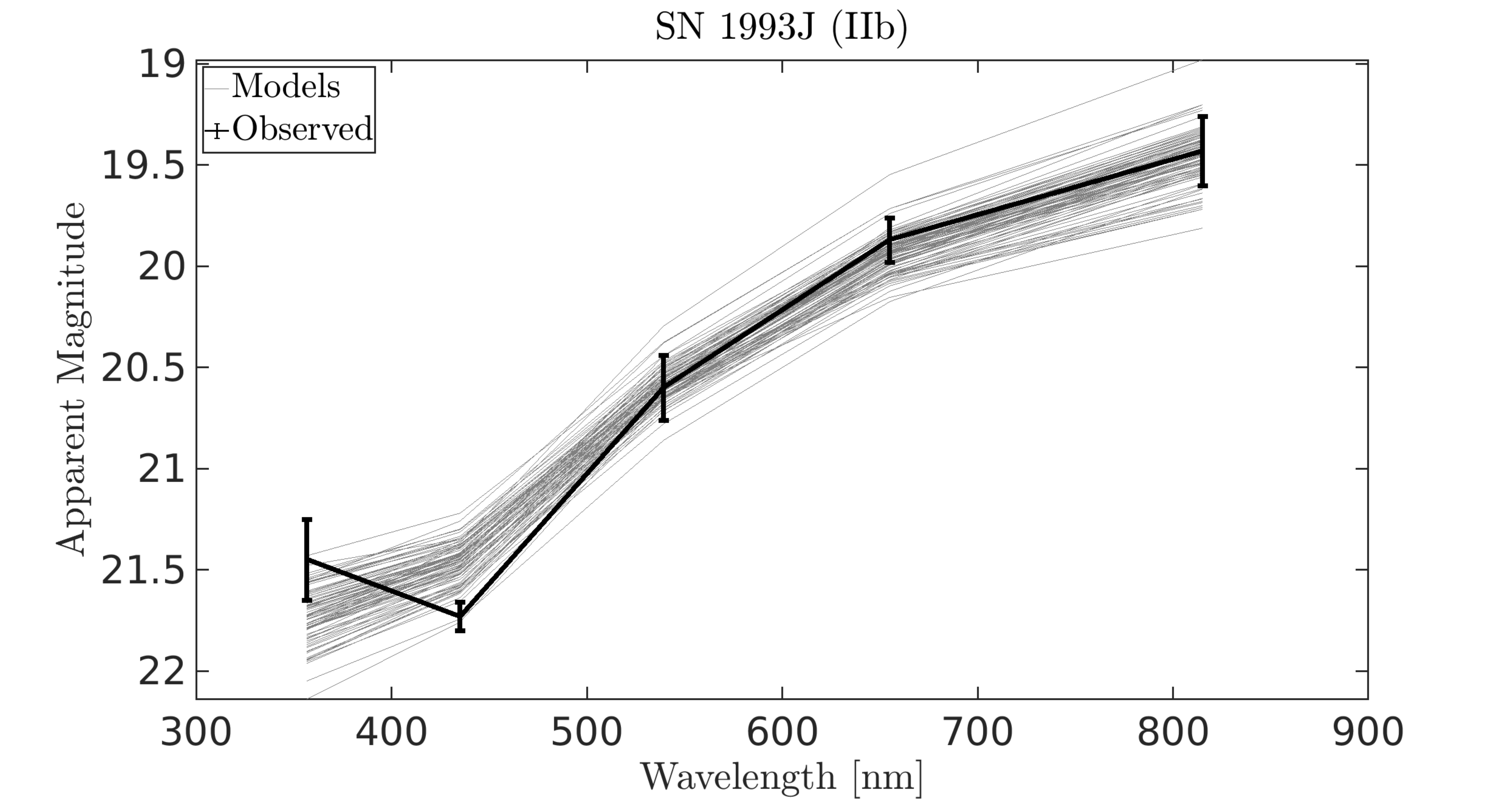}
    \label{fig:MC_SN1993J}
  \end{subfigure}
  \begin{subfigure}{0.49\textwidth}
    \centering
    \includegraphics[width=\textwidth]{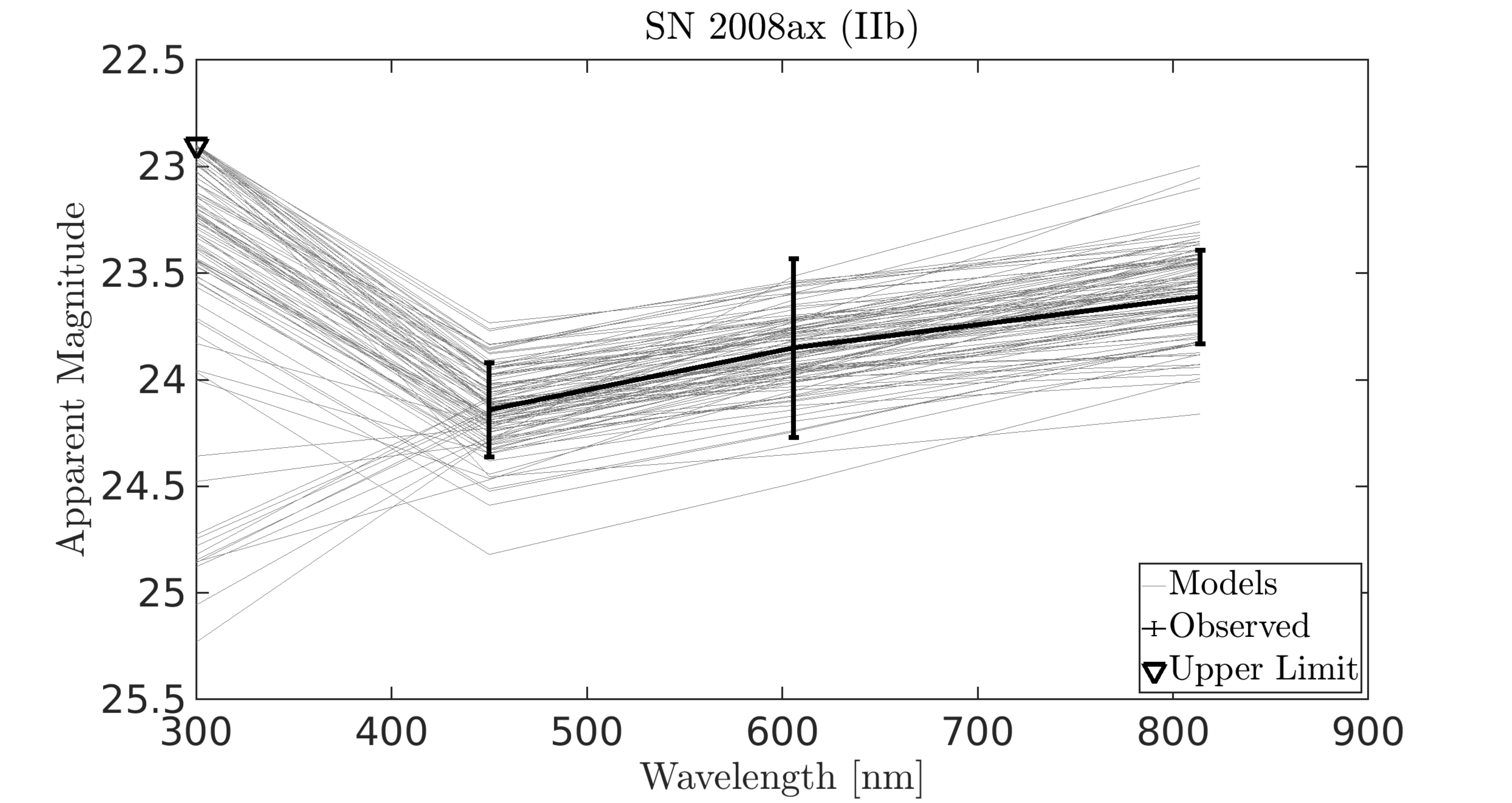}
    \label{fig:MC_SN2008ax}
  \end{subfigure}\\
    \begin{subfigure}{0.49\textwidth}
    \centering
    \includegraphics[width=\textwidth]{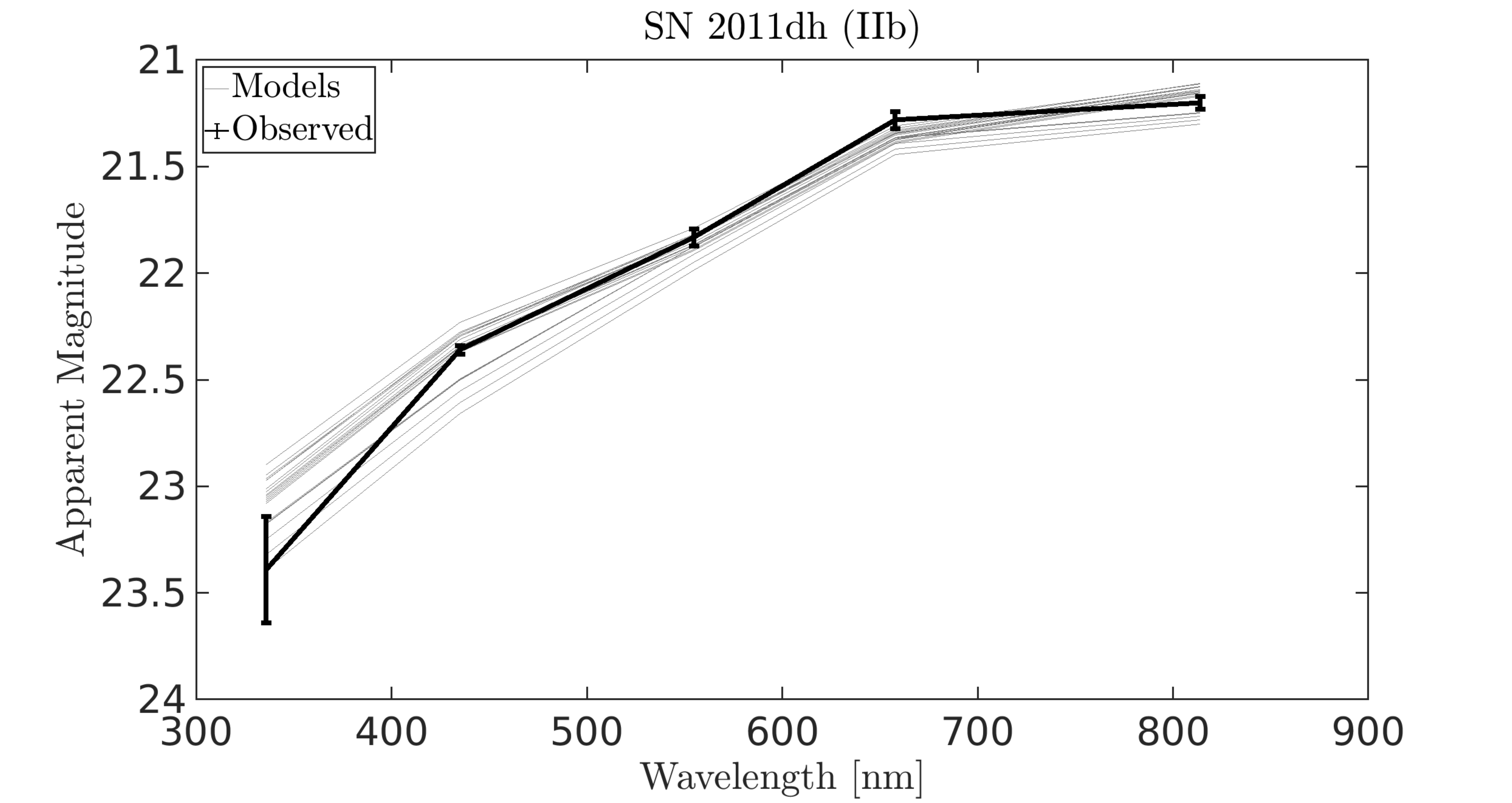}
    \label{fig:MC_SN2011dh}
  \end{subfigure}
  \centering
    \begin{subfigure}{0.49\textwidth}
    \centering
    \includegraphics[width=\textwidth]{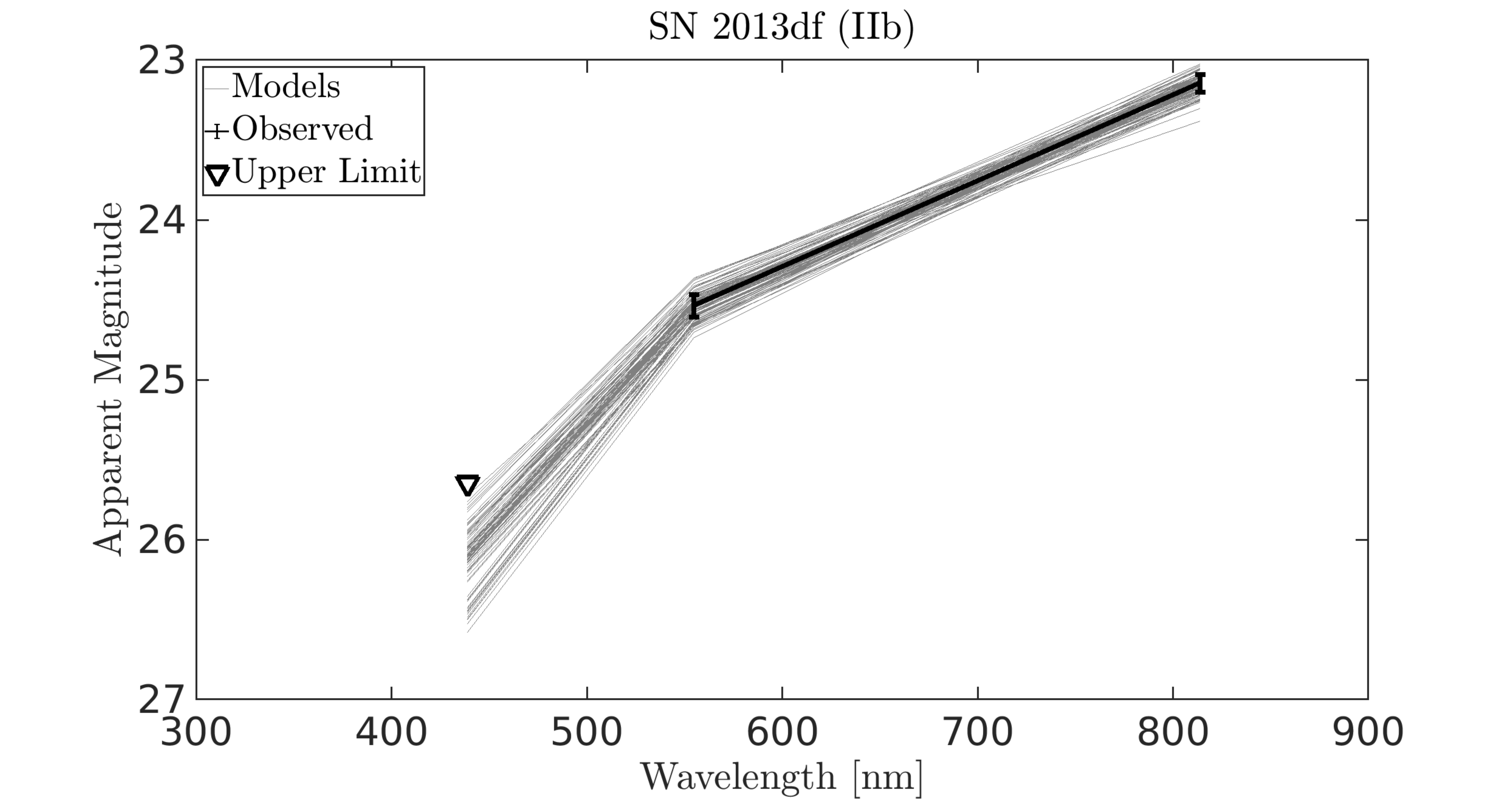}
    \label{fig:MC_SN2013df}
  \end{subfigure}\\
    \begin{subfigure}{0.49\textwidth}
    \centering
    \includegraphics[width=\textwidth]{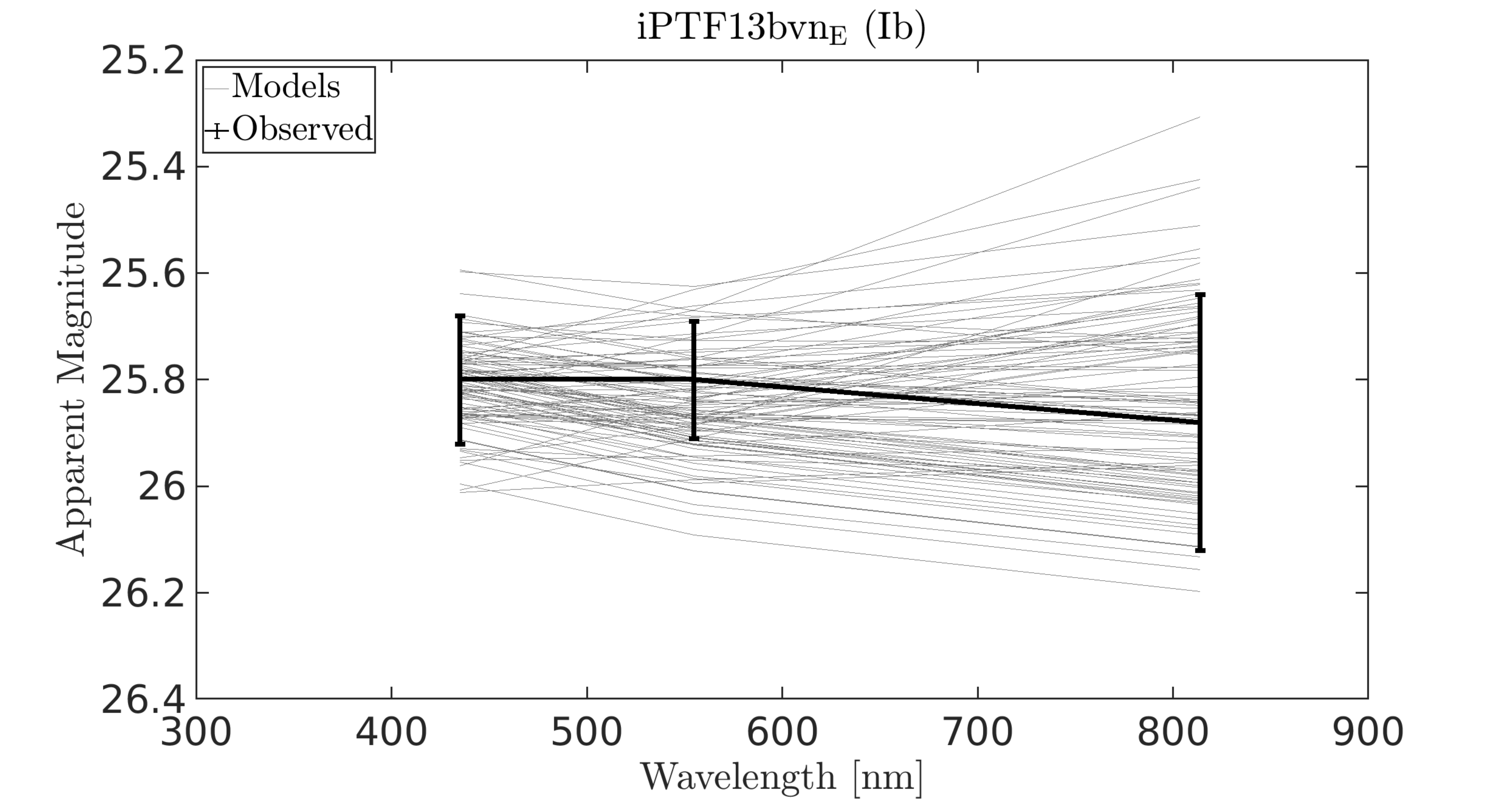}
    \label{fig:MC_iPTF13bvnE}
  \end{subfigure}
  \centering
    \begin{subfigure}{0.49\textwidth}
    \centering
    \includegraphics[width=\textwidth]{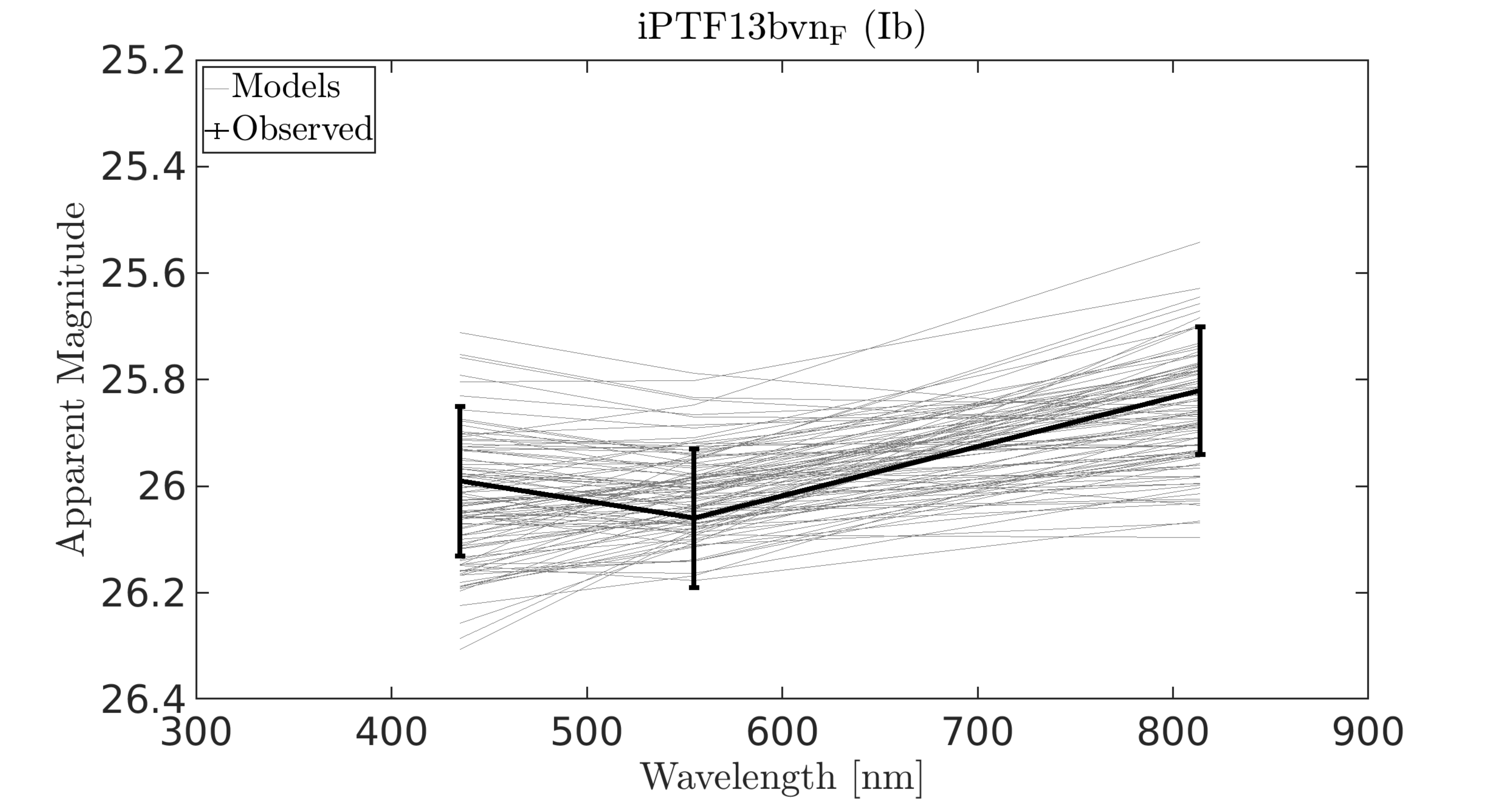}
    \label{fig:MC_iPTF13bvnF}
  \end{subfigure}\\
    \begin{subfigure}{0.49\textwidth}
    \centering
    \includegraphics[width=\textwidth]{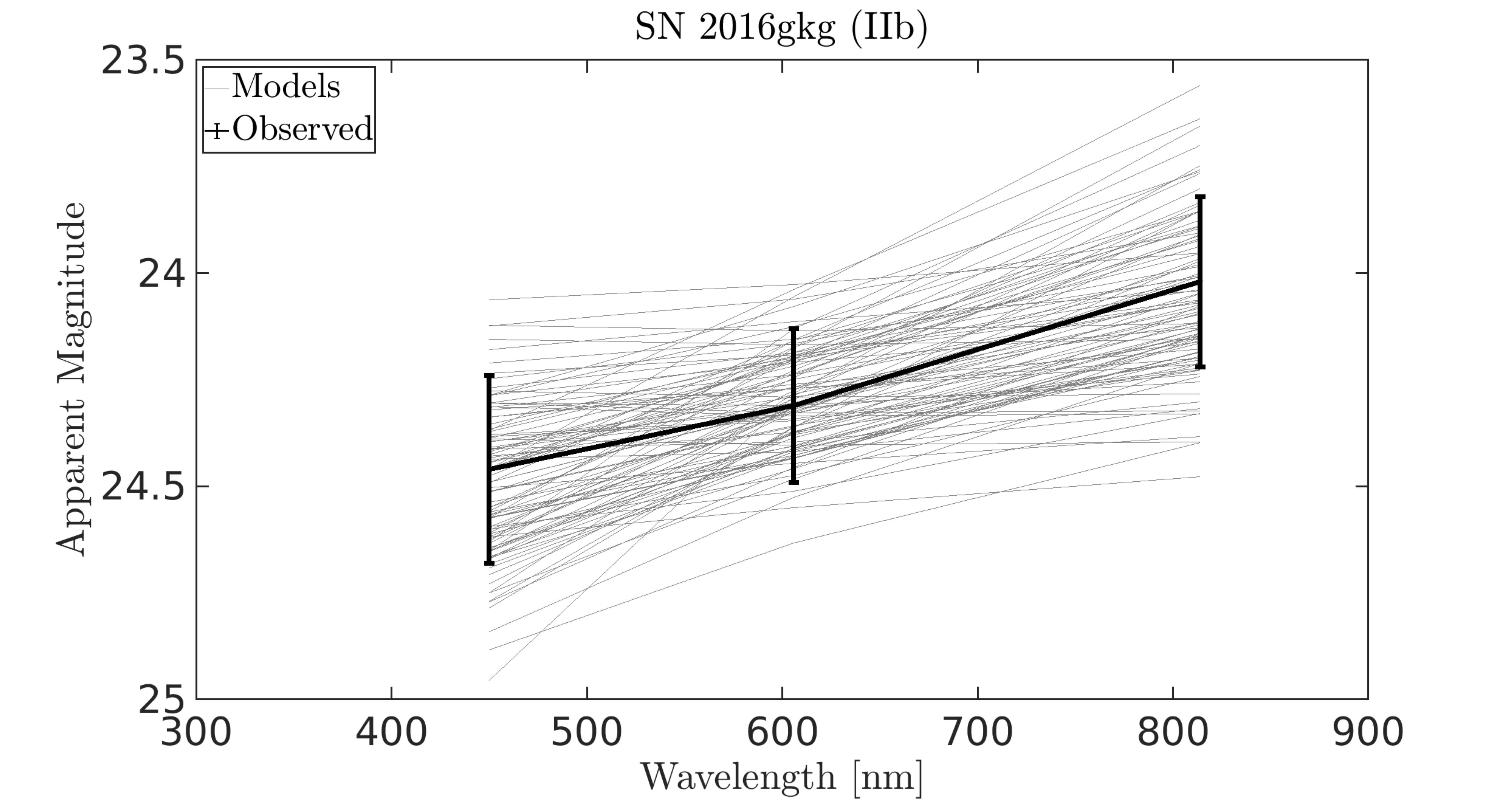}
    \label{fig:MC_SN2016gkg}
  \end{subfigure}
  \begin{subfigure}{0.49\textwidth}
    \centering
    \includegraphics[width=\textwidth]{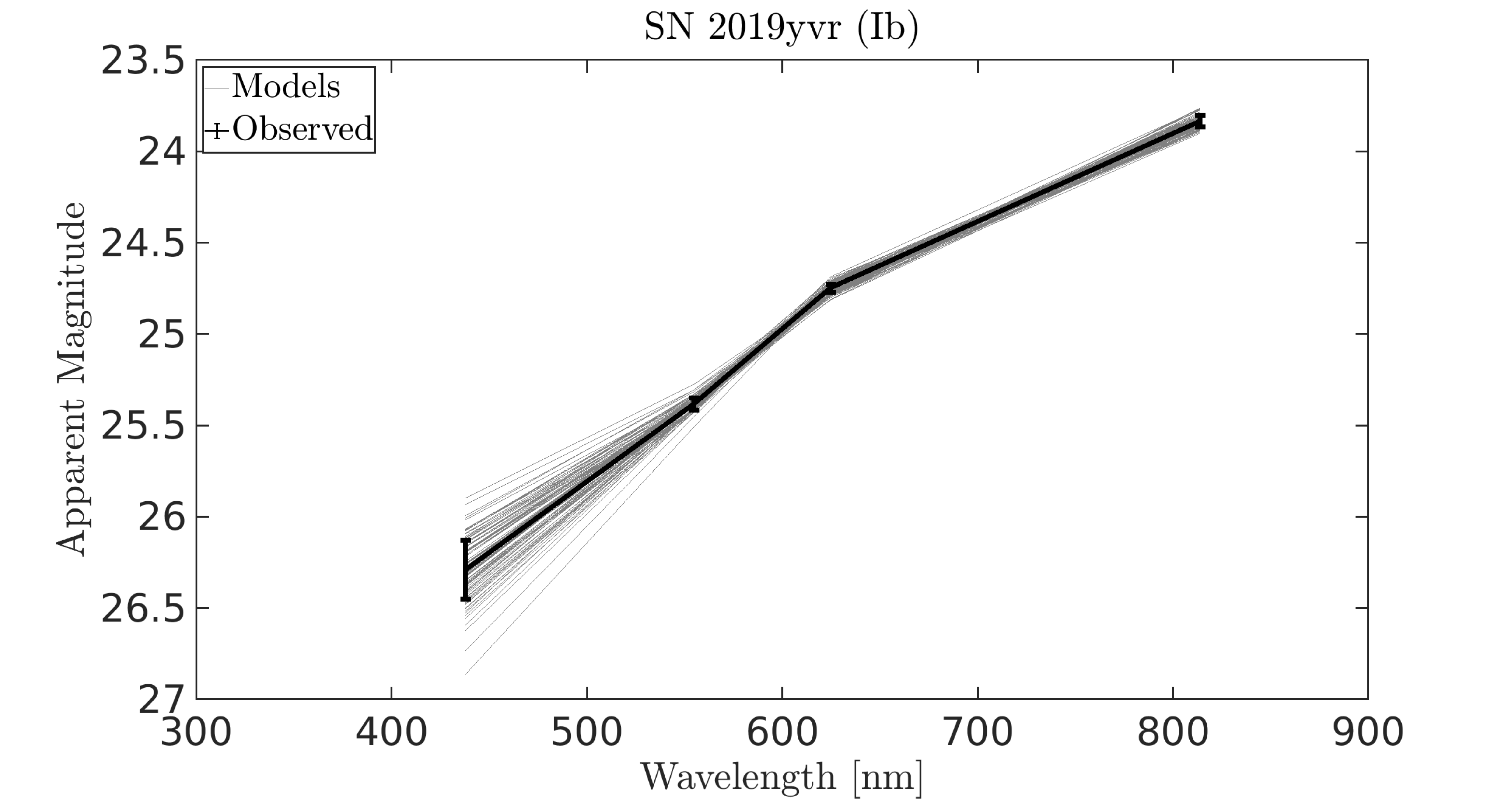}
    \label{fig:MC_SN2019yvr}
  \end{subfigure}\\
  \caption{Observed pre-explosion magnitudes (thick black lines with error bars) and upper limits (black triangles) compared to best-fitting models to mock observations generated by Monte Carlo realisations (thin grey lines).}
  \label{fig:MC_magnitudes}
\end{figure*}
\begin{figure}
  \includegraphics[width=0.48\textwidth]{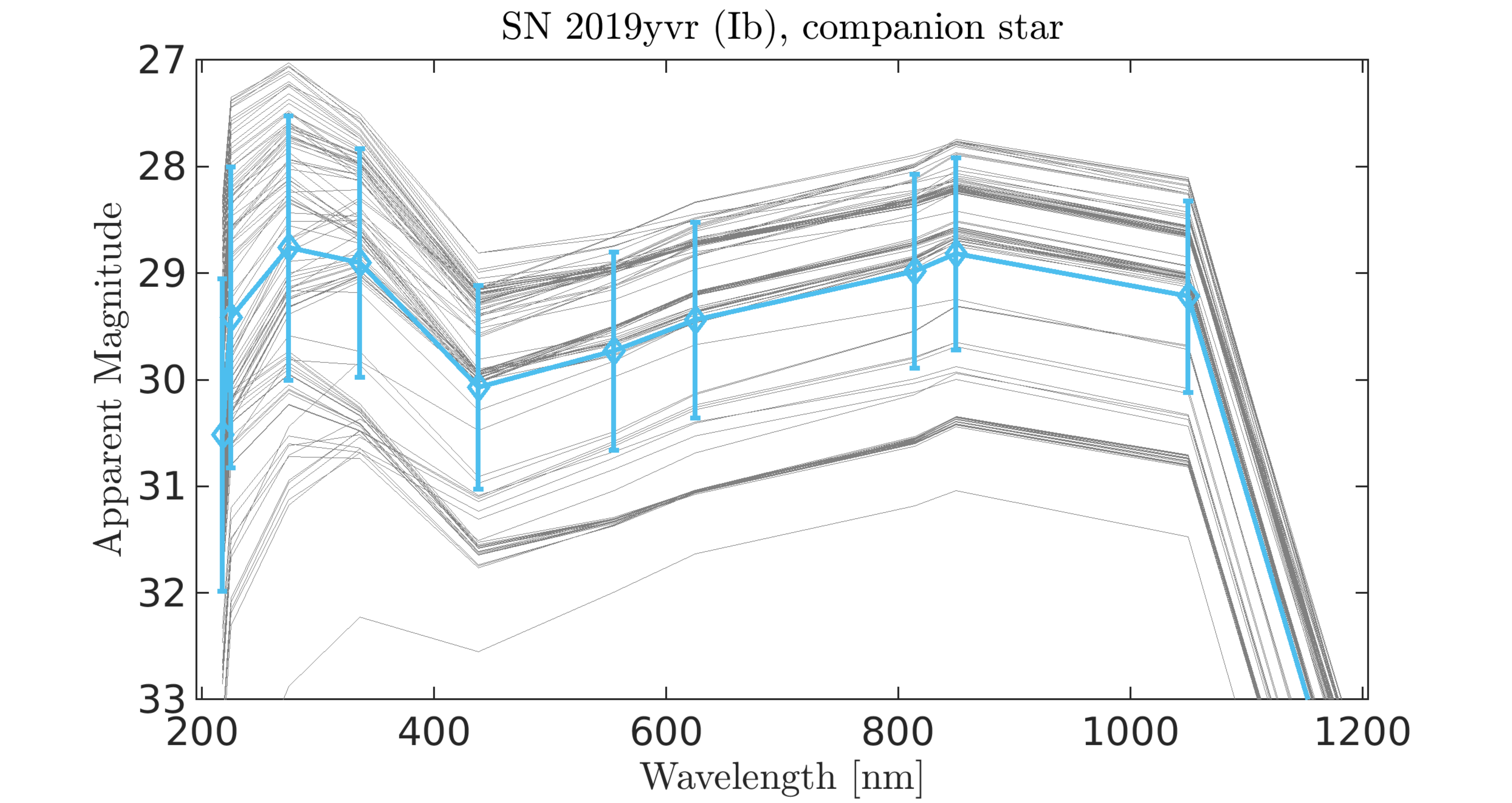}
  \caption{Computed magnitudes of the companion star from our Monte Carlo best-fitting models for SN~2019yvr, obtained with variable $E\left(B-V\right)$ and $R_V$. The mean computed surviving magnitudes of the companion and their standard deviation are shown in light blue.}
  \label{fig:MC_magnitudes_post_SN2019yvr}
\end{figure}
\begin{figure*}
  \centering
  \begin{subfigure}{\textwidth}
    \centering
    \includegraphics[width=\textwidth]{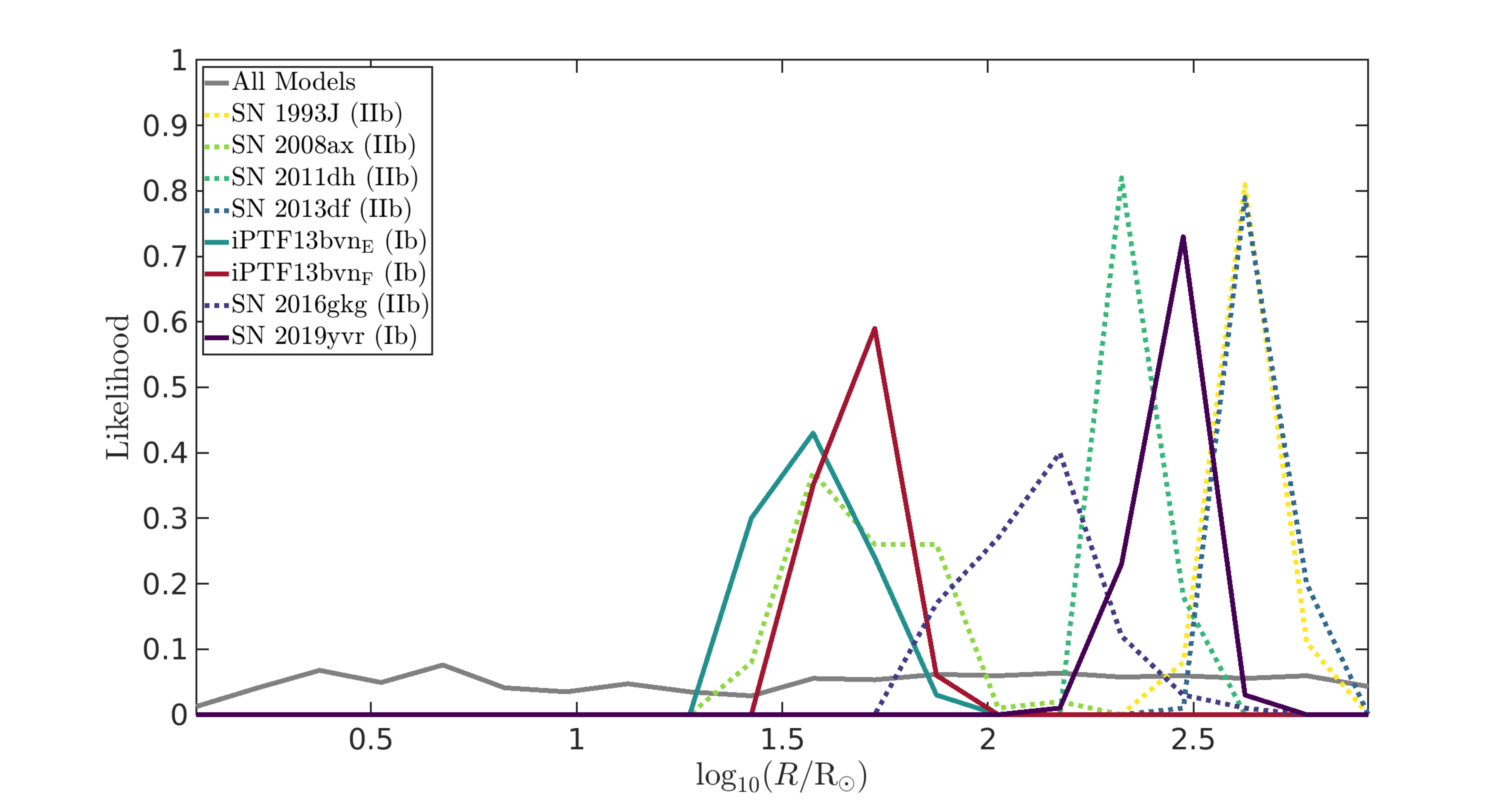}
    \label{fig:MC_like_R}
  \end{subfigure}\\
  \begin{subfigure}{\textwidth}
    \centering
    \includegraphics[width=\textwidth]{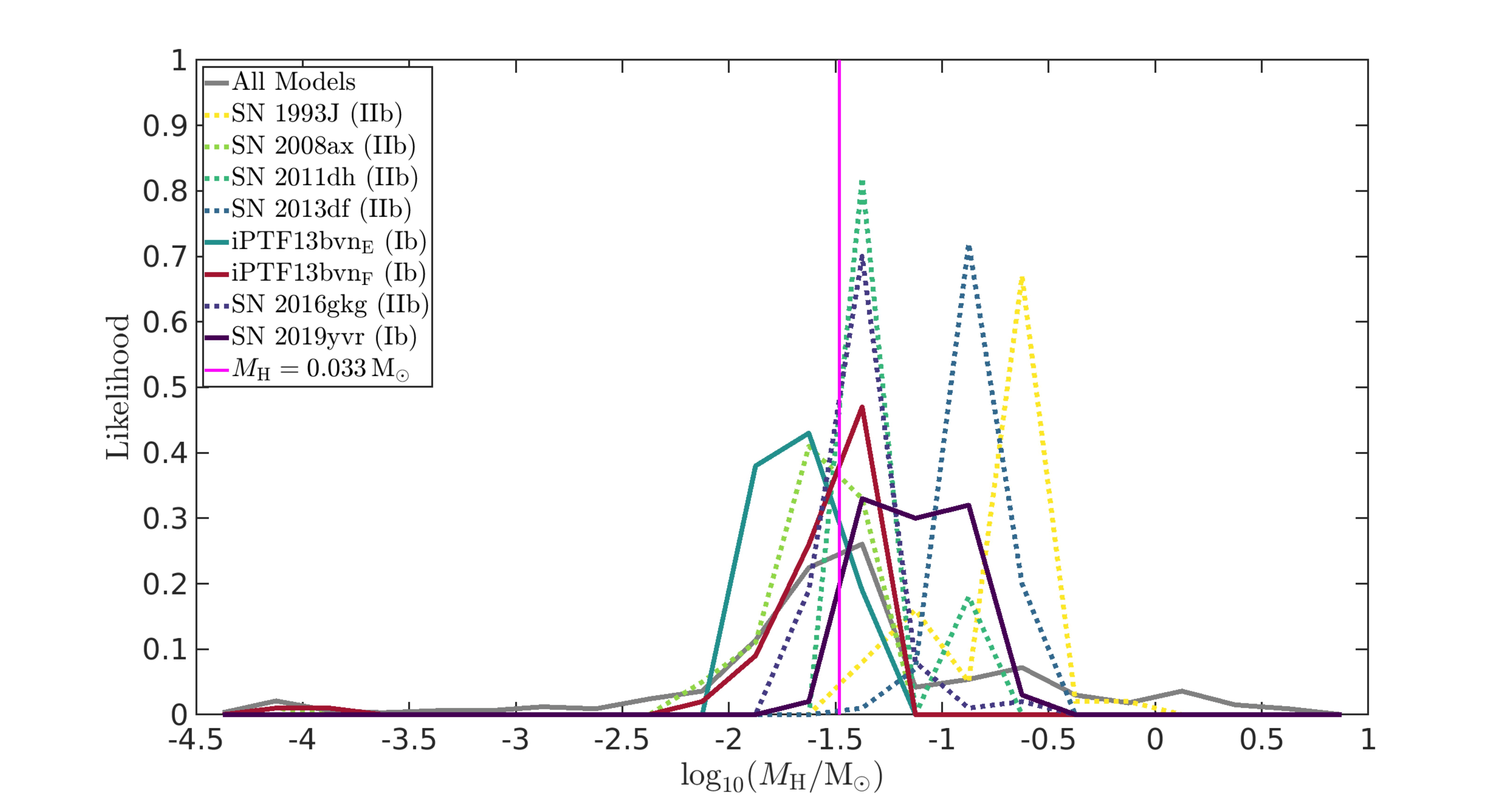}
    \label{fig:MC_like_MH}
  \end{subfigure}\\
  \caption{Likelihood distributions of the stellar radius (top) and hydrogen mass (bottom). The likelihood distributions are generated by counting the number of models in $\log_{10}\left(R/ \mathrm{R}_\odot\right)$ bins of width $0.15$ and $\log_{10}\left(M_\mathrm{H}/ \mathrm{M}_\odot\right)$ bins of width $0.25$.}
  \label{fig:MC_like_R_and_MH}
\end{figure*}
\begin{figure*}
  \includegraphics[width=\textwidth]{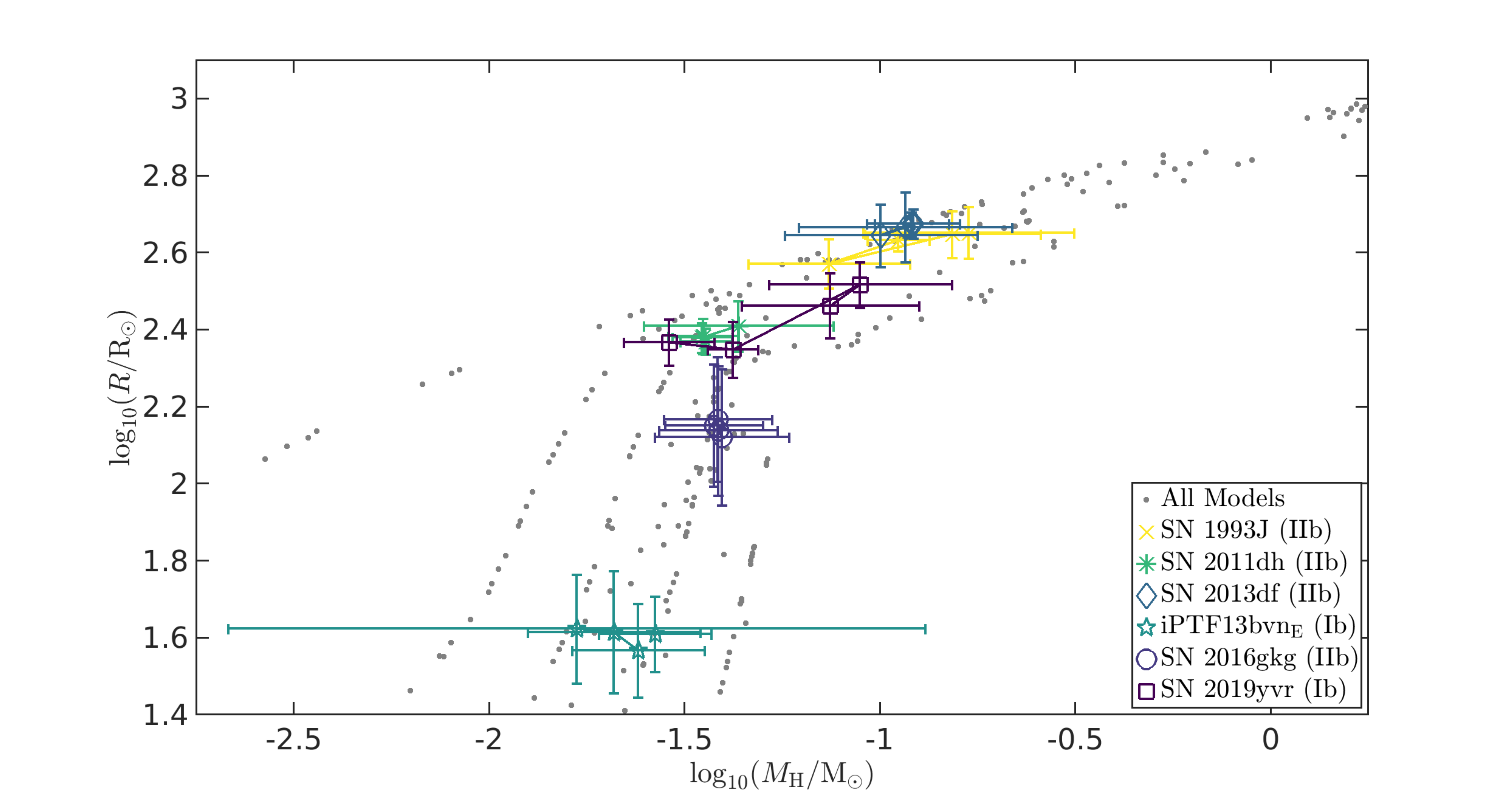}
  \caption{Mean progenitor stellar radius vs. mean hydrogen mass and their standard deviations. The primary components in all computed stellar models are also shown, for reference.}
  \label{fig:MC_like_R_and_MH_norm}
\end{figure*}
\begin{figure*}
  \includegraphics[width=\textwidth]{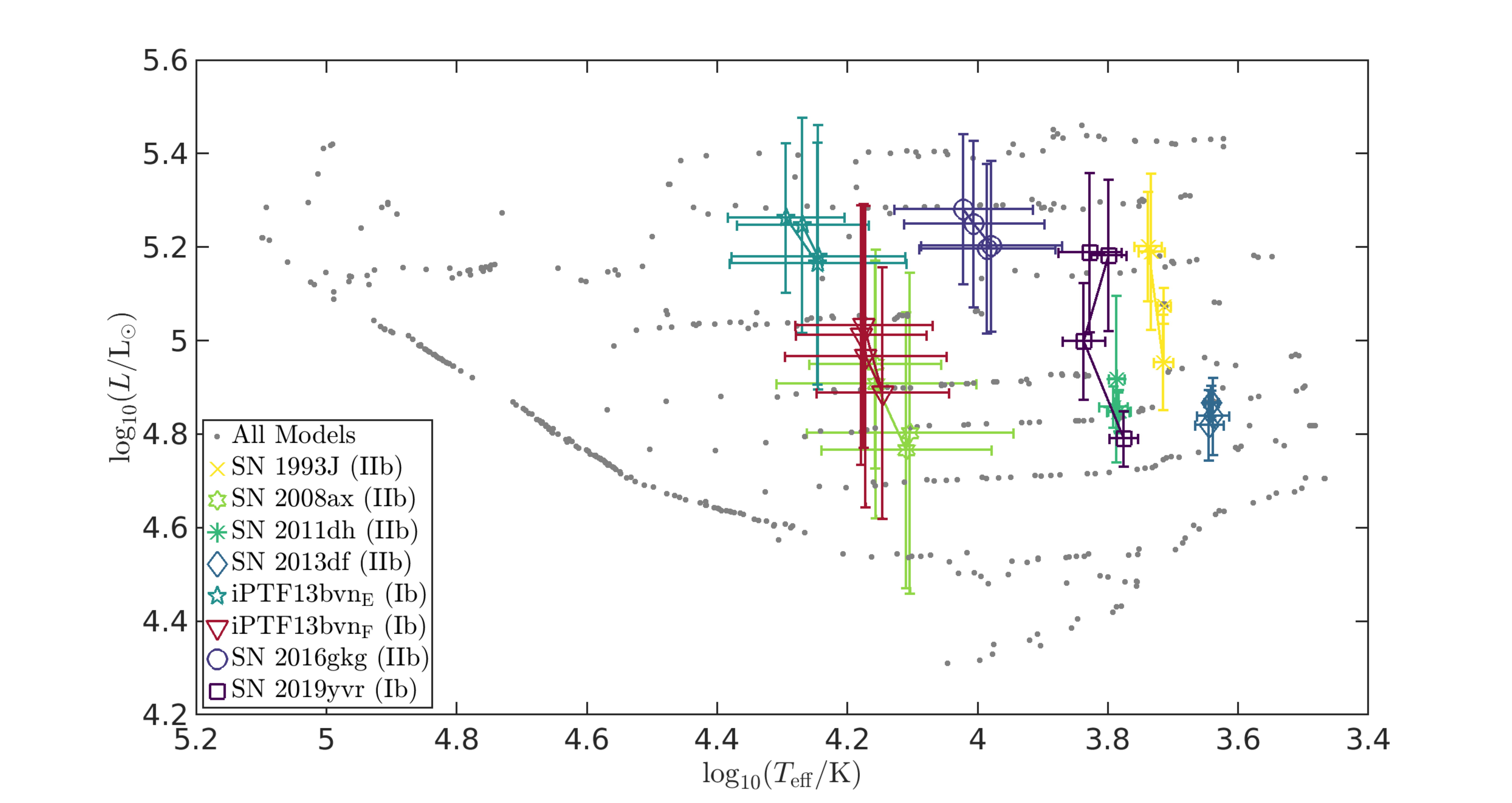}
  \caption{Mean luminosity and mean effective surface temperature with their standard deviations for all SN progenitors. The primary components in all computed stellar models are also shown, for reference.}
  \label{fig:MC_HRD}
\end{figure*}
\begin{table*}
\centering
\caption{Effective surface temperature, luminosity, total hydrogen mass and stellar radius and their standard deviations, and occurrences of stellar types (defined in Section \ref{subsec:EvolutionEndPoints}) and wind schemes in fits to Monte Carlo realisations.}
\begin{threeparttable}
\begin{tabular}{c|cccccccccccc}
\hline
SN & $\log_{10}(T_\mathrm{eff}/ \mathrm{K})$ & $\log_{10}(L / \mathrm{L}_\odot)$ & $\log_{10}(M_\mathrm{H} / \mathrm{M}_\odot)$ & $\log_{10}(R / \mathrm{R}_\odot)$ & RSG & YSG & BSG & HeG & WR & $X_\mathrm{s}\ge 0.4$ & V17 & NL00 \\ 
\hline
1993J & $3.71\pm 0.01$ & $5.07\pm 0.04$ & $-0.95\pm 0.08$ & $2.63\pm 0.03$ & $0$ & $100$ & $0$ & $0$ & $0$ & $98$ & $2$ & $0$\\ 
1993J & $3.71\pm 0.01$ & $4.95\pm 0.1$ & $-1.13\pm 0.21$ & $2.57\pm 0.06$ & $0$ & $100$ & $0$ & $0$ & $0$ & $99$ & $1$ & $0$\\ 
1993J & $3.74\pm 0.02$ & $5.2\pm 0.12$ & $-0.81\pm 0.22$ & $2.65\pm 0.06$ & $0$ & $100$ & $0$ & $0$ & $0$ & $96$ & $4$ & $0$\\ 
1993J & $3.73\pm 0.02$ & $5.19\pm 0.17$ & $-0.77\pm 0.27$ & $2.65\pm 0.07$ & $1$ & $99$ & $0$ & $0$ & $0$ & $99$ & $1$ & $0$\\  
\hline
2008ax & $4.1\pm 0.16$ & $4.8\pm 0.34$ & $-5.26\pm 6.38$ & $1.72\pm 0.17$ & $0$ & $0$ & $75$ & $25$ & $0$ & $22$ & $53$ & $25$\\ 
2008ax & $4.11\pm 0.13$ & $4.77\pm 0.3$ & $-4.65\pm 3.98$ & $1.69\pm 0.16$ & $0$ & $0$ & $72$ & $28$ & $0$ & $18$ & $54$ & $28$\\ 
2008ax & $4.16\pm 0.15$ & $4.91\pm 0.29$ & $-3.32\pm 4.26$ & $1.67\pm 0.19$ & $0$ & $0$ & $84$ & $16$ & $0$ & $16$ & $68$ & $16$\\ 
2008ax & $4.16\pm 0.1$ & $4.95\pm 0.22$ & $-2.39\pm 2.52$ & $1.68\pm 0.15$ & $0$ & $0$ & $91$ & $9$ & $0$ & $19$ & $72$ & $9$\\ 
\hline
2011dh & $3.79\pm 0.02$ & $4.86\pm 0.04$ & $-1.45\pm 0.07$ & $2.37\pm 0.03$ & $0$ & $100$ & $0$ & $0$ & $0$ & $1$ & $99$ & $0$\\ 
2011dh & $3.78\pm 0.02$ & $4.85\pm 0.05$ & $-1.45\pm 0.09$ & $2.38\pm 0.05$ & $0$ & $100$ & $0$ & $0$ & $0$ & $3$ & $97$ & $0$\\ 
2011dh & $3.78\pm 0.02$ & $4.85\pm 0.04$ & $-1.46\pm 0.08$ & $2.38\pm 0.04$ & $0$ & $100$ & $0$ & $0$ & $0$ & $2$ & $98$ & $0$\\ 
2011dh & $3.79\pm 0.01$ & $4.92\pm 0.18$ & $-1.36\pm 0.24$ & $2.41\pm 0.07$ & $0$ & $100$ & $0$ & $0$ & $0$ & $18$ & $82$ & $0$\\ 
\hline
2013df & $3.64\pm 0.02$ & $4.84\pm 0.08$ & $-0.93\pm 0.27$ & $2.67\pm 0.09$ & $100$ & $0$ & $0$ & $0$ & $0$ & $100$ & $0$ & $0$\\ 
2013df & $3.64\pm 0.02$ & $4.82\pm 0.08$ & $-1\pm 0.25$ & $2.64\pm 0.08$ & $100$ & $0$ & $0$ & $0$ & $0$ & $100$ & $0$ & $0$\\ 
2013df & $3.64\pm 0.01$ & $4.87\pm 0.03$ & $-0.92\pm 0.09$ & $2.67\pm 0.03$ & $100$ & $0$ & $0$ & $0$ & $0$ & $100$ & $0$ & $0$\\ 
2013df & $3.64\pm 0.01$ & $4.87\pm 0.04$ & $-0.91\pm 0.12$ & $2.67\pm 0.04$ & $99$ & $1$ & $0$ & $0$ & $0$ & $100$ & $0$ & $0$\\ 
\hline
$\mathrm{iPTF13bvn}_\mathrm{E}$ & $4.27\pm 0.1$ & $5.25\pm 0.23$ & $-1.58\pm 0.14$ & $1.61\pm 0.1$ & $0$ & $0$ & $100$ & $0$ & $0$ & $69$ & $31$ & $0$\\ 
$\mathrm{iPTF13bvn}_\mathrm{E}$ & $4.24\pm 0.13$ & $5.18\pm 0.28$ & $-1.78\pm 0.89$ & $1.62\pm 0.14$ & $0$ & $0$ & $99$ & $1$ & $0$ & $68$ & $31$ & $1$\\ 
$\mathrm{iPTF13bvn}_\mathrm{E}$ & $4.25\pm 0.14$ & $5.16\pm 0.26$ & $-1.68\pm 0.22$ & $1.61\pm 0.16$ & $0$ & $0$ & $100$ & $0$ & $0$ & $65$ & $35$ & $0$\\ 
$\mathrm{iPTF13bvn}_\mathrm{E}$ & $4.29\pm 0.09$ & $5.26\pm 0.16$ & $-1.62\pm 0.17$ & $1.57\pm 0.12$ & $0$ & $0$ & $100$ & $0$ & $0$ & $78$ & $22$ & $0$\\ 
\hline
$\mathrm{iPTF13bvn}_\mathrm{F}$ & $4.17\pm 0.12$ & $4.97\pm 0.32$ & $-2.41\pm 2.08$ & $1.66\pm 0.1$ & $0$ & $0$ & $94$ & $6$ & $0$ & $38$ & $56$ & $6$\\ 
$\mathrm{iPTF13bvn}_\mathrm{F}$ & $4.15\pm 0.1$ & $4.89\pm 0.27$ & $-2.21\pm 2.46$ & $1.68\pm 0.09$ & $0$ & $0$ & $98$ & $2$ & $0$ & $19$ & $79$ & $2$\\ 
$\mathrm{iPTF13bvn}_\mathrm{F}$ & $4.17\pm 0.11$ & $5.03\pm 0.26$ & $-1.94\pm 1.75$ & $1.69\pm 0.1$ & $0$ & $0$ & $96$ & $4$ & $0$ & $26$ & $70$ & $4$\\ 
$\mathrm{iPTF13bvn}_\mathrm{F}$ & $4.18\pm 0.1$ & $5.01\pm 0.28$ & $-2.75\pm 3.03$ & $1.67\pm 0.09$ & $0$ & $0$ & $86$ & $14$ & $0$ & $26$ & $60$ & $14$\\ 
\hline
2016gkg & $3.99\pm 0.1$ & $5.2\pm 0.18$ & $-1.42\pm 0.13$ & $2.15\pm 0.16$ & $0$ & $13$ & $87$ & $0$ & $0$ & $76$ & $18$ & $6$\\ 
2016gkg & $3.98\pm 0.11$ & $5.2\pm 0.18$ & $-1.41\pm 0.14$ & $2.17\pm 0.16$ & $0$ & $13$ & $87$ & $0$ & $0$ & $76$ & $8$ & $16$\\ 
2016gkg & $4.01\pm 0.11$ & $5.25\pm 0.18$ & $-1.41\pm 0.15$ & $2.14\pm 0.17$ & $0$ & $11$ & $89$ & $0$ & $0$ & $83$ & $17$ & $0$\\ 
2016gkg & $4.02\pm 0.11$ & $5.28\pm 0.16$ & $-1.4\pm 0.17$ & $2.12\pm 0.18$ & $0$ & $6$ & $94$ & $0$ & $0$ & $89$ & $10$ & $1$\\  
\hline
2019yvr & $3.78\pm 0.02$ & $4.79\pm 0.06$ & $-1.54\pm 0.12$ & $2.37\pm 0.06$ & $0$ & $100$ & $0$ & $0$ & $0$ & $0$ & $100$ & $0$\\ 
2019yvr & $3.84\pm 0.03$ & $5\pm 0.12$ & $-1.38\pm 0.06$ & $2.35\pm 0.07$ & $0$ & $91$ & $9$ & $0$ & $0$ & $41$ & $38$ & $21$\\ 
2019yvr & $3.8\pm 0.03$ & $5.18\pm 0.16$ & $-1.05\pm 0.23$ & $2.52\pm 0.06$ & $0$ & $97$ & $3$ & $0$ & $0$ & $86$ & $14$ & $0$\\ 
2019yvr & $3.83\pm 0.05$ & $5.19\pm 0.17$ & $-1.13\pm 0.23$ & $2.46\pm 0.08$ & $0$ & $83$ & $17$ & $0$ & $0$ & $85$ & $14$ & $1$\\ 
\hline
\hline
\end{tabular}
\footnotesize
\begin{tablenotes}
\textit{Note.} For each SN the first line is for fixed $R_V$ and $E\left(B-V\right)$, the second line is for fixed $R_V$ and variable $E\left(B-V\right)$, the third line is for variable $R_V$ and fixed $E\left(B-V\right)$ and the fourth line is for variable $R_V$ and $E\left(B-V\right)$.
\end{tablenotes}
\end{threeparttable}
\label{tab:MC_TABLE}
\end{table*}
We employ a Monte Carlo realisations approach to estimate the sensitivity of the observations to the progenitor properties. For each SN progenitor, we generate $100$ mock observations by assuming a normal distribution for each observable with the nominal value as the mean and the error as the standard deviation. For each generated mock observation, we find the evolutionary endpoint and distance which minimise $\sum\limits_{\lambda} \left(m^\mathrm{obs,mock}_\lambda - m^\mathrm{calc}_\lambda\right)^2$, where $m^\mathrm{obs,mock}_\lambda$ are the mock observation magnitudes. Similarly to Section \ref{subsec:chi2best}, we employ four different approaches to dust extinction. The best-fitting models obtained using the fourth dust extinction approach, where both $R_V$ and $E\left(B-V\right)$ are allowed to vary, are presented in Figure \ref{fig:MC_magnitudes}. Each panel shows the observed magnitudes as well as all $100$ synthetic photometry models that were found to fit best the mock observations generated as described above. Each synthetic photometry model represents one evolutionary endpoint, with adjustments according to dust and distance. One endpoint can appear numerous times, with different distance and dust values. By counting the number of appearances of an evolutionary endpoint (out of $100$) we derive statistics for the derived progenitor properties, such as effective surface temperature, luminosity, total hydrogen mass and radius, as well as evolutionary parameters like mass transfer efficiency and wind mass-loss rates. For example, we find that the progenitors of SN~1993J and SN~2016gkg prefer a high mass transfer efficiency ($\approx 0.9$), while SN~2008ax, SN~2011dh and SN~2013df prefer a lower mass transfer efficiency ($\la 0.35$). No strong preference was found for the progenitors of the Type Ib events iPTF13bvn or SN~2019yvr.

All model magnitudes plotted in Figure \ref{fig:MC_magnitudes} are the combinations of contributions from the primary star and from its companion. The contribution of the companion alone is of interest when considering post-SN observations, whether these have already been done, or will be obtained in the future. We plot the computed companion magnitudes in the best-fitting models for SN~2019yvr in Figure \ref{fig:MC_magnitudes_post_SN2019yvr} (the computed companion magnitudes for the five SNe with post-SN upper limits are presented in Appendix \ref{sec:appendixb}), and show the mean and error of our predicted magnitudes for the companion in each filter wavelength. We predict the surviving companion to have an apparent magnitude between $32$ and $27$ Vega magnitudes.

The likelihood distributions for the stellar radius and the hydrogen mass, for the fourth dust extinction approach as described in Section \ref{subsec:chi2best}, are presented in Figure \ref{fig:MC_like_R_and_MH}. The likelihood distribution of all models is also shown, for reference. This is simply the number of computed models in each bin, i.e., all models are assumed equally likely. We do not convolve these results with the various observational estimates of the probability distributions of initial masses \citep[e.g.,][]{Salpeter1955,Kroupa2001}, or binary parameters such as mass ratios or orbital periods \citep[e.g.,][]{Sanaetal2012,MoeDiStefano2017}. The likelihood distributions of the progenitor properties are computed by counting the number of best-fitting models out of the $100$ obtained from fitting the Monte Carlo realisations of the observations.

In Figure \ref{fig:MC_like_R_and_MH_norm} we present the stellar radii and hydrogen masses and their standard deviations for all four dust extinction approaches and for all SN progenitors except SN~2008ax and $\mathrm{iPTF13bvn_\mathrm{F}}$, for which the possibility of hydrogen-free progenitors (Table \ref{tab:MC_TABLE}) causes an extremely large variance in the hydrogen mass estimate. The luminosity and effective surface temperature and their standard deviations are presented in Figure \ref{fig:MC_HRD}. Each SN progenitor is plotted four times, for the four different dust extinction approaches. The progenitors of SN~1993J and SN~2013df are the coolest and largest, while those of SN~2008ax and iPTF13bvn are the hottest and most compact. The progenitor of SN~2016gkg is relatively compact, in agreement with \cite{Arcavietal2017}. The properties derived for the progenitor candidate of SN~2019yvr show the highest sensitivity to the assumptions on dust, especially the luminosity and the total hydrogen mass, with the derived effective surface temperature and radius less affected. The effective surface temperature, luminosity, leftover hydrogen mass and stellar radius (and their standard deviations), and the numbers of occurrences of stellar types and wind schemes, are listed in Table \ref{tab:MC_TABLE}.

\begin{figure}
  \includegraphics[width=0.48\textwidth]{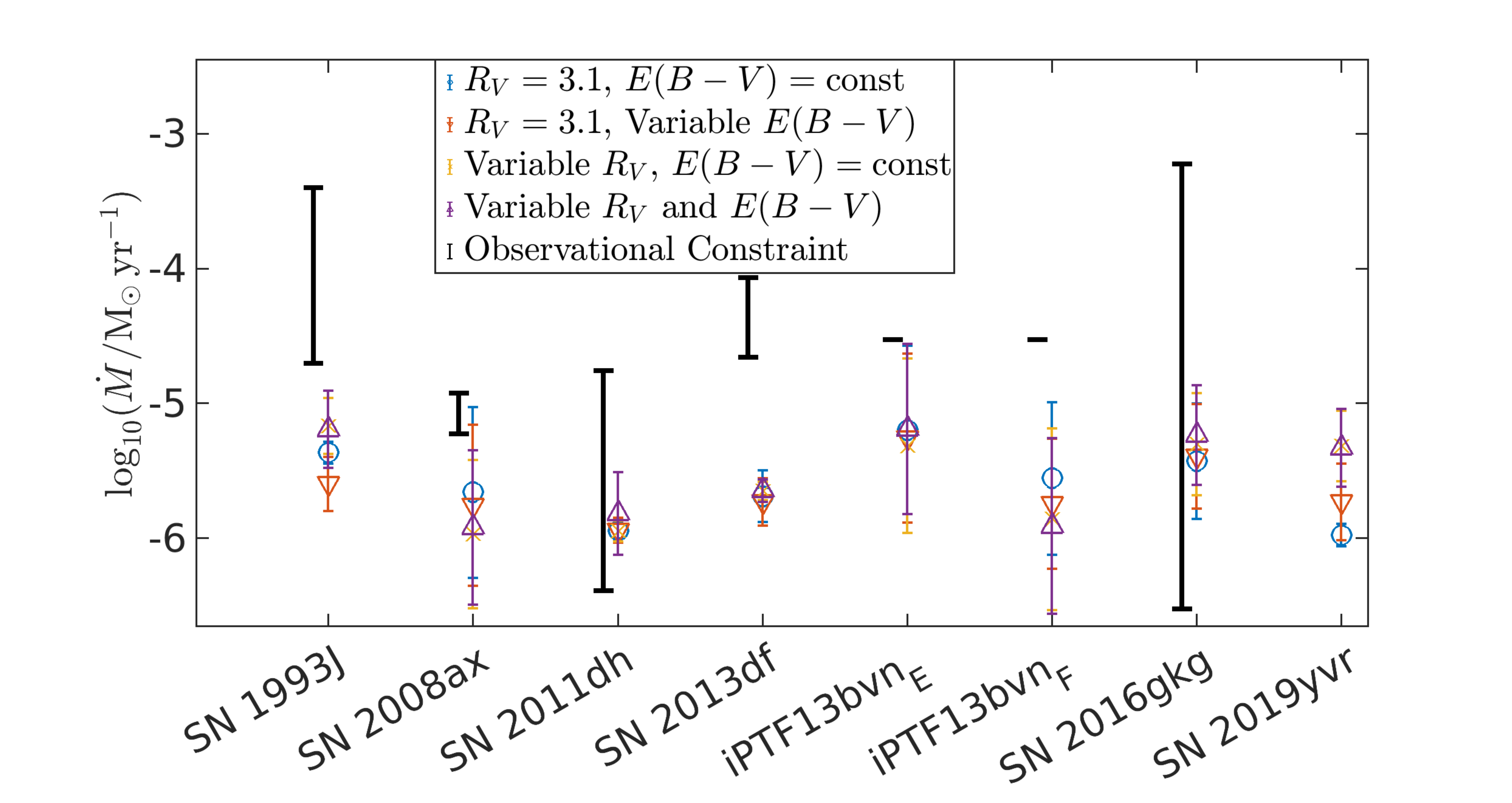}
  \caption{The mass-loss rate (averaged over the last $1000\,\mathrm{yr}$ of evolution) of our best-fitting models compared to observational constraints for SN~1993J (\citealt{VanDyketal1994}; \citealt*{Franssonetal1996,Immleretal2001}), SN~2008ax \citep{Romingetal2009}, SN~2011dh \citep{Kraussetal2012,Maedaetal2014,Kunduetal2019}, SN~2013df \citep{Maedaetal2015} and iPTF13bvn \citep{Caoetal2013}. The observational estimate for the pre-SN mass-loss rate of SN~2016gkg covers a range from the lower value derived for late-time interaction \citep{Kilpatricketal2022} up to the higher value suggested for early-time interaction \citep{Berstenetal2018}.}
  \label{fig:final_mass_loss}
\end{figure}
For each endpoint we computed the mass lost from the system in the last $\approx 1000\,\mathrm{yr}$ of the evolution to derive an average mass-loss rate. In Figure \ref{fig:final_mass_loss} we compare the distribution of the average mass-loss rate in the best-fitting models for each SN to constraints on mass lost in the last $10$-$10000\,\mathrm{yr}$ before the explosion obtained from observations. Most of the mass lost in this final stage of evolution is from the wind of the primary, while the mass transfer rate from RLOF is either much lower or non-existent (if the model is smaller than its Roche-lobe radius), or the mass transfer efficiency is high (and mass is not lost from the system because of RLOF). The wind mass-loss rate is computed according to the \citetalias{NL00}, \citetalias{Vink2001} or \citetalias{Vink2017} prescription\footnote{The wind mass-loss rate at the end of the evolution for cases where $T_\mathrm{eff} > 10000\,\mathrm{K}$ is computed by the \citetalias{Vink2017} or \citetalias{NL00} prescription if $X_\mathrm{s} < 0.4$ and by the \citetalias{Vink2001} prescription otherwise (see Sections \ref{subsubsec:winds} and \ref{subsec:EvolutionEndPoints}).} for SN~2008ax and iPTF13bvn, and according to the \citetalias{deJager1988} prescription for the rest. The estimated mass-loss rate for SN~2019yvr, for variable $R_V$ and $E\left(B-V\right)$, is $\log_{10}\left(\dot{M} / \mathrm{M}_\odot\,\mathrm{yr}^{-1}\right)=-5.33\pm 0.29$.

\section{Discussion}
\label{sec:discussion}

\subsection{Progenitor properties}
\label{subsec:discuss_progenitor_properties}

\begin{table*}
\centering
\caption{Effective surface temperature and luminosity for Type Ib and IIb SN progenitors in previous literature and in our study.}
\begin{threeparttable}
\begin{tabular}{c|cccc}
\hline
SN & $\log_{10}(T_\mathrm{eff}/ \mathrm{K})$ & $\log_{10}(L / \mathrm{L}_\odot)$ & $\log_{10}(T_\mathrm{eff}/ \mathrm{K})_\mathrm{literature}$ & $\log_{10}(L / \mathrm{L}_\odot)_\mathrm{literature}$ \\ 
\hline
SN 1993J (IIb) & $3.73\pm 0.02$ & $5.19\pm 0.17$ & $3.63\pm 0.05$ & $5.1\pm 0.3$ \\ 
SN 2008ax (IIb) & $4.16\pm 0.1$ & $4.95\pm 0.22$ & $4.09\pm 0.21$ & $4.855\pm 0.445$ \\ 
SN 2011dh (IIb) & $3.79\pm 0.01$ & $4.92\pm 0.18$ & $3.779\pm 0.02$ & $4.945\pm 0.045$ \\ 
SN 2013df (IIb) & $3.64\pm 0.01$ & $4.87\pm 0.04$ & $3.628\pm 0.01$ & $4.94\pm 0.06$ \\
SN 2016gkg (IIb) & $4.02\pm 0.11$ & $5.28\pm 0.16$ & $3.978^{+0.215}_{-0.158}$ & $4.985\pm 0.335$ \\ 
\hline
$\mathrm{iPTF13bvn}_\mathrm{E}$ (Ib) & $4.29\pm 0.09$ & $5.26\pm 0.16$ & $4.06\pm 0.04$ & $4.6\pm 0.1$ \\ 
$\mathrm{iPTF13bvn}_\mathrm{F}$ (Ib) & $4.18\pm 0.1$ & $5.01\pm 0.28$ & $4.06\pm 0.04$ & $4.6\pm 0.1$ \\ 
SN 2019yvr (Ib) & $3.83\pm 0.05$ & $5.19\pm 0.17$ & $3.833^{+0.025}_{-0.013}$ & $5.3\pm 0.2$ \\ 
\hline
\hline
\end{tabular}
\footnotesize
\begin{tablenotes}
\textit{Notes.} The literature values for the five Type~IIb progenitors are taken from \cite{Yoonetal2017}, who compiled results from several sources \citep{Alderingetal1994,Maundetal2004,Maundetal2011,VanDyketal2011,VanDyketal2014,Berstenetal2012,Folatellietal2015,Arcavietal2017,Kilpatricketal2017,Tartagliaetal2017}. The literature values for iPTF13bvn are taken from \cite{EldridgeMaund2016}. Our higher temperature for the progenitor of this event is more consistent with that of \cite{Berstenetal2014} and our higher luminosity is due in part to our larger assumed distance. The literature values for SN~2019yvr are taken from \cite{Kilpatricketal2021}. Our $T_\mathrm{eff}$ and $L$ estimates are for variable $E\left(B-V\right)$ and $R_V$ (fourth row for each SN in Table \ref{tab:MC_TABLE}).
\end{tablenotes}
\end{threeparttable}
\label{tab:compare_to_literature}
\end{table*}
We find that the effective surface temperatures of the SN progenitors are well constrained by the photometry, in general agreement with earlier studies (Table \ref{tab:compare_to_literature}). \cite{Yoonetal2017}, for example, classify the progenitors of SN~2008ax and SN~2016gkg as BSGs, the progenitor of SN~2011dh as a YSG, and the progenitors of SN~1993J and SN~2013df as RSGs. Our results are almost the same (Table \ref{tab:bestfitdetails}), with the only exception the classification of SN~1993J as a YSG. However, the effective surface temperature for SN~1993J is found very near the boundary between the definitions of RSGs and YSGs, and within the uncertainty a classification of RSG is acceptable (Table \ref{tab:MC_TABLE}). The progenitor of iPTF13bvn is classified as a BSG, and it is consistently hotter than all other progenitors. For SN~2019yvr, we find a YSG progenitor, with an effective surface temperature almost identical to that of the progenitor of SN~2011dh (Tables \ref{tab:bestfitdetails} and \ref{tab:MC_TABLE}).

While the effective surface temperature is not much affected by the assumptions on dust extinction, the luminosity experiences more variance. This is especially the case for SN~2019yvr, which is the most heavily extinguished SN. Because of the non-trivial relation between the hydrogen mass and the progenitor size (Figure \ref{fig:MHR1}), the uncertainty in luminosity results also in an uncertainty in the hydrogen mass. Within the uncertainties, it is plausible that the two progenitors of Type~Ib SNe (iPTF13bvn and SN~2019yvr) contain less hydrogen in their envelopes compared to all the Type~IIb progenitors. If this is the case, the tension between the Type~Ib classification of SN~2019yvr and its progenitor properties, as claimed by \cite{Kilpatricketal2021}, is lessened.

The binary progenitor models considered by \cite{Kilpatricketal2021} were computed with the \textsc{bpass} code \citep{BPASS2017}, with the best-fitting models resulting from CEE of systems in which the companion has a significantly lower mass than the primary star. Our models do not cover this part of the parameter space, and have initial companion masses $M_2 > 0.35 M_1$. However, we do find reasonable fits in the part of the parameter space that we do cover, where mass transfer is stable. The disparate results obtained by using two separate stellar evolution codes might arise from the relation between the hydrogen mass and the envelope radius being sensitive to certain aspects of the code, like the EOS and opacities. We defer the expansion of our parameter space to include CEE to a future study.

\cite{Kilpatricketal2021} discuss the possibility that the progenitor of SN~2019yvr had a radiation dominated inflated envelope. Our models include the MLT++ treatment of \textsc{mesa}, which suppresses super-Eddington regions stabilized by density inversions in an inflated envelope. If MLT++ is disabled, an inflated envelope can form near the end of the computed stellar evolution, with a large radius and a smaller hydrogen mass compared to that obtained with the MLT++ treatment. This further reduces the tension between the progenitor properties of SN~2019yvr and its Type~Ib classification, though the feasibility of this scenario depends on the stability of density inversions in stellar envelopes.

Our analysis and hydrogen mass estimates are derived solely from the progenitor photometry, and are independent from the explosion characteristics and the observed SN spectra. However, we did check the implications of limiting the progenitor models included in finding a best fit according to theoretical SN computations. When taking the threshold of $M_\mathrm{H,min,IIb}\approx 0.033\,\mathrm{M}_\odot$ given by \cite{Hachinger2012}, we find no significant reduction in the quality of the fits for iPTF13bvn and SN~2019yvr. However, when taking a threshold of $M_\mathrm{H,min,IIb}\approx 0.001\,\mathrm{M}_\odot$ (consistent with \citealt{Dessartetal2011}), we find it difficult to fit any model to the progenitor of SN~2019yvr, and the quality of the fits for the progenitor of iPTF13bvn are reduced. We can therefore claim that our analysis favors the higher value for the threshold hydrogen mass differentiating between Type~Ib and Type~IIb SNe. Speculating further, it is possible that most Type~Ib SNe are actually hydrogen deficient but not hydrogen free, and that theoretical ``pure helium stars'' might not be required for progenitors of Type~Ib SNe.

\subsection{Mass transfer efficiency}
\label{subsec:discuss_mass_transfer}

Overall, when considering a population of SN progenitors resulting from binary evolution, we find that a mass transfer efficiency which is not fixed at an arbitrary value is beneficial. Our results are in agreement with evolutionary scenarios proposed for SN~1993J in which significant accretion onto the companion star occurs \citep{Podsiadlowskietal1993,Nomotoetal1993,Woosleyetal1994,Maundetal2004}. For SN~2011dh, \cite*{Benvenutoetal2013} proposed a scenario where the mass transfer efficiency is zero, with no accretion by the companion star. Our Monte Carlo analysis shows the strongest tendency toward low mass transfer efficiency for the case of SN~2011dh, in agreement with \cite{Benvenutoetal2013}. Fitting the photometry of the SN~2011dh progenitor was challenging, with good fits obtained only when allowing the largest freedom in the dust extinction parameters. It is possible that allowing an even lower mass transfer efficiency than that resulting from our prescription would have eased the fitting procedure. 

\subsection{Wind mass loss}
\label{subsec:discuss_wind_mass_loss}

We find that the best-fit models have a preference not to include the \citetalias{NL00} mass-loss prescription (Table \ref{tab:bestfitdetails} and Table \ref{tab:MC_TABLE}). In one case for SN~2008ax where the best-fit model was evolved with the \citetalias{NL00} prescription, the leftover hydrogen mass is too low to qualify as a Type~IIb SN. Optically thin winds are more favorable in producing Type~Ib and Type~IIb SN progenitors, and optically thick WR-like winds do not result in progenitor properties as observed for these SNe. However, the sample size is rather small, with only five Type~IIb SN progenitors identified, and only two for Type~Ib SNe (with SN~2019yvr still too recent to confirm its progenitor). Still, the role of the assumed mass-loss rate during the post-RLOF stage is crucial, as already asserted by \cite{GilkisVinkEldridgeTout2019}.

\cite{Bjorklund2021} give a revised mass-loss prescription for O-type stars in the Galaxy and the Magellanic Clouds, which is generally lower than \cite{Vink2001}, and also with no bi-stability jump. For WR stars, the rate prescription given by \cite{NL00} is outdated compared to the prescriptions given by \cite{Hainich2014} and \cite*{TSK2016}, which have been adopted by \cite{Yoon2017} and \cite{Woosley2019}. For RSGs, \cite{Beasoretal2020} derive mass-loss rates which are significantly lower than the \citetalias{deJager1988} prescription that we use. Our models include binary interactions and in many cases mass loss occurs through RLOF during cool phases, but a lower RSG mass-loss rate might still have an effect. However, we expect this effect to be most pronounced for hydrogen-rich Type~IIP SNe, which we do not discuss here.

Late-time post-SN observations indicate that interaction with circumstellar material takes place, and the mass loss preceding the SN can be estimated. We compare the observational constraints for the six earlier SNe to the derived mass-loss rate (averaged over the last $1000\,\mathrm{yr}$ of evolution) in our best-fitting Monte Carlo models (Figure \ref{fig:final_mass_loss}) and find a general agreement in most cases. Our models tend to underestimate the mass-loss rate, and this is especially the case for SN~1993J and SN~2013df. The mass lost in the late evolutionary stages is mostly from stellar winds, and not from RLOF. Significant mass loss from the system by inefficient RLOF mass transfer occurs only at early evolutionary stages, when the mass of the envelope is large. Although many evolutionary endpoints fill their Roche lobe, by this stage the mass ratio has usually been inverted, and the mass transfer rate is low and almost fully conservative according to our prescription for the mass transfer efficiency.

\subsection{Metallicity}
\label{subsec:discuss_metallicity}

As stated in Section \ref{subsec:StellarEvolution}, the metallicity in all our stellar evolution simulations is $Z=0.019$, which is close to Solar. Metallicity can affect the evolution through opacity, with lower metallicity models reaching smaller radii and therefore losing less of their envelope through RLOF (\citealt*{Gotbergetal2017}). Remaining hydrogen can lead to a large radial expansion prior to core collapse \citep{Laplaceetal2020}, with a tenuous low-mass envelope, as suggested for some Type~IIb SNe \citep[e.g., SN~2011dh; ][]{Berstenetal2012}. Another effect of a lower metallicity is reduced mass-loss rates, with all three hot-wind rates used in our simulations (\citetalias{NL00}, \citetalias{Vink2001}, \citetalias{Vink2017}) strongly depending on metallicity.

Solar metallicity is inferred for the progenitors of SN~2011dh \citep{Maundetal2011}, SN~2013df \citep{VanDyketal2014}, iPTF13bvn \citep{Fremlingetal2016} and SN~2016gkg \citep{Berstenetal2018}. The metallicity of the progenitor of SN~1993J is suggested to be between Solar and slightly super-Solar \citep{Alderingetal1994}, while for the progenitor of SN~2008ax, \cite{Crockettetal2008} conclude that the metallicity of the progenitor was between that of the Large Magellanic Cloud and Solar. For SN~2019yvr, \cite{Kilpatricketal2021} assume Solar metallicity for the progenitor, while \cite{Sunetal2021} derive a slightly lower metallicity. We conclude that the metallicity used in our simulations is consistent with what is known for the SN progenitors that we study.

\section{Summary and conclusions}
\label{sec:summary}

We have systematically investigated the properties of Type~Ib and Type~IIb SN progenitors by using detailed binary stellar evolution simulations, synthetic photometry, and Monte Carlo random realisations of the observations. We validate our analysis by comparing the derived progenitor properties to previous studies, finding a general agreement with massive stars partially stripped by binary interaction as the progenitors of Type~Ib and Type~IIb SNe. We apply a uniform analysis to all seven known Type~Ib and Type~IIb SN progenitors (six confirmed, one candidate) to establish a coherent picture of their evolution.

Several key points are emphasised:
\begin{enumerate}
    \item A small amount (mass) of hydrogen has a significant effect on the progenitor radius (Figure \ref{fig:MHR1}). The relation between $M_\mathrm{H}$ and $R$ is non-trivial, with less luminous stellar models reaching larger radii for a certain hydrogen mass. Owing to the uncertainty in both the distance to the host galaxy and the amount of dust extinction, there is therefore a non-negligible error on the derived hydrogen mass (Figure \ref{fig:MC_like_R_and_MH_norm}).
    \item According to our analysis it is plausible that the progenitors of iPTF13bvn and SN~2019yvr contained less hydrogen than all known Type~IIb SN progenitors, alleviating the proposed tension between the Type~Ib classification of SN~2019yvr and its progenitor properties \citep{Kilpatricketal2021}, and mitigating the need for special scenarios to get rid of the progenitor envelope in the short duration between the pre-explosion observation and the SN itself. Our results are in tentative agreement with a mass threshold for a Type~IIb appearance close to $M_\mathrm{H,min,IIb}\approx 0.033\,\mathrm{M}_\odot$ \citep{Hachinger2012}, but inconsistent with the findings of \cite{Dessartetal2011} that even $0.001\,\mathrm{M}_\odot$ of hydrogen in the exploding envelope would be enough to result in a type~IIb SN.
    \item The best-fitting models prefer evolutionary pathways which include optically thin winds \citep{Vink2001,Vink2017} rather than optically thick WR-like winds \citep{NL00}, and a non-negligible leftover hydrogen mass. Even for the Type~Ib SN progenitors, the progenitor is not a ``pure helium star'', but is probably more accurately classified as a CSG. The assumptions on post-RLOF wind mass loss are therefore crucial for understanding the evolution towards Type~Ib and Type~IIb SNe \citep{GilkisVinkEldridgeTout2019}.
    \item We predict that a faint ($V$-band apparent magnitude of $\approx 30$) companion will remain and that the YSG progenitor candidate will disappear in future observations of SN~2019yvr. We note that there are alternative scenarios, such as a hot progenitor reddened by an optically-thick wind \citep*{JungYoonKim2021}, which will also result in the progenitor candidate disappearing in the future, or that the YSG is a companion to the true progenitor \citep{Sunetal2021}, in which case a luminous counterpart will remain at the SN site in future observations.
\end{enumerate}

The differences between progenitors of Type~Ib and Type~IIb SNe might offer a valuable opportunity to test key points in SN physics and in the evolution of massive stars. Processes such as stellar winds, mass transfer in binaries and CEE might be crucial for analysing the progenitors of CCSNe. In our study we made use of one set of stellar models, where only one key aspect, post-RLOF winds, was explored (and to a lesser extent also the mass transfer efficiency). Even so, we find distinct implications for the assumed mass loss prescription. In future studies our systematic approach can be expanded to investigate additional stellar evolution issues, to cover a larger parameter space of initial conditions, and also to include constraints from the SN explosion and additional post-SN constraints.

\section*{Acknowledgments}

We thank T. Shenar and C. Kilpatrick for helpful comments and discussions. We thank the
anonymous referee for constructive comments on the manuscript. AG and IA acknowledge support from the European Research Council (ERC) under the European Union’s Horizon 2020 research and innovation program (grant agreement number 852097). IA is a CIFAR Azrieli Global Scholar in the Gravity and the Extreme Universe Program and acknowledges support from that program, from the Israel Science Foundation (grant number 2752/19), from the United States - Israel Binational Science Foundation (BSF), and from the Israeli Council for Higher Education Alon Fellowship.

\section*{Data Availability Statement}

The code and input files necessary to reproduce our simulations and associated data products are available at
\href{https://doi.org/10.5281/zenodo.5897214}{https://doi.org/10.5281/zenodo.5897214}.

\bibliographystyle{mnras}
\input{IbIIbSNe.bbl}

\appendix

\section{Coverage of spectral grids}
\label{sec:appendixa}

\begin{figure*}
\centering
\includegraphics[width=\textwidth]{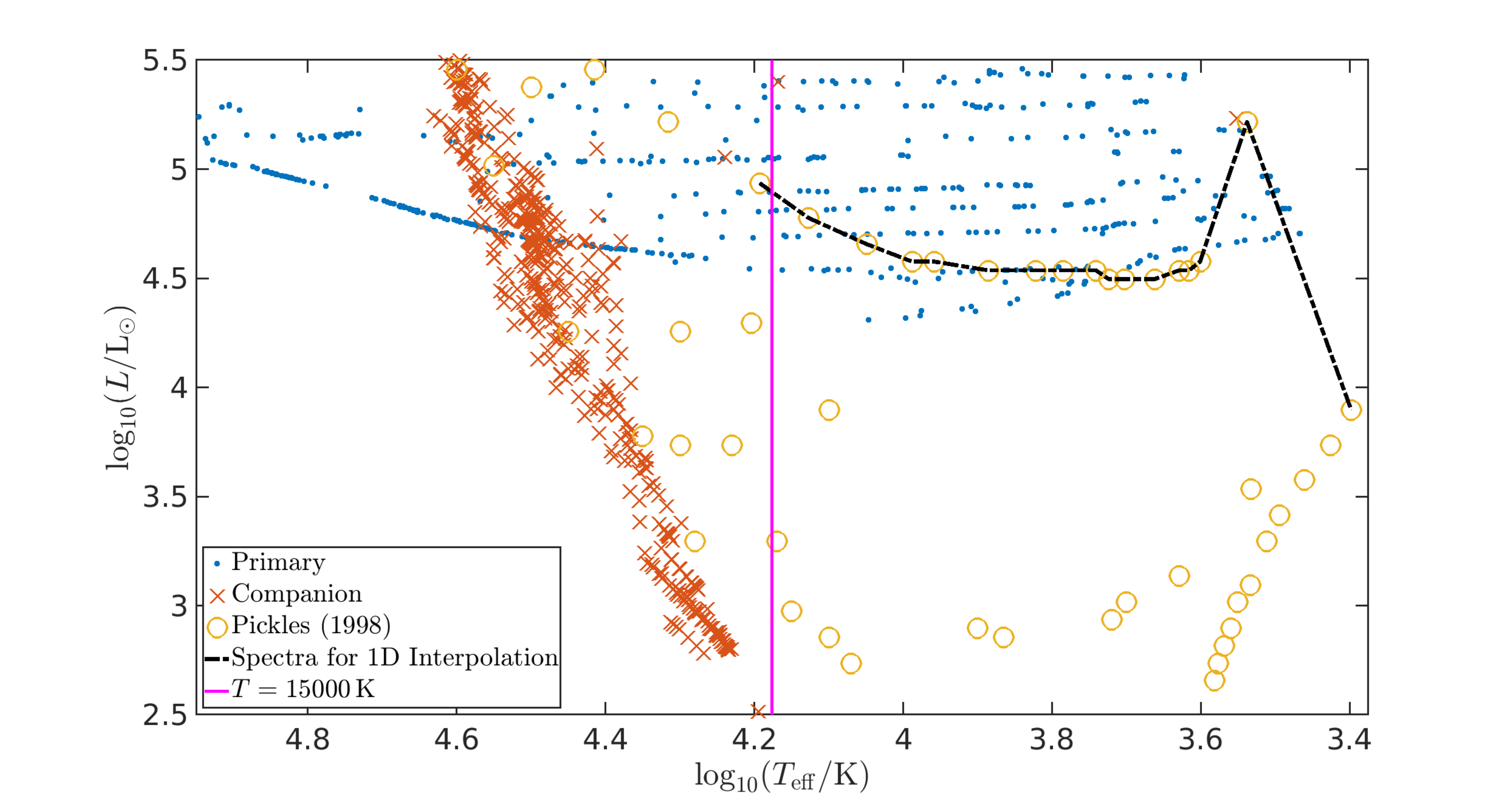} \\
\caption{Luminosity and effective surface temperature for which spectra are available \citep[from][shown as yellow circles]{Pickles1998} and for all computed models with $T_\mathrm{eff}\le 10^{4.95}\,\mathrm{K}$ (primaries shown as blue dots and companions as orange crosses). The dash-dotted black line connects the sequence of spectra used for synthetic photometry interpolation.}
\label{fig:SpecGridPickles}
\end{figure*}

\begin{figure*}
\centering
\includegraphics[width=\textwidth]{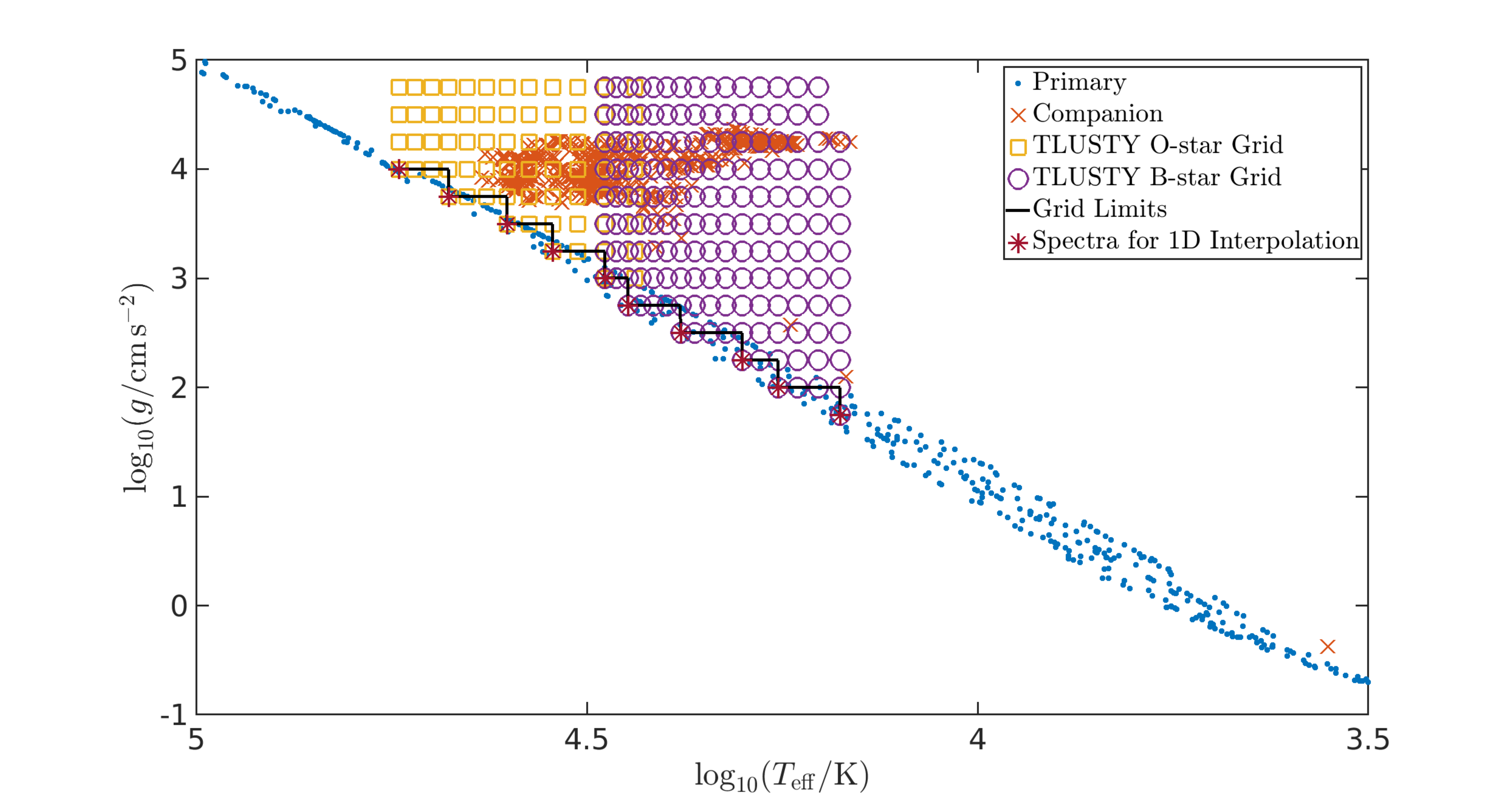} \\
\caption{Gravity and effective surface temperature for which \textsc{tlusty} synthetic spectra are available (O-star grid points shown as yellow squares, B-star grid points shown as purple circles) and for all computed models in the range $10^{3.5}\,\mathrm{K}\le T_\mathrm{eff}\le 10^5\,\mathrm{K}$ (primaries shown as blue dots and companions as orange crosses). Models beyond the grid limits (black line) with $15\,\mathrm{kK} \le T_\mathrm{eff} \le 55\,\mathrm{kK}$ are assigned synthetic photometry by interpolation between the grid points marked by maroon asterisks.}
\label{fig:SpecGridTLUSTY}
\end{figure*}

\begin{figure*}
\centering
\includegraphics[width=\textwidth]{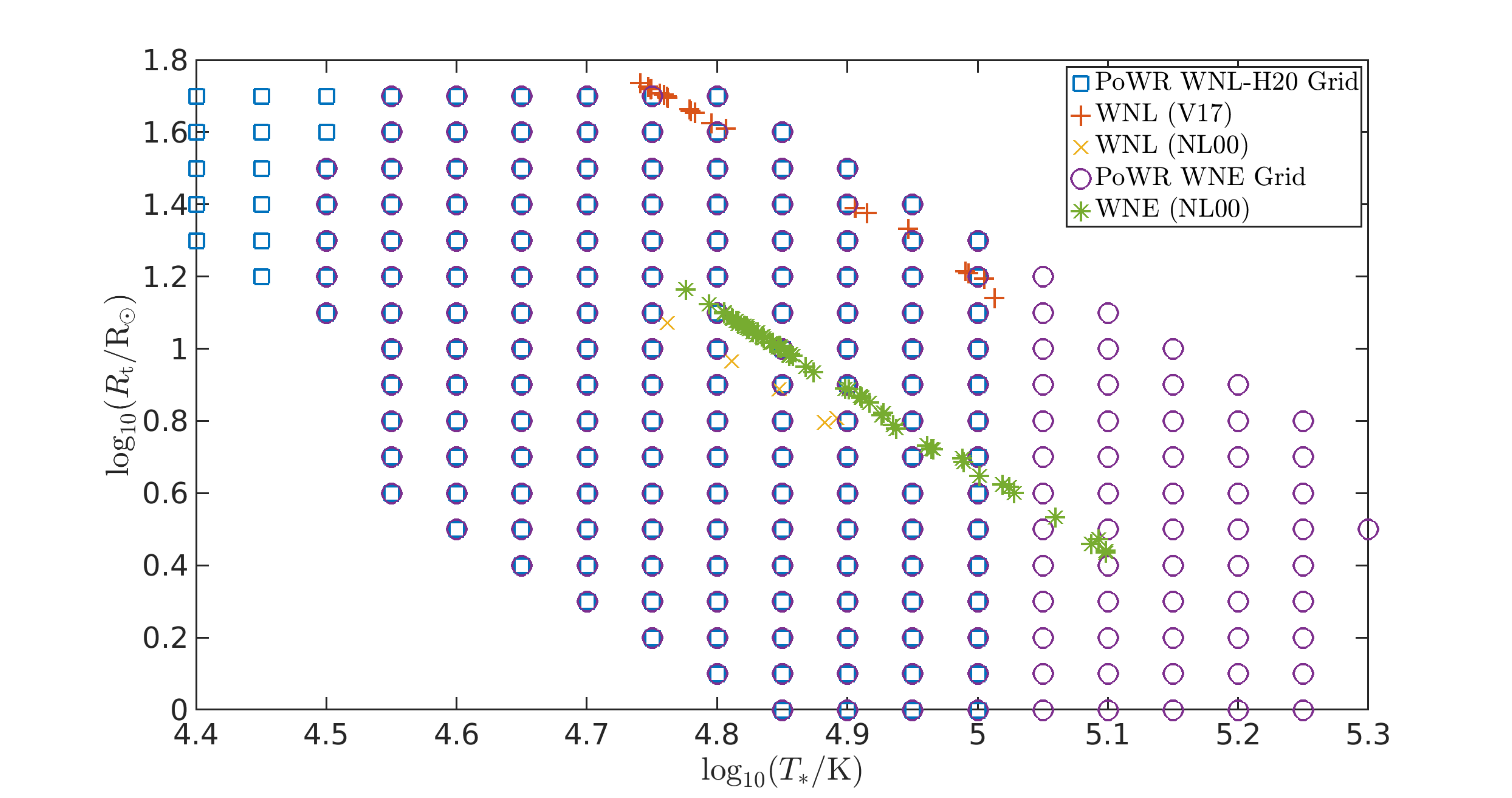} \\
\caption{Transformed radius (Eq.~\ref{eq:Rt}) and effective surface temperature for which \textsc{powr} synthetic spectra are available and for progenitor models with $T_\mathrm{eff} > 55\,\mathrm{kK}$ (models with $X_\mathrm{s} \ge 0.05$ obtained with the \citetalias{Vink2017} mass-loss rate marked as orange pluses, models with $X_\mathrm{s} \ge 0.05$ obtained with the \citetalias{NL00} mass-loss rate marked as yellow crosses and models with $X_\mathrm{s} < 0.05$ marked as green asterisks).}
\label{fig:SpecGridPoWR}
\end{figure*}

Here we present the physical properties associated with all spectra used for the generation of synthetic photometry. In Figure \ref{fig:SpecGridPickles} we show the luminosity and effective surface temperature for which \cite{Pickles1998} spectra are available and mark the points used to generate synthetic photometry. In Figure \ref{fig:SpecGridTLUSTY} we show the gravity and effective surface temperature for which we have \textsc{tlusty} spectra and specify the region for which a one-dimensional interpolation as described in Section \ref{subsec:SyntheticPhotometry} is performed. In Figure \ref{fig:SpecGridPoWR}  we show the transformed radius and effective surface temperature for which we have \textsc{powr} synthetic spectra and the stellar models for which WR spectra are used for synthetic photometry generation.

\section{Companion magnitudes}
\label{sec:appendixb}

\begin{figure*}
  \centering
    \begin{subfigure}{0.49\textwidth}
    \centering
    \includegraphics[width=\textwidth]{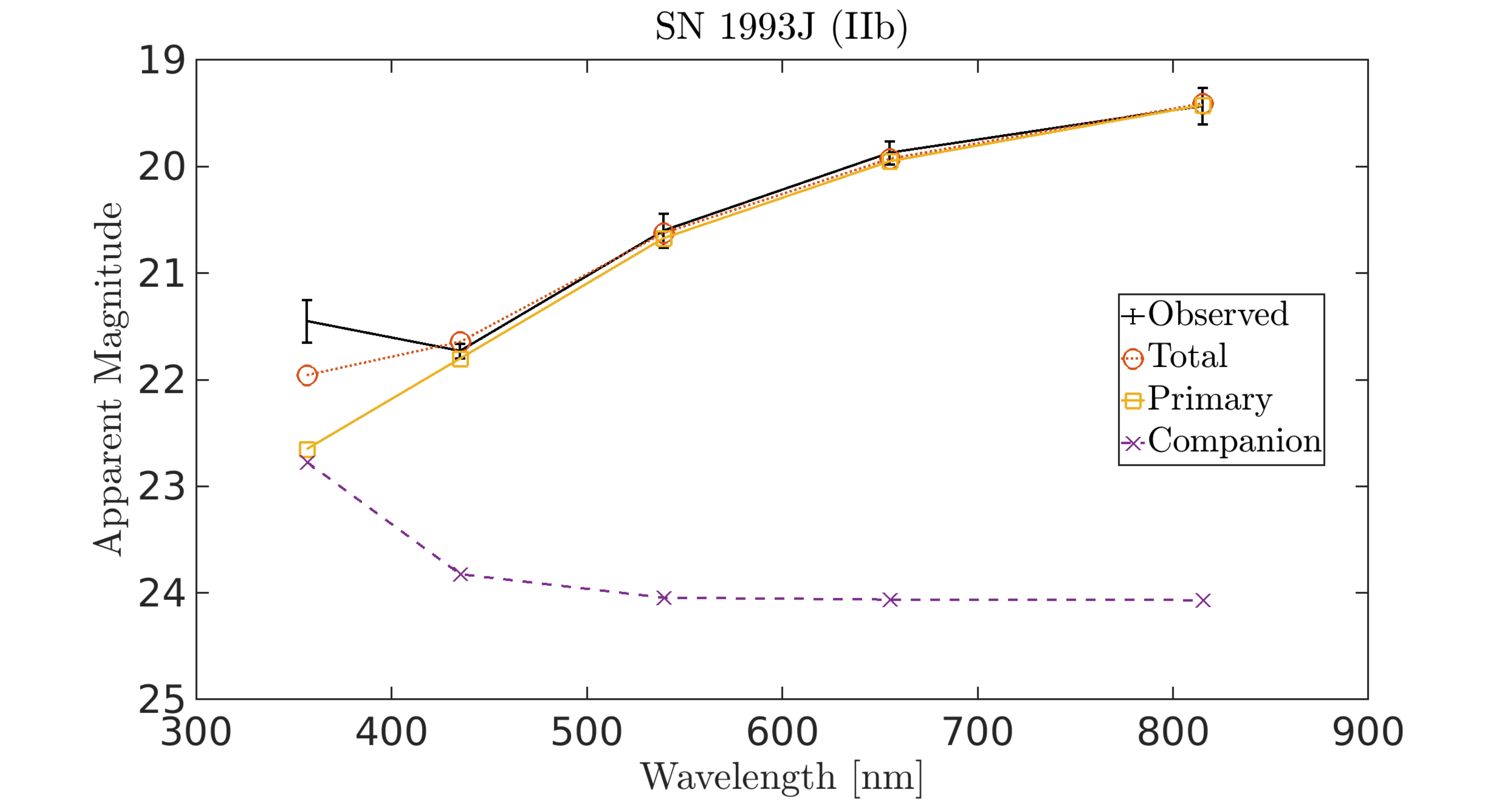}
    \label{fig:chi2mag2SN1993J}
  \end{subfigure}
  \begin{subfigure}{0.49\textwidth}
    \centering
    \includegraphics[width=\textwidth]{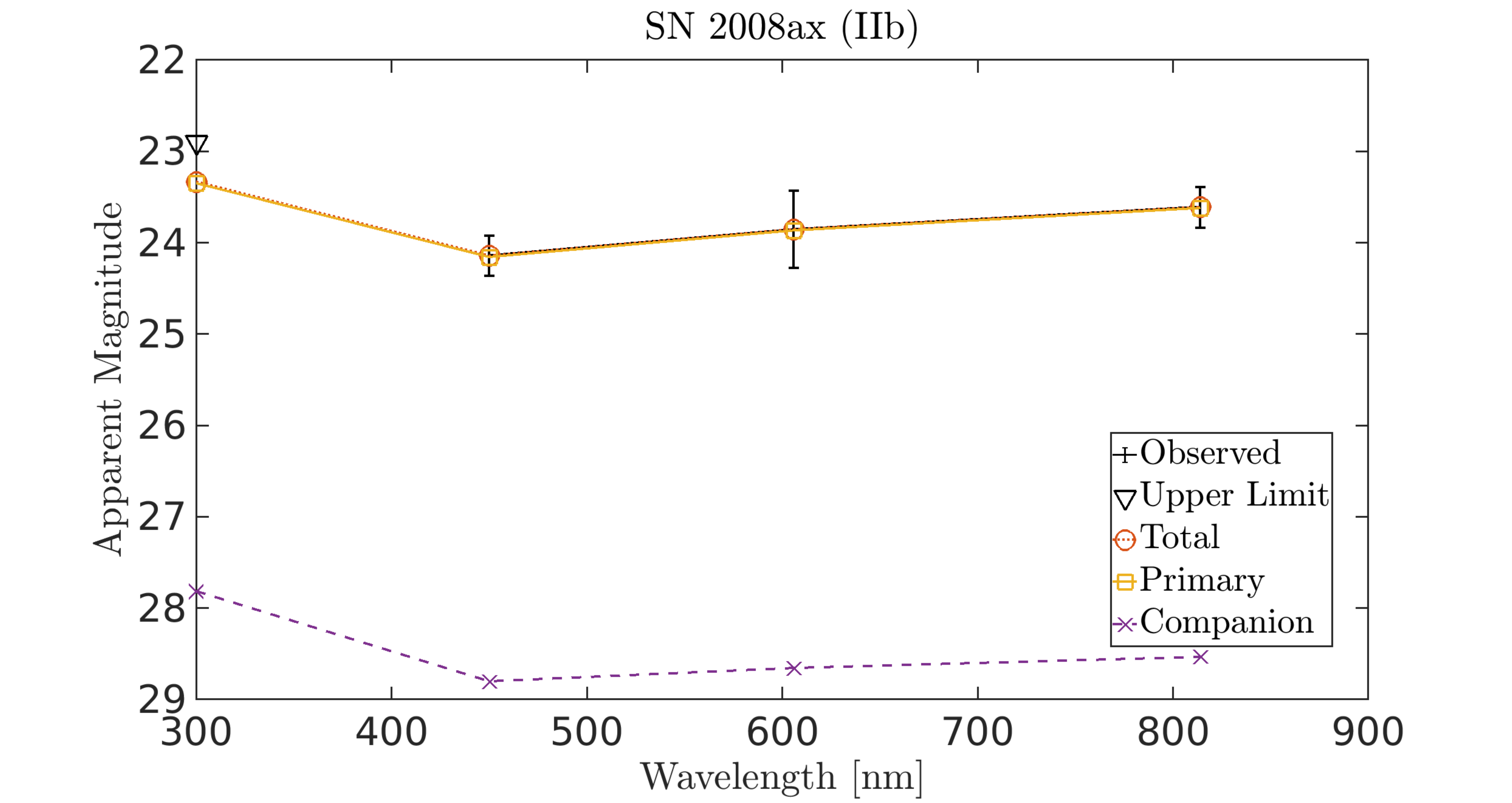}
    \label{fig:chi2mag2SN2008ax}
  \end{subfigure}\\
    \begin{subfigure}{0.49\textwidth}
    \centering
    \includegraphics[width=\textwidth]{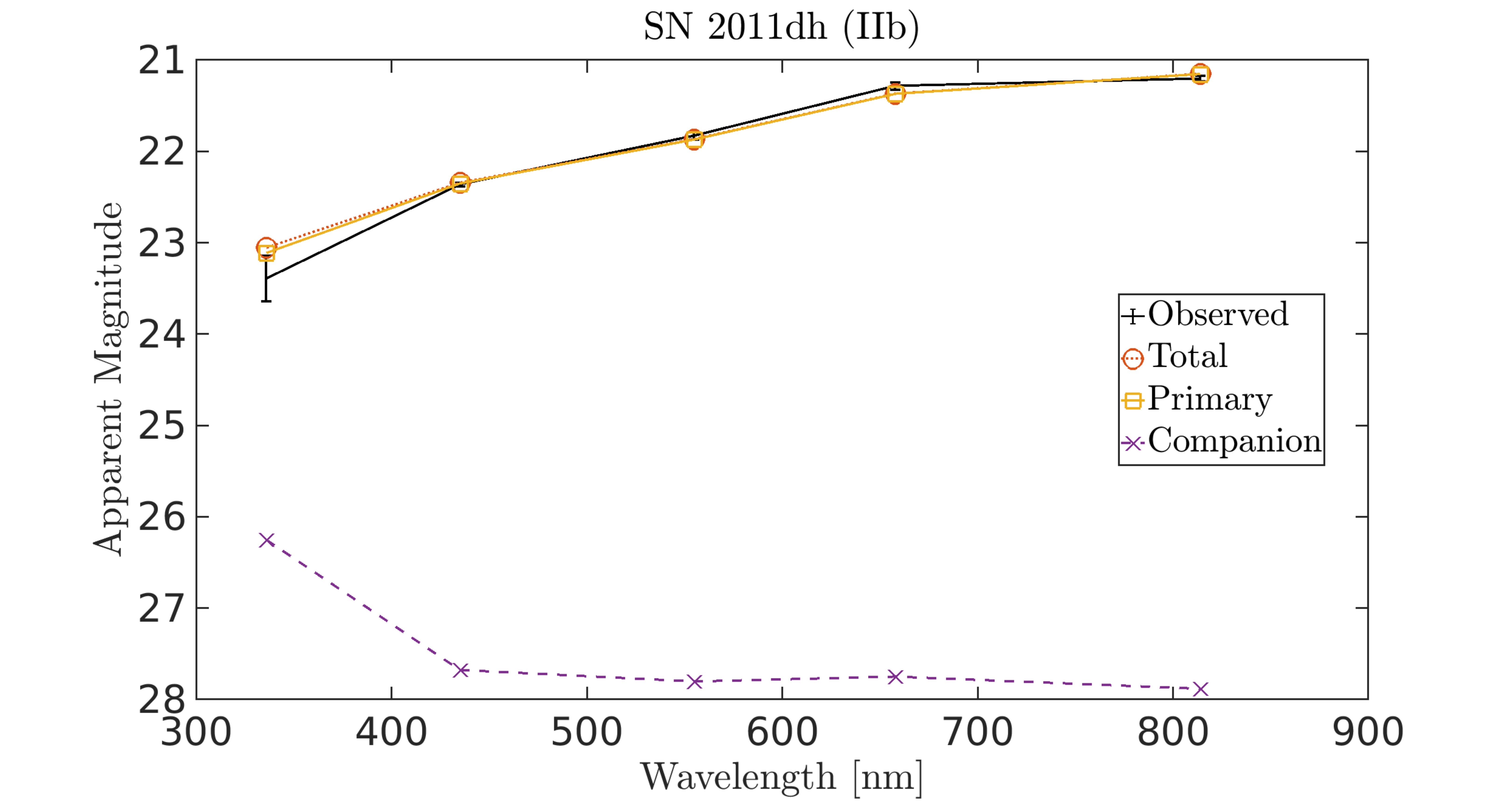}
    \label{fig:chi2mag2SN2011dh}
  \end{subfigure}
  \centering
    \begin{subfigure}{0.49\textwidth}
    \centering
    \includegraphics[width=\textwidth]{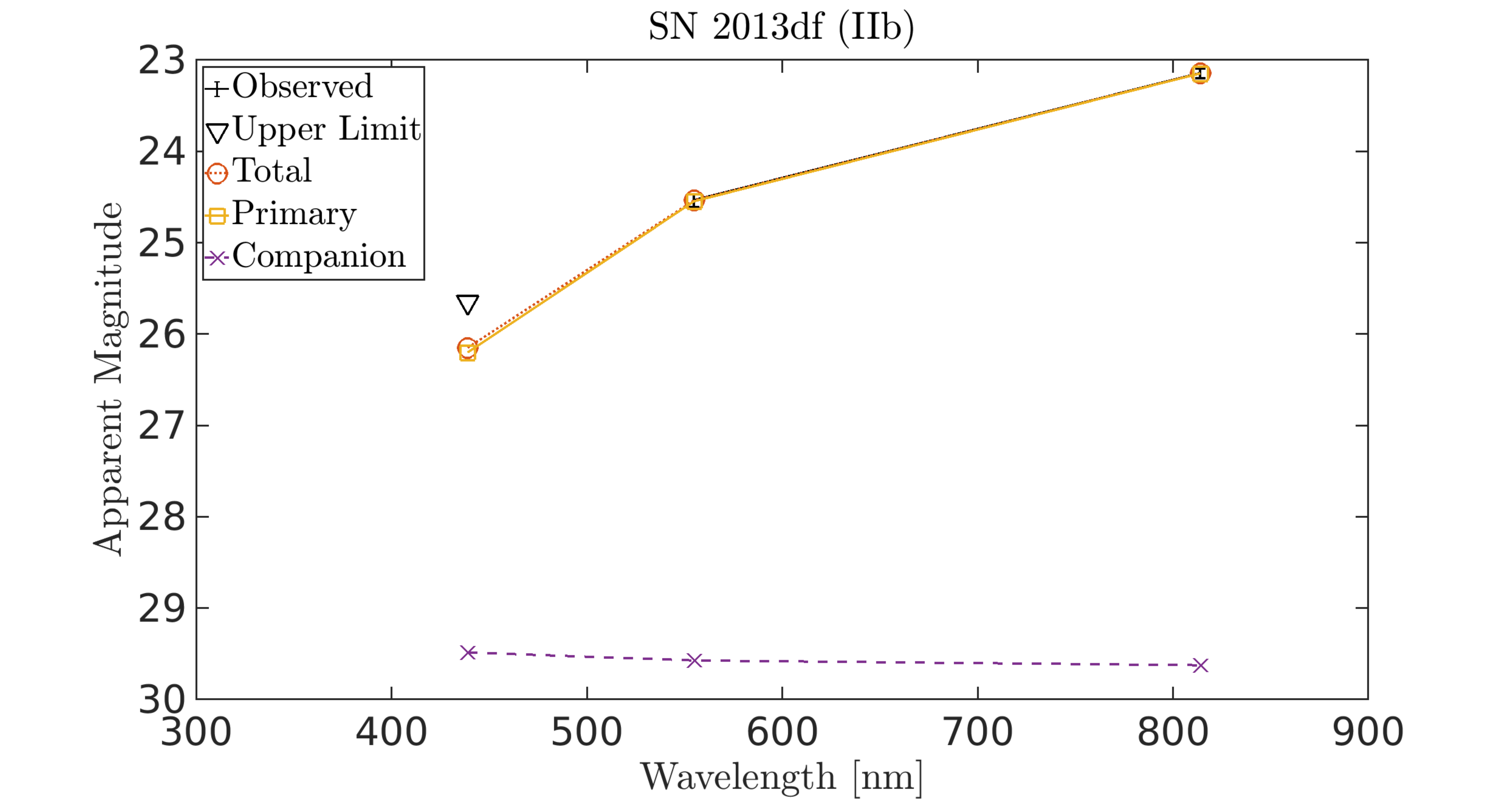}
    \label{fig:chi2mag2SN2013df}
  \end{subfigure}\\
  \begin{subfigure}{0.49\textwidth}
    \centering
    \includegraphics[width=\textwidth]{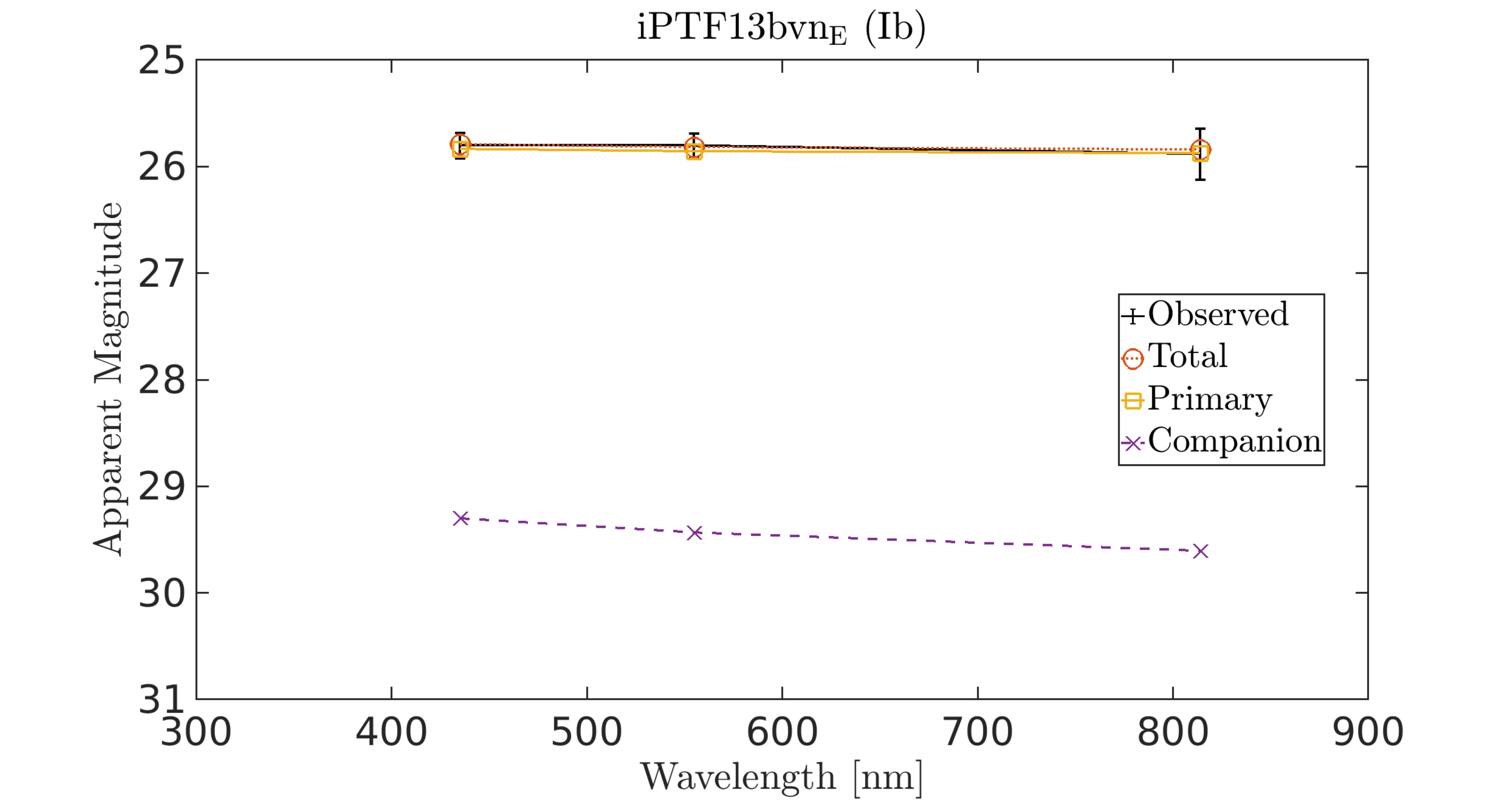}
    \label{fig:chi2mag2iPTF13bvnE}
  \end{subfigure}
    \begin{subfigure}{0.49\textwidth}
    \centering
    \includegraphics[width=\textwidth]{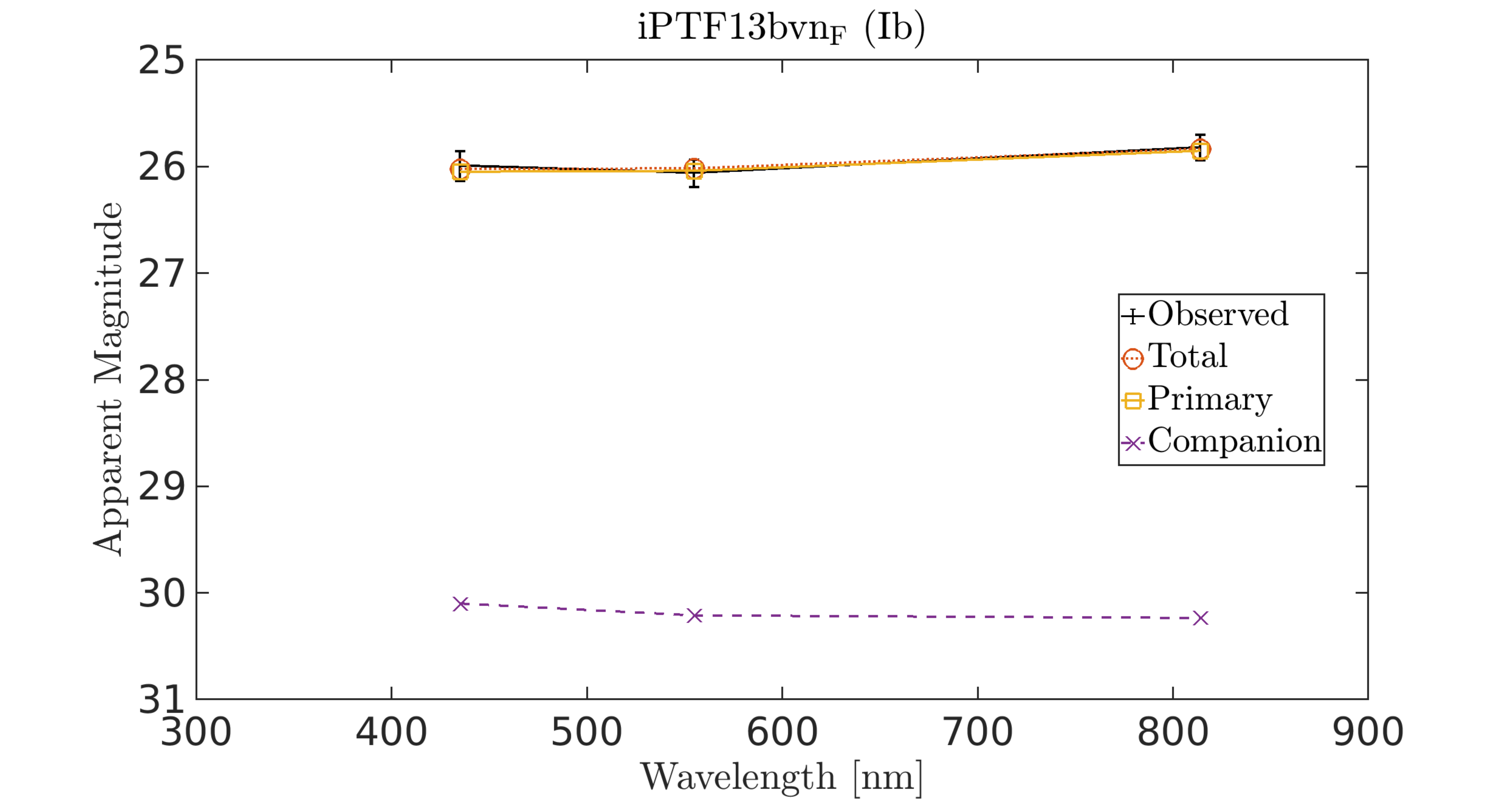}
    \label{fig:chi2mag2iPTF13bvnF}
  \end{subfigure}\\
    \begin{subfigure}{0.49\textwidth}
    \centering
    \includegraphics[width=\textwidth]{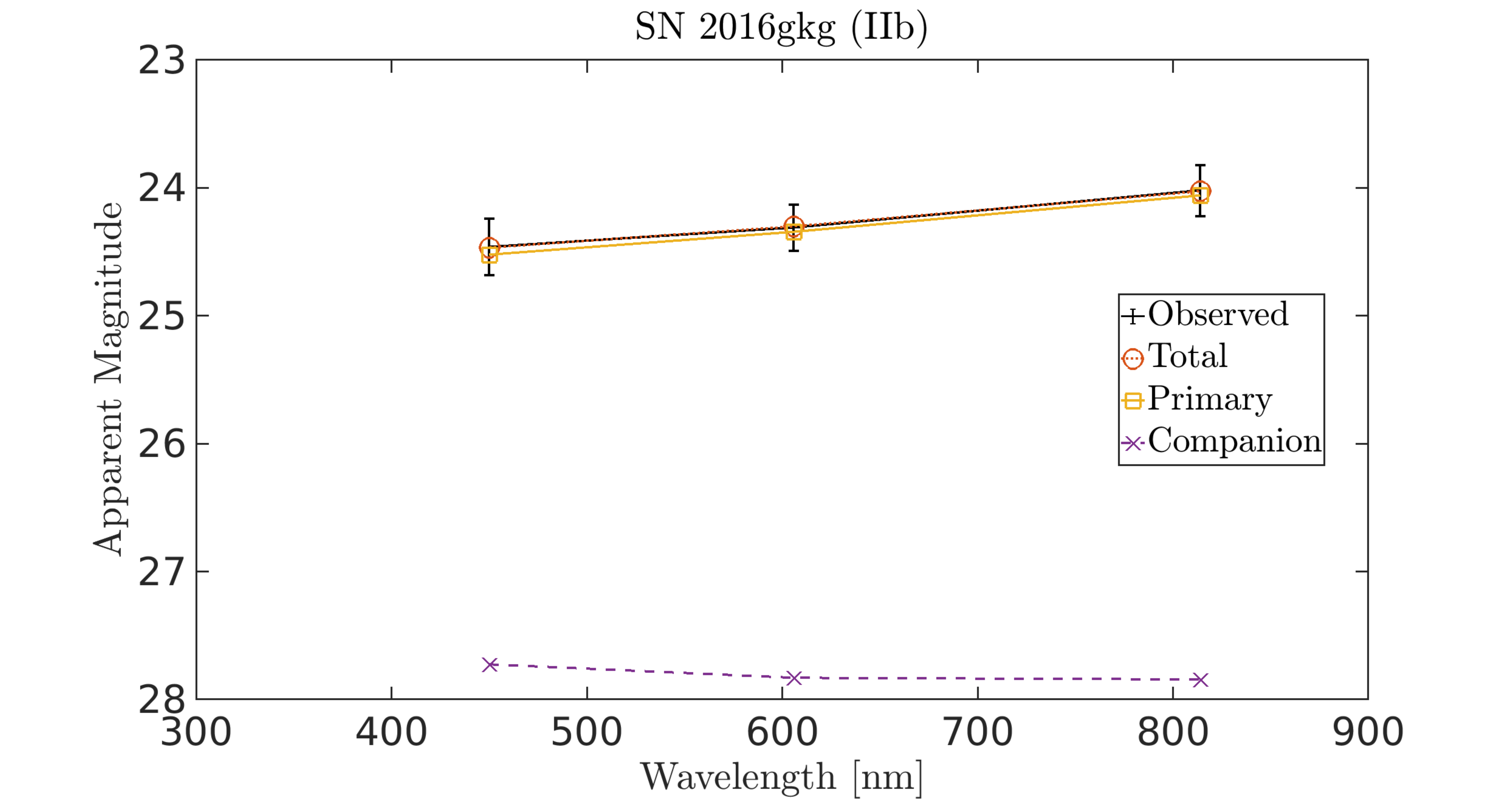}
    \label{fig:chi2mag2SN2016gkg}
  \end{subfigure}
  \begin{subfigure}{0.49\textwidth}
    \centering
    \includegraphics[width=\textwidth]{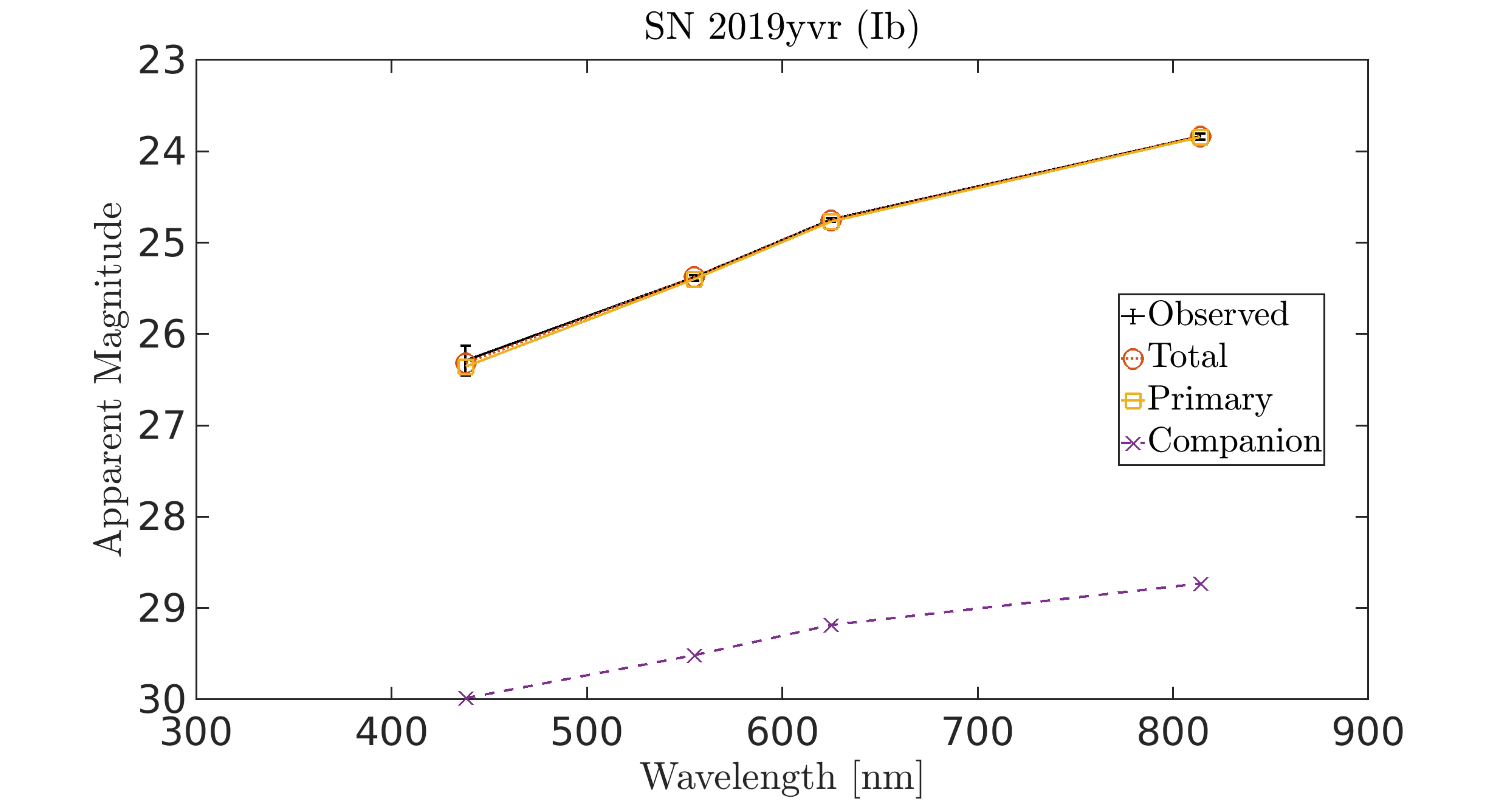}
    \label{fig:chi2mag2SN2019yvr}
  \end{subfigure}\\    
  \caption{Computed magnitudes for best-fitting models, obtained with variable $E\left(B-V\right)$ and $R_V$, decomposed to the two stellar components.} 
  \label{fig:chi2_mag2}
\end{figure*}

\begin{figure*}
  \centering
    \begin{subfigure}{0.49\textwidth}
    \centering
    \includegraphics[width=\textwidth]{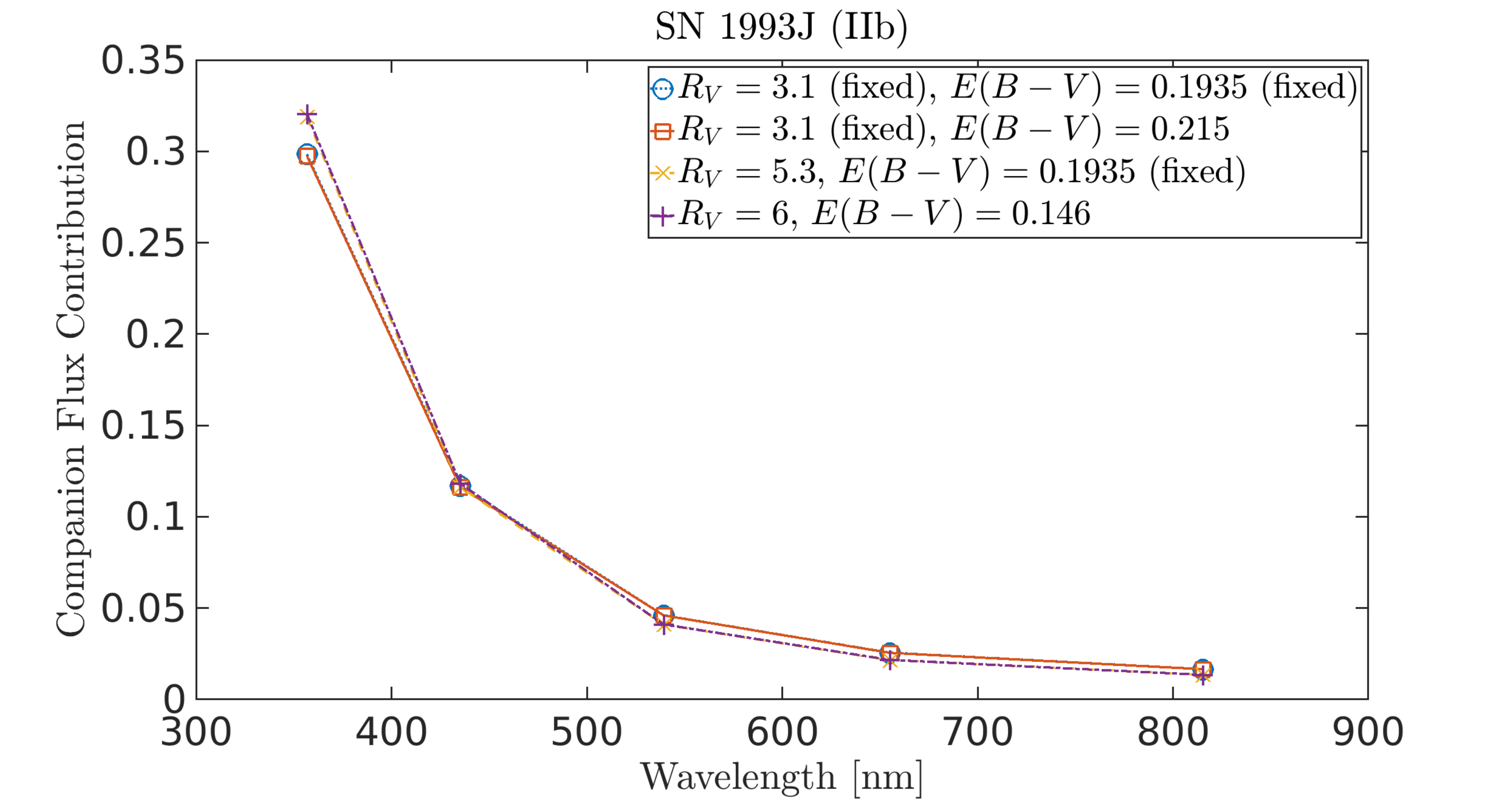}
    \label{fig:chi2flux2SN1993J}
  \end{subfigure}
  \begin{subfigure}{0.49\textwidth}
    \centering
    \includegraphics[width=\textwidth]{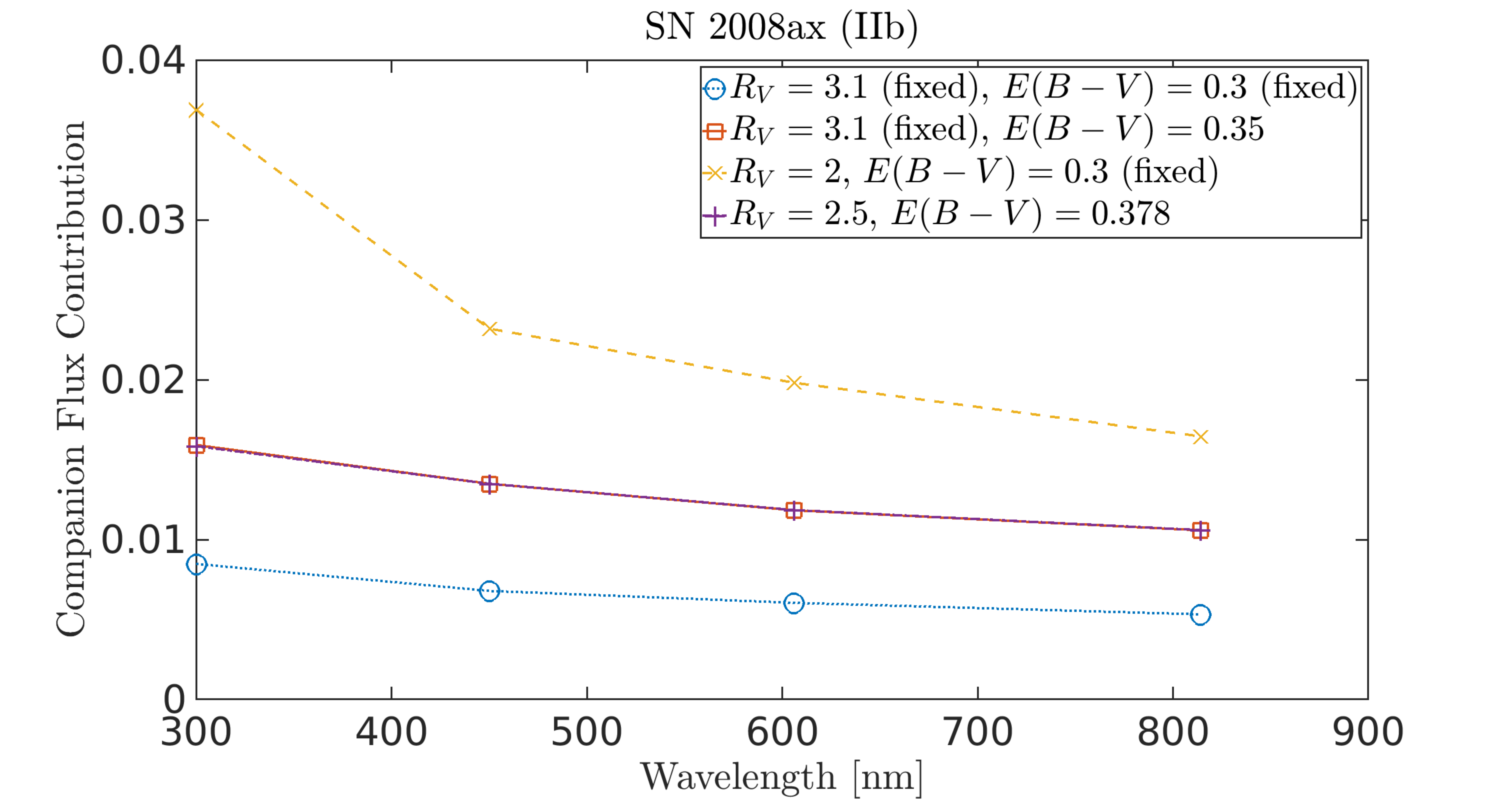}
    \label{fig:chi2flux2SN2008ax}
  \end{subfigure}\\
    \begin{subfigure}{0.49\textwidth}
    \centering
    \includegraphics[width=\textwidth]{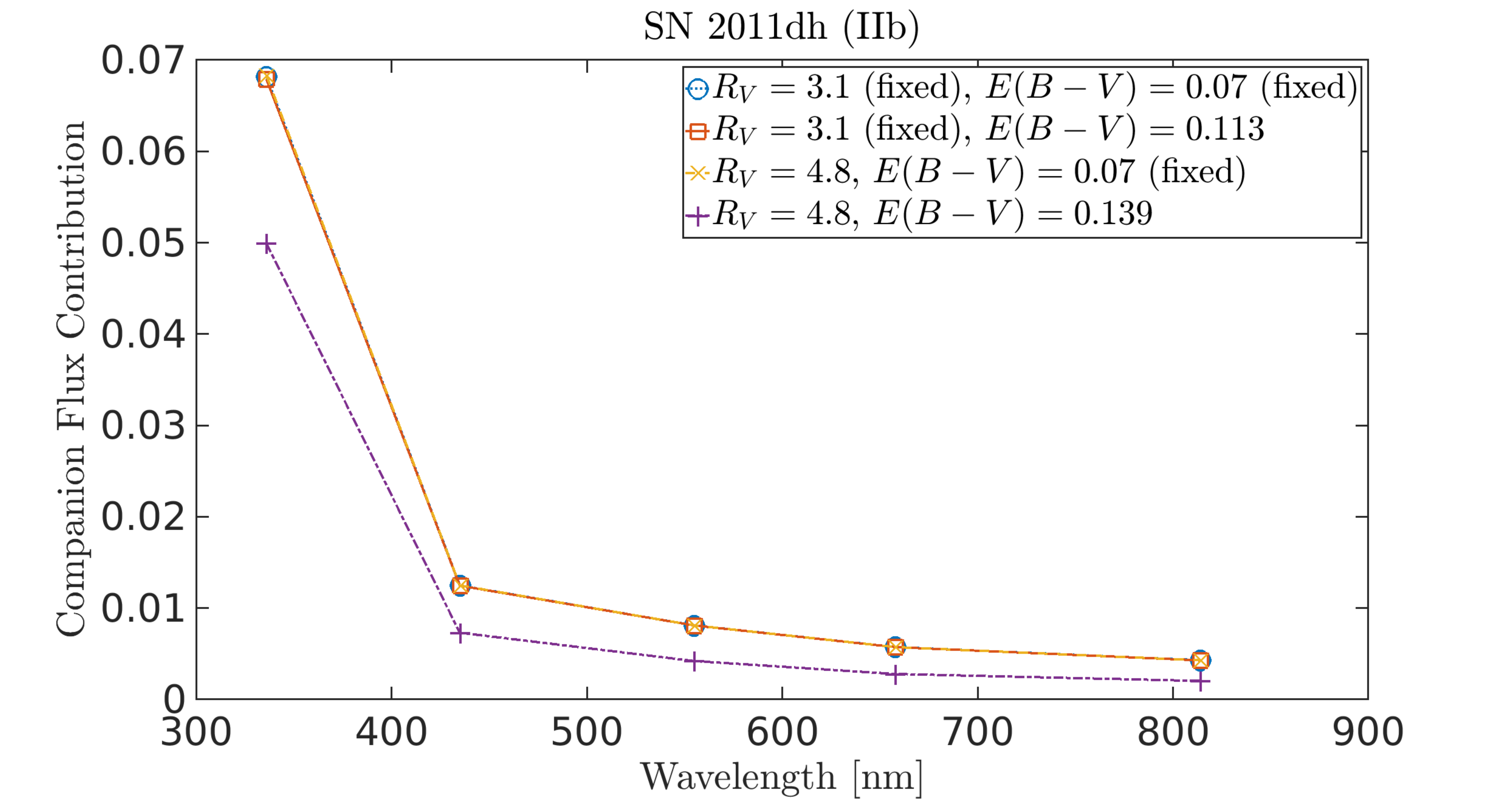}
    \label{fig:chi2flux2SN2011dh}
  \end{subfigure}
  \centering
    \begin{subfigure}{0.49\textwidth}
    \centering
    \includegraphics[width=\textwidth]{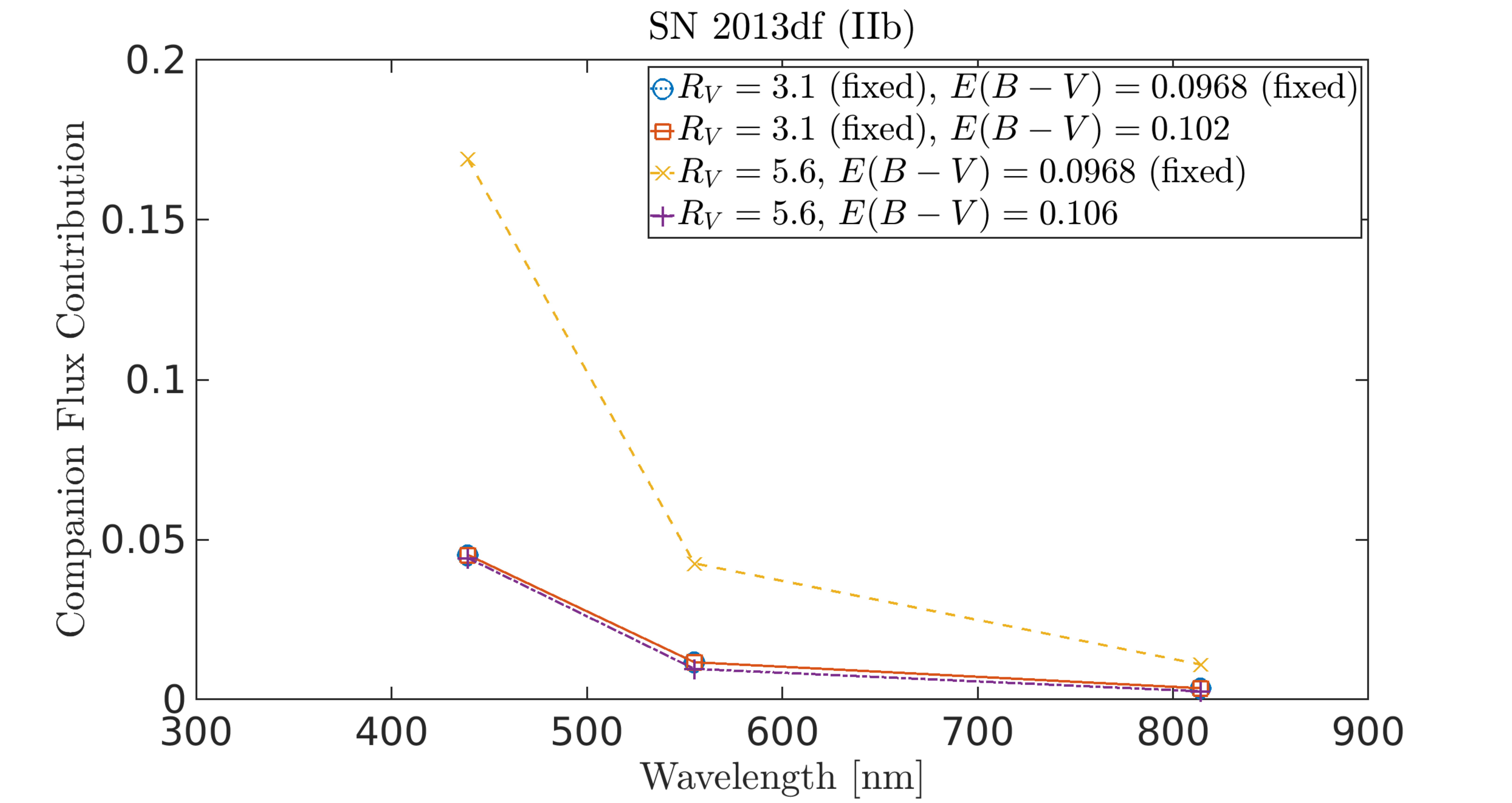}
    \label{fig:chi2flux2SN2013df}
  \end{subfigure}\\
  \begin{subfigure}{0.49\textwidth}
    \centering
    \includegraphics[width=\textwidth]{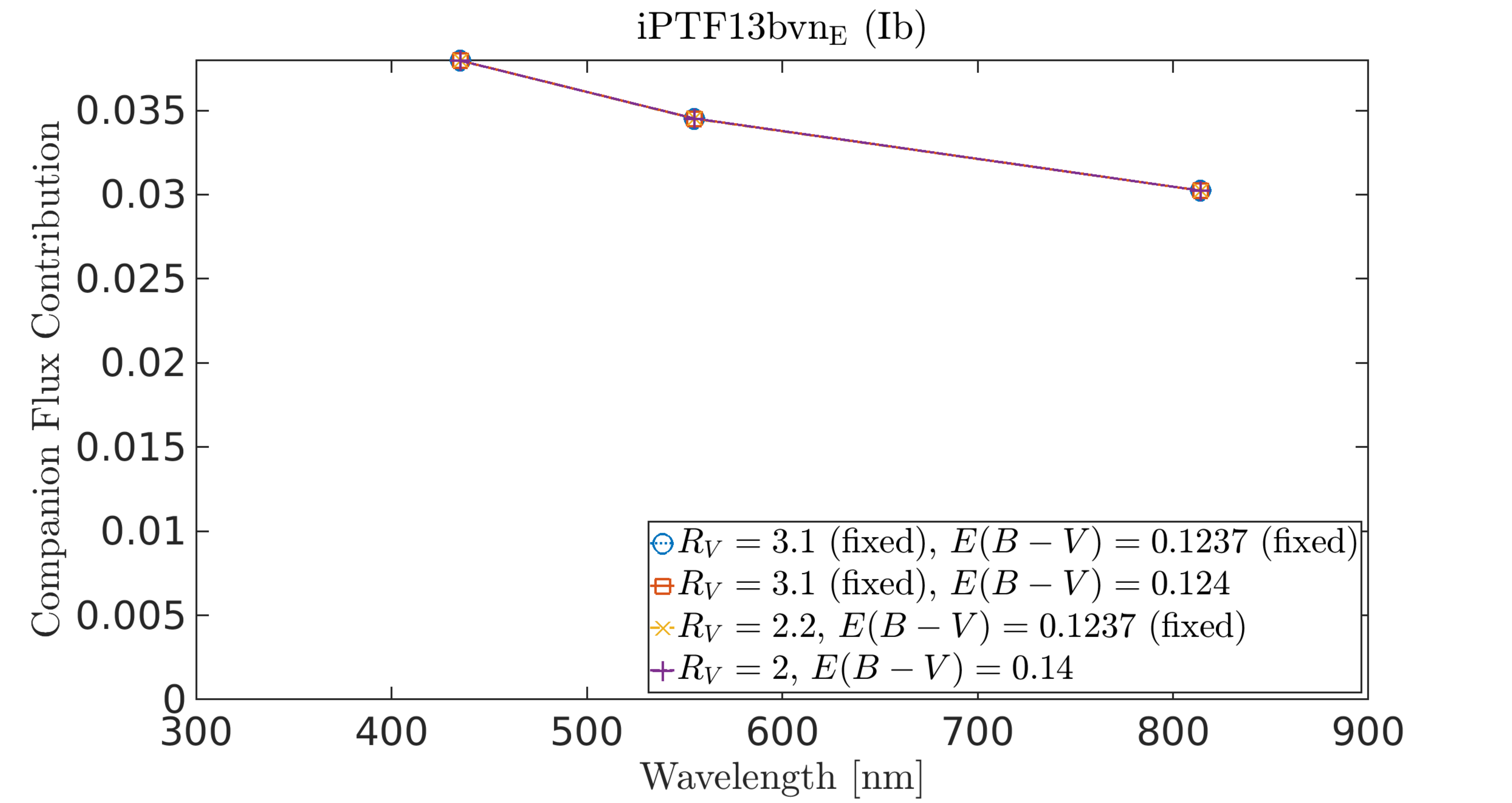}
    \label{fig:chi2flux2iPTF13bvnE}
  \end{subfigure}
    \begin{subfigure}{0.49\textwidth}
    \centering
    \includegraphics[width=\textwidth]{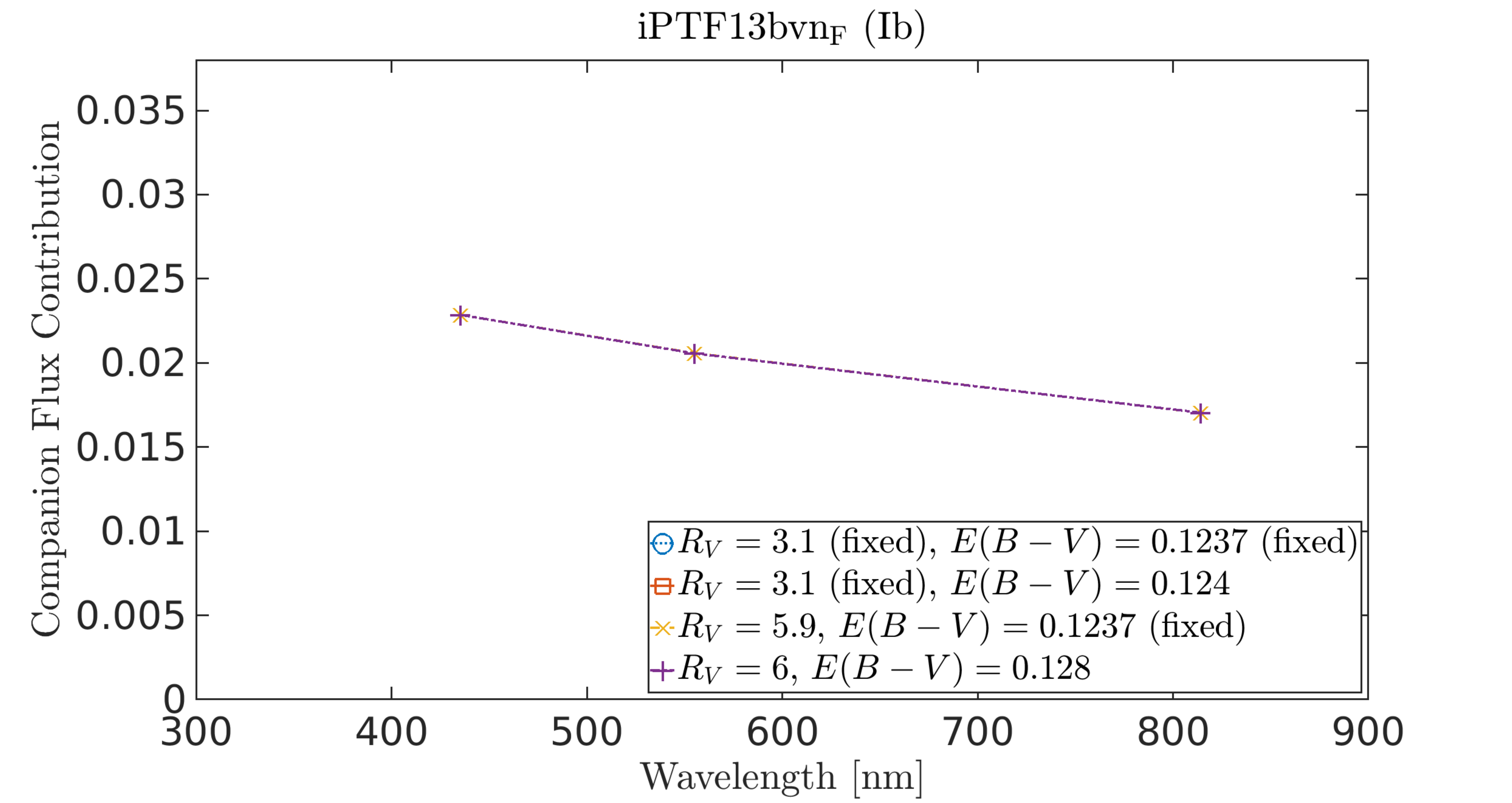}
    \label{fig:chi2flux2iPTF13bvnF}
  \end{subfigure}\\
    \begin{subfigure}{0.49\textwidth}
    \centering
    \includegraphics[width=\textwidth]{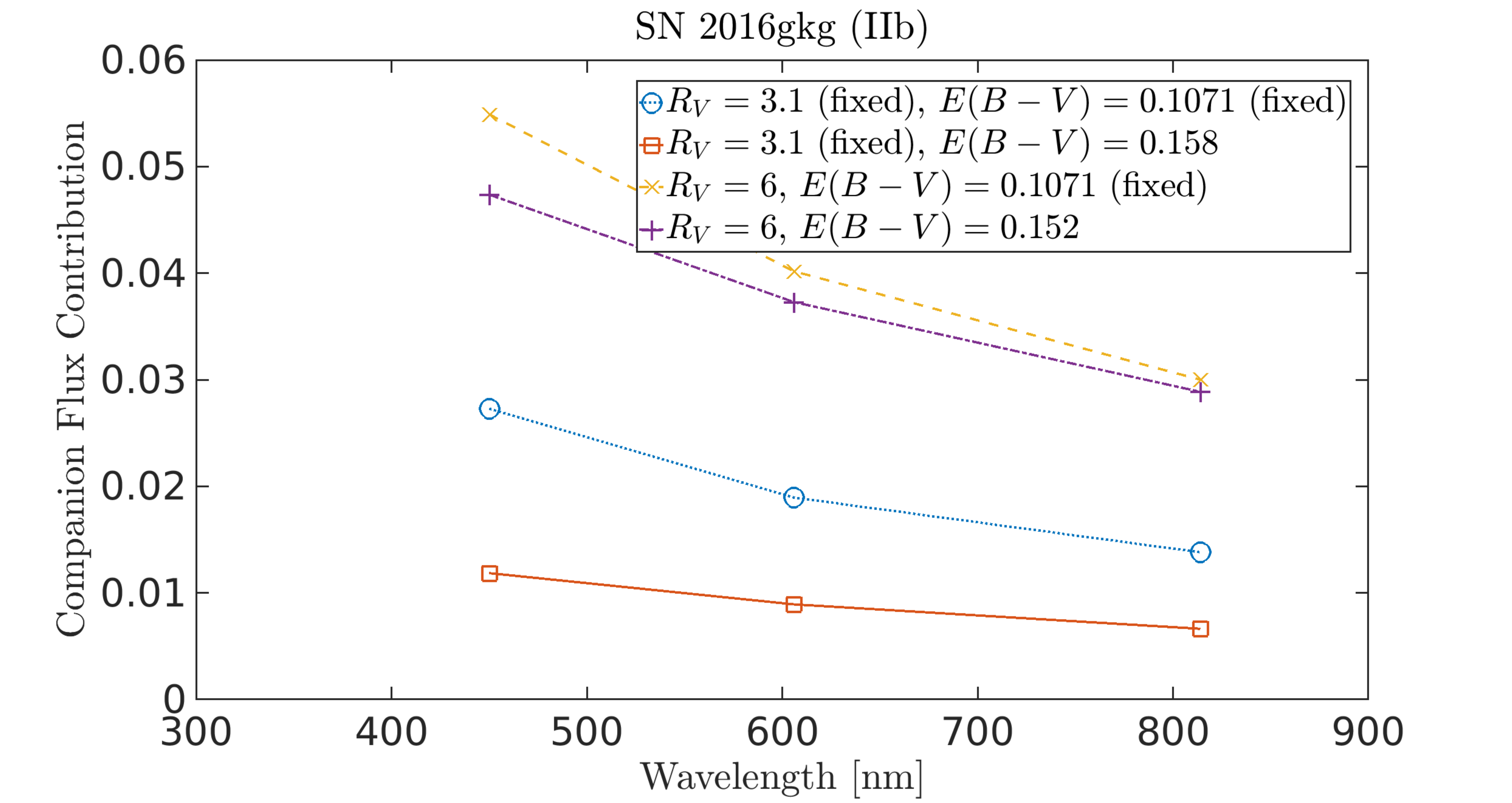}
    \label{fig:chi2flux2SN2016gkg}
  \end{subfigure}
  \begin{subfigure}{0.49\textwidth}
    \centering
    \includegraphics[width=\textwidth]{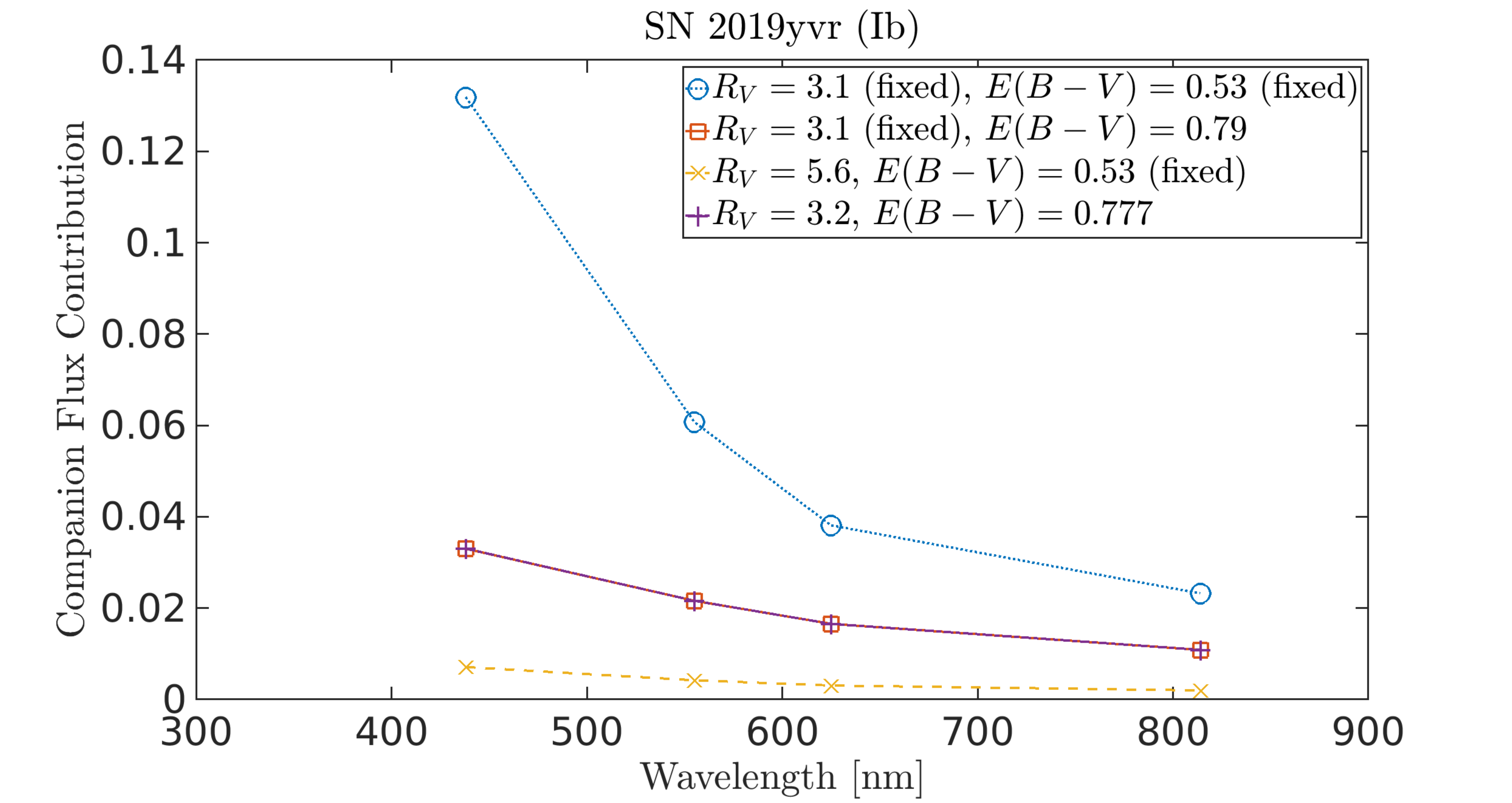}
    \label{fig:chi2flux2SN2019yvr}
  \end{subfigure}\\    
  \caption{The flux contribution from the companion in best-fitting models.} 
  \label{fig:chi2_flux2}
\end{figure*}

Here we present the flux contribution of the companion star in our models. In Figure \ref{fig:chi2_mag2} we present the computed magnitudes (generated from the synthetic photometry) decomposed to the two stellar components in each best-fitting model. The flux contribution of the companion star is usually a few percent of the total flux in all filters, with the highest contribution in the bluest filter. For the case of SN~1993J the contribution in the \textit{U} filter reaches about a third of the total flux (Figure \ref{fig:chi2_flux2}).

\begin{figure*}
  \centering
    \begin{subfigure}{0.49\textwidth}
    \centering
    \includegraphics[width=\textwidth]{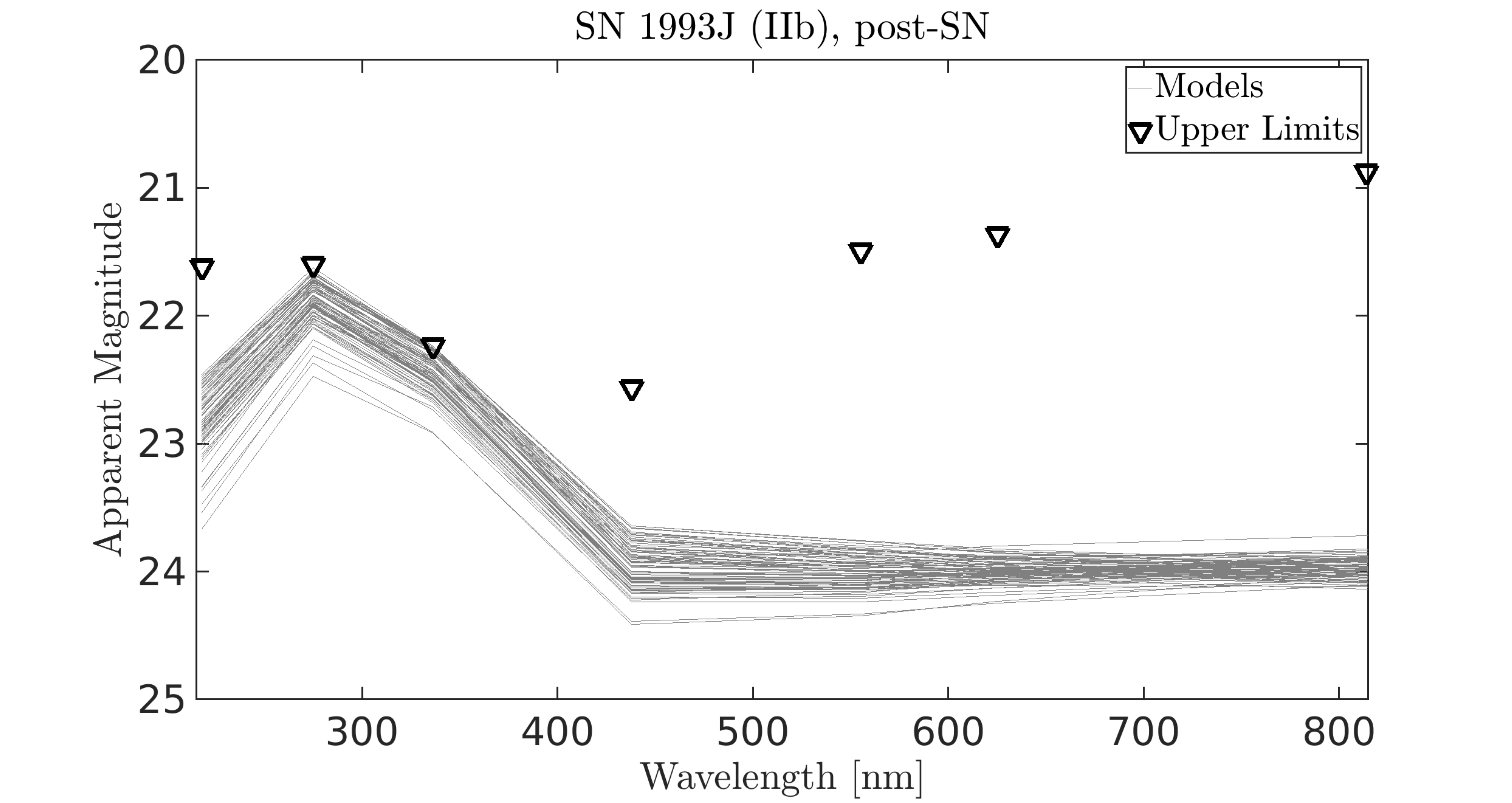}
    \label{fig:MCmag2SN1993J}
  \end{subfigure}
  \begin{subfigure}{0.49\textwidth}
    \centering
    \includegraphics[width=\textwidth]{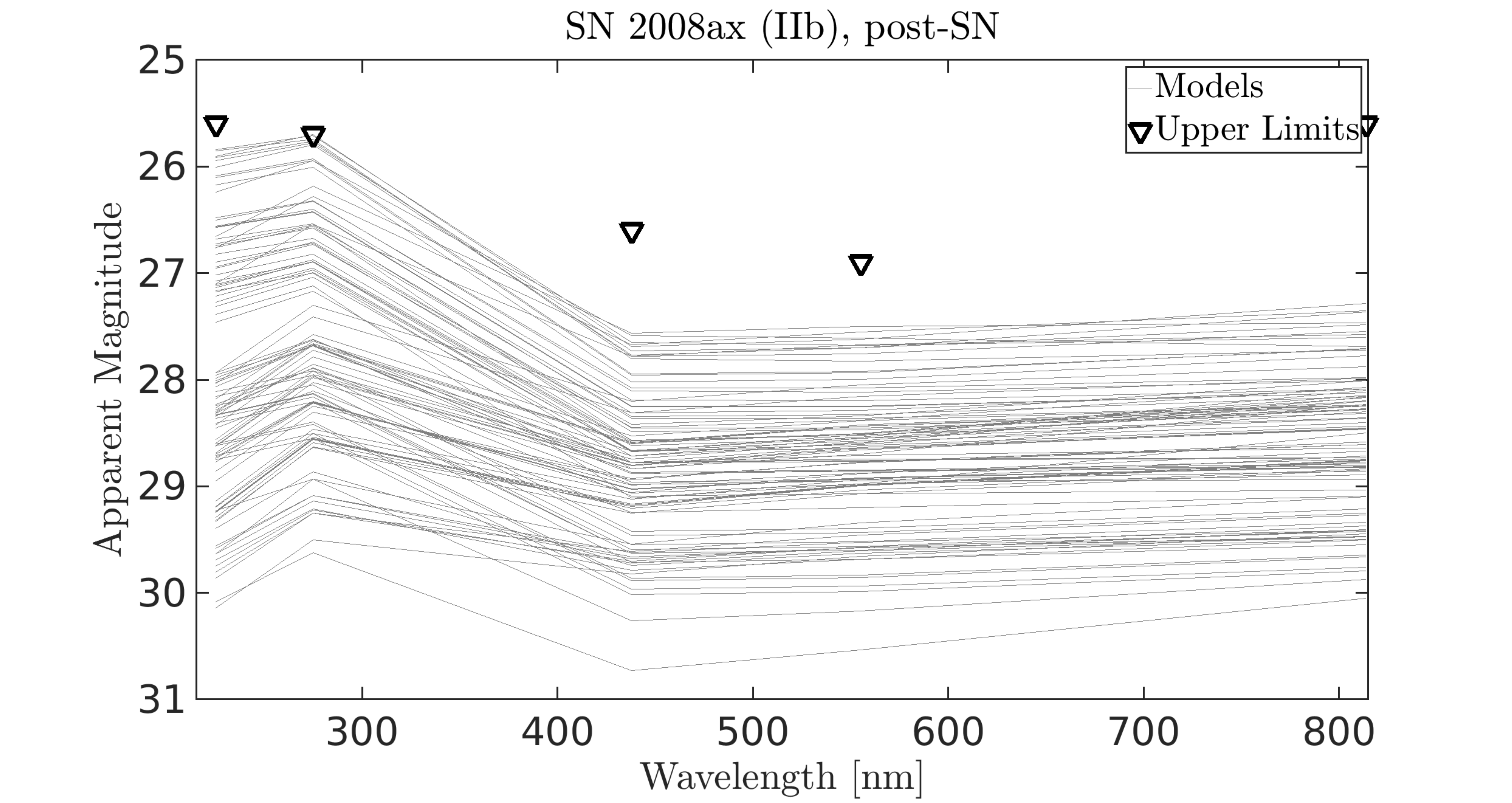}
    \label{fig:MCmag2SN2008ax}
  \end{subfigure}\\
  \begin{subfigure}{0.49\textwidth}
    \centering
    \includegraphics[width=\textwidth]{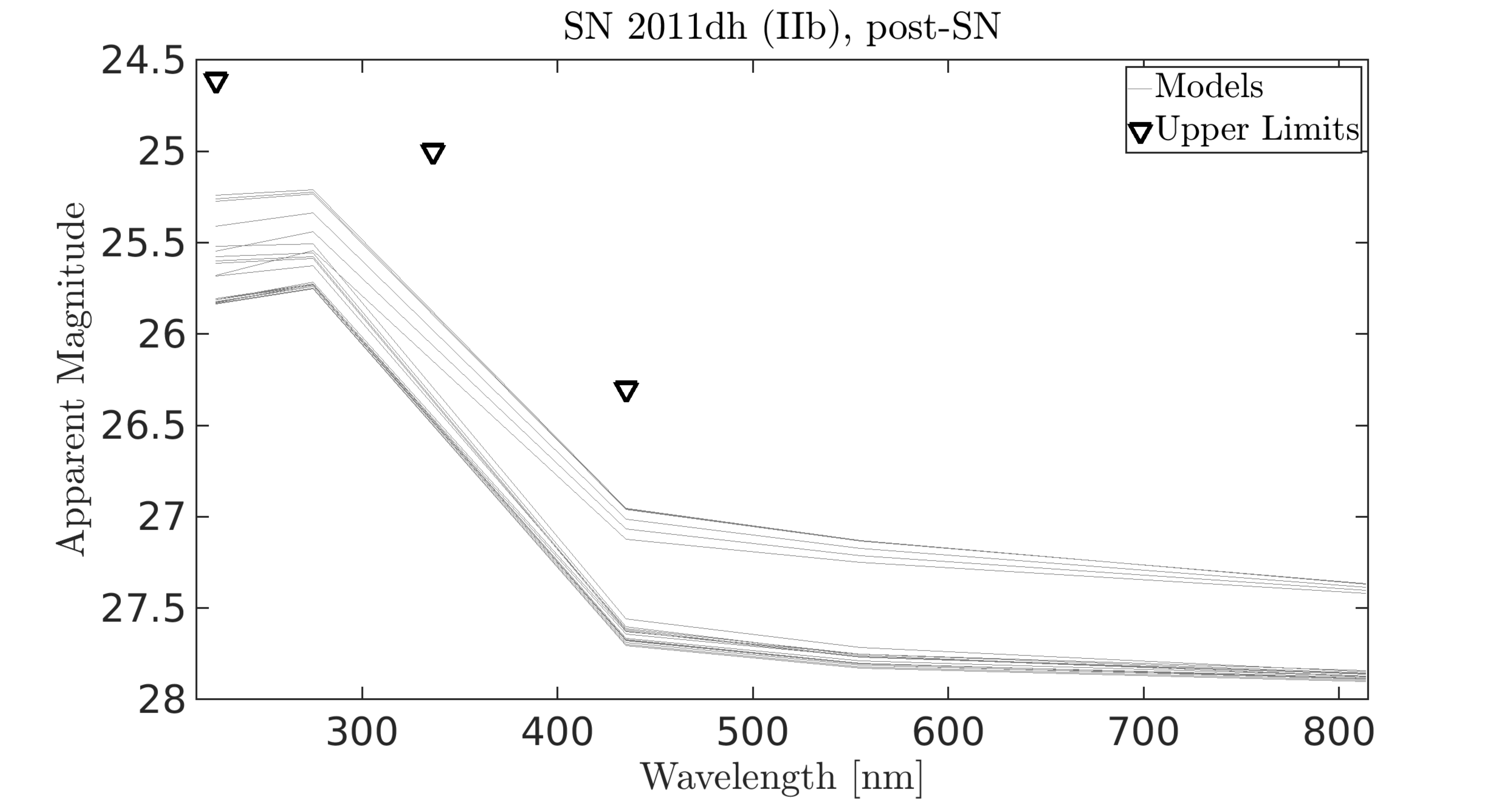}
    \label{fig:MCmag2SN2011dh}
  \end{subfigure}
    \begin{subfigure}{0.49\textwidth}
    \centering
    \includegraphics[width=\textwidth]{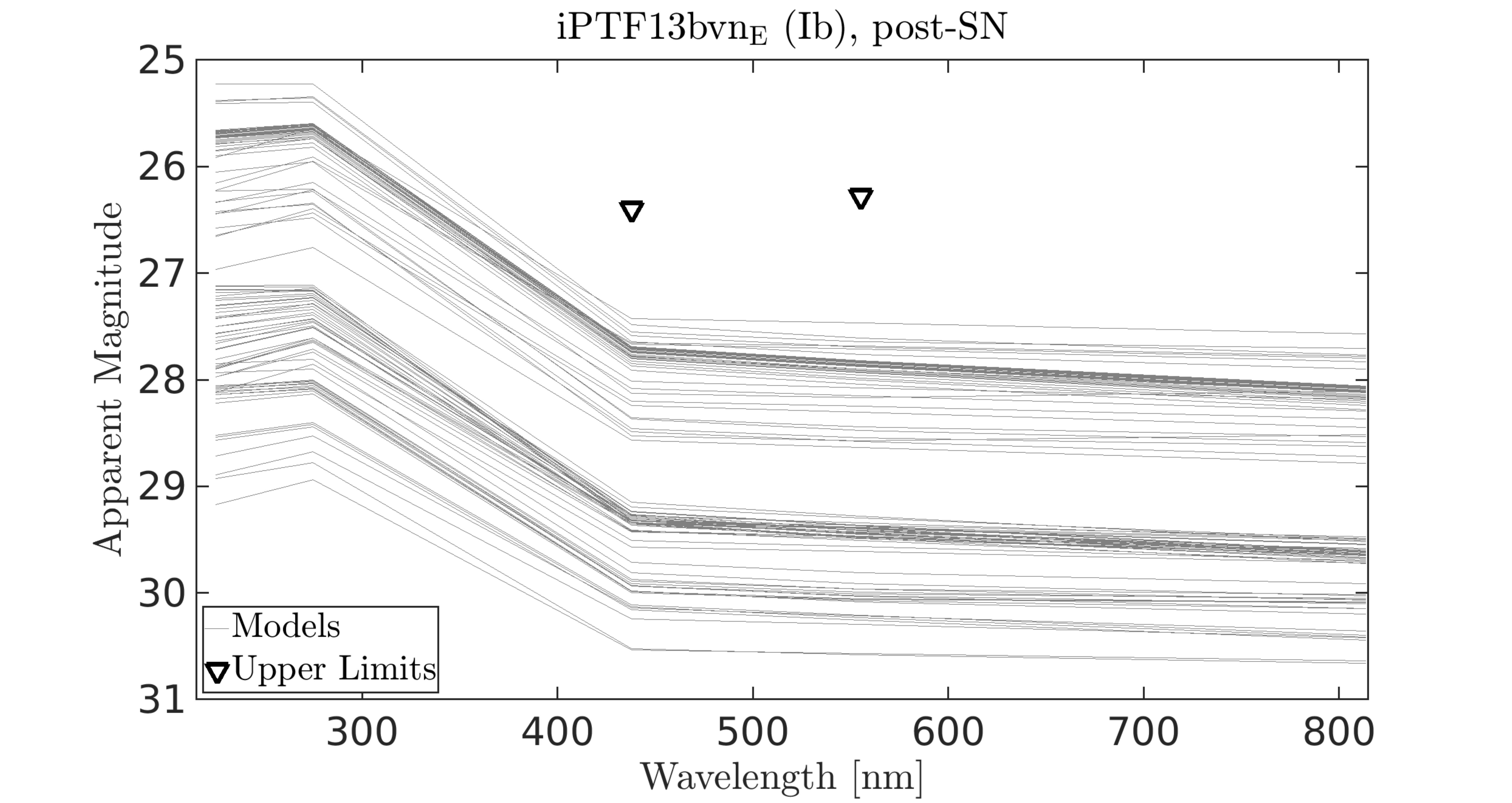}
    \label{fig:MCmag2SNiPTF13bvnE}
  \end{subfigure}\\
    \begin{subfigure}{0.49\textwidth}
    \centering
    \includegraphics[width=\textwidth]{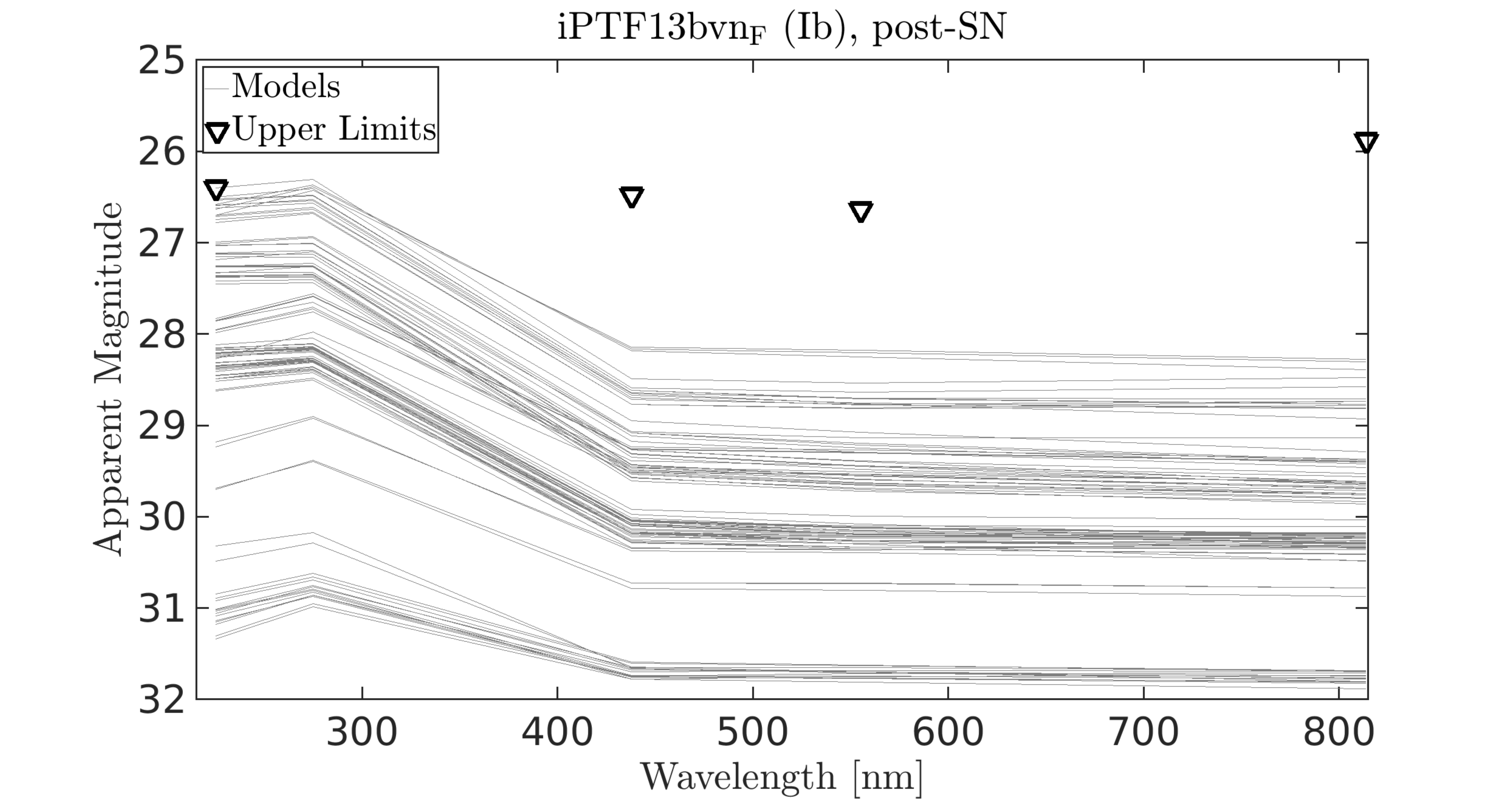}
    \label{fig:MCmag2SNiPTF13bvnF}
  \end{subfigure}
  \begin{subfigure}{0.49\textwidth}
    \centering
    \includegraphics[width=\textwidth]{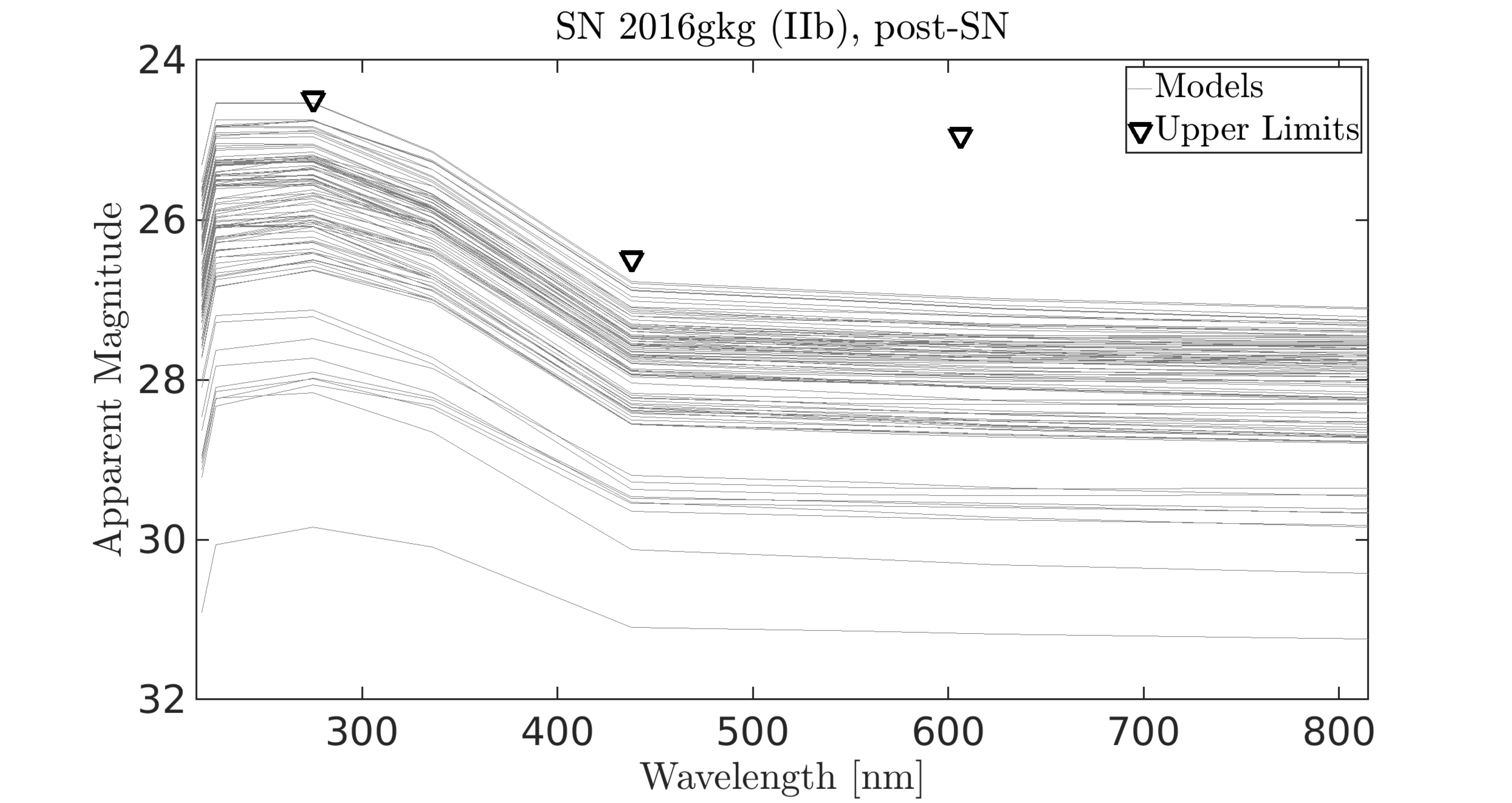}
    \label{fig:MCmag2SNiPTF13bvn}
  \end{subfigure}\\
  \centering
  \caption{Computed magnitudes of the companion star from our Monte Carlo best-fitting models, obtained with variable $E\left(B-V\right)$ and $R_V$, and observational data (where available). Most of the observed late-time flux of SN~1993J is attributed to the (still fading) SN remnant \citep{Foxetal2014}, with perhaps only some of the UV flux resulting from the companion.}
  \label{fig:MC_mag2}
\end{figure*}

In Figure \ref{fig:MC_mag2} we plot the magnitudes of the companion star from our Monte Carlo best-fitting models compared to the available post-SN limits. We do this for the SNe with post-SN measurements listed in Table \ref{tab:postmagnitudes}.

\label{lastpage}

\end{document}